\documentstyle[12pt, preprint, aps, floats, epsfig, amsmath, amsbsy]{revtex}
\tighten
\newcommand{\qr}{q\!\cdot\!r}

\newcommand{\rsl}{r \hspace{-5pt} / }

\newcommand{\psl}{p \hspace{-5pt} / }
\newcommand{\pssl}{p' \hspace{-7pt} / }

\newcommand{\eff}{{\text{eff}}}
\newcommand{\tree}{{\text{tree}}}
\newcommand{\virt}{{\text{virt}}}
\newcommand{\brems}{{\text{brems}}}

\newcommand{\IR}{{\text{IR}}}

\newcommand{\s}{\hat{s}}

\newcommand{\zz}{{z}}

\newcommand{\cb}{\bar{c}}
\renewcommand{\sb}{\bar{s}}
\newcommand{\Order}{{\mathcal O}}
\newcommand{\bra}{\langle}
\newcommand{\ket}{\rangle}

\newcommand{\OSevenTree}{\bra s\ell^+\ell^- |O_7|b\ket_{\text{tree}}}
\newcommand{\OTwoOneLoop}{\bra s\ell^+\ell^- |O_2|b\ket_{\text{1-loop}}}
\newcommand{\OOneOneLoop}{\bra s\ell^+\ell^- |O_1|b\ket_{\text{1-loop}}}

\newcommand{\OFourOneLoop}{\bra s\ell^+\ell^- |O_4|b\ket_{\text{1-loop}}}

\newcommand{\OEleven}{\bra s\ell^+\ell^- |O_{11}|b\ket_{\text{1-loop}}}
\newcommand{\OTwelve}{\bra s\ell^+\ell^- |O_{12}|b\ket_{\text{1-loop}}}
\renewcommand{\to}{\rightarrow}
\renewcommand{\Re}{\mbox{Re}}

\newcommand{\mubeps}{\left( \frac{\mu}{m_b} \right)^{2\epsilon}}

\def\Journal#1#2#3#4{{#1} {\bf #2}, #3 (#4)}

\def\NPB{{\em Nucl. Phys.} B}
\def\PLB{{\em Phys. Lett.}  B}
\def\PRL{\em Phys. Rev. Lett.}
\def\PRD{{\em Phys. Rev.} D}

\begin{document}
\thispagestyle{empty}

\preprint{
    \noindent
    \hfill
    \begin{minipage}[t]{6in}
        \begin{flushright}
            BUTP--01/19     \\
            hep-ph/0109140  \\
            \vspace*{1.0cm}
        \end{flushright}
    \end{minipage}
}

\draft

\title{Calculation of two-loop virtual corrections to
{\bf $b\! \to \! s \ell^+ \ell^-$} \\ in the standard model
\footnote{Work partially supported by Schweizerischer
Nationalfonds and SCOPES program}}
\vspace{2.0cm}
\author{H.H. Asatryan$^a$, H.M. Asatrian$^a$, C. Greub$^b$ and M. Walker$^b$}
\vspace{2.0cm}
\address{
    a) Yerevan Physics Institute, 2 Alikhanyan Br.,
    375036 Yerevan, Armenia; \\
    b) Institut f\"ur Theoretische Physik, Universit\"at Bern, \\
    CH--3012 Bern, Switzerland.
}
\maketitle
\thispagestyle{empty}
\setcounter{page}{0}

\vspace*{1truecm}
\begin{abstract}
    We present in detail the calculation of the virtual $\Order(\alpha_s)$
corrections to the inclusive semi-leptonic rare decay
    $b \to s \ell^+ \ell^-$.
We also include those $\Order(\alpha_s)$ bremsstrahlung
contributions which cancel the infrared and mass
    singularities showing up in the virtual corrections.
In order to avoid large
resonant contributions, we
    restrict the invariant mass squared $s$ of the lepton pair to the range
$0.05\leq s/m_b^2\leq 0.25$.
    The analytic results are represented
    as expansions in the small parameters
$\s = s/m_b^2$, $z = m_c^2/m_b^2$
    and $s/(4 m_c^2)$. The new contributions drastically reduce the
renormalization scale dependence of the decay spectrum.
    For the corresponding branching ratio (restricted to the above $\s$-range)
the renormalization scale uncertainty gets
    reduced from $\sim\pm 13\%$ to $\sim\pm 6.5\%$.
\end{abstract}
\vfill

\setlength{\parskip}{1.2ex}

\section{Introduction}
\label{intro}
Rare $B$-decays are an extremely helpful tool for examining the standard model
(SM) and searching for new physics. Within
the SM, they provide checks on the one-loop structure of the theory and allow one
to retrieve information on the
Cabibbo-Kobayashi-Maskawa (CKM) matrix elements $V_{ts}$ and $V_{td}$, which
cannot be measured directly.

The first measurement of the exclusive rare decay $B\to K^*\gamma$ was obtained
in 1992 by the CLEO collaboration
\cite{CLEOrare1}. Somewhat later, also the inclusive transition
$B\to X_s\gamma$ was observed by the same collaboration
\cite{Alam:1995aw}. Although challenging for the experimentalists, the
inclusive decays are clean from the
theoretical point of view, as they are well approximated by the underlying
partonic transitions, up to small and calculable power corrections
which start at $\Order(\Lambda_{\rm{QCD}}^2/m_b^2)$ \cite{Bigi1,Bigi2}.

The measured photon energy spectrum \cite{CLEO01} and the branching ratio
for the decay $B\to X_s\gamma$ \cite{Alam:1995aw,ALEPH,BELLE}
are in good agreement with the next-to-leading logarithmic (NLL)
standard model predictions
(see e.g. \cite{spektrum,matching,Giudice98,Chetyrkin,matel,matelnew,Gambino01}).
Consequently, the decay $B \to X_s \gamma$ places stringent constraints on
the extensions of the SM,
such as two-Higgs doublet models \cite{Giudice98,BG98,Asatryan:2001kt},
supersymmetric models
\cite{Bertolini91,Giudice99,Misiaksusy,BGHW00,BGH01,Asatrian:1999sg}, etc.

$B\to X_s\ell^{+}\ell^{-}$ is another interesting rare decay mode
which has been extensively considered in the
literature in the framework of the SM and its extensions (see e.g
\cite{Grinstein:1989,Misiak:1993bc,ALI95,DESPHANDE,CHO,Buras:1995dj}).
This decay has not been observed so far,
but it is expected to be measured at the operating $B$-factories
after a few years of data taking (for upper limits on its branching ratio
we refer to \cite{Glenn:1998gh,Abe:2001}).
The measurement of various kinematical distributions of the decay
$B\to X_s \ell^{+} \ell^{-}$, combined with
improved data on $B\to X_s\gamma$,
will tighten the constraints
on the extensions of the SM or perhaps even reveal some deviations.

The main problem of the theoretical description of $B\to X_s\ell^{+}\ell^{-}$
is due to the long-distance contributions
{}from $\bar{c}c$ resonant states. When the invariant mass
$\sqrt{s}$ of the lepton pair is close to the
mass of a resonance, only model dependent predictions for such long distance
contributions are available today. It is
therefore unclear whether the theoretical uncertainty can be reduced to less
than $\pm 20\%$ when
integrating over these domains \cite{Ligeti:1996yz}.

However, restricting $\sqrt s$ to a region below the resonances, the long
distance effects are under control.
The corrections to the pure perturbative picture can be analyzed within the
heavy quark effective theory (HQET). In particular,
all available studies indicate that for the region $0.05<\hat{s}=s/m_b^2<0.25$
the non-perturbative effects are
below 10$\%$
\cite{Falk:1994dh,Ali:1997bm,chen:1997,Buchalla:1998ky,Buchalla:1998mt,Krueger:1996}.
Consequently,
the differential decay rate for $B \to X_s \ell^+ \ell^-$
can be precisely predicted in this region using
renormalization group improved perturbation theory.
It was pointed out in the literature that the differential decay rate and the
forward-backward asymmetry are particularly sensitive to
new physics in this kinematical window \cite{Ball,Lunghi,Silvestrini}.

Calculations of the
next-to-leading logarithmic (NLL) corrections to the process
$B \to X_s \ell^+ \ell^-$ have
been performed in refs. \cite{Misiak:1993bc} and \cite{Buras:1995dj}.
It turned out that the NLL result suffers from a relatively large
($\pm 16\%$) dependence on the matching scale $\mu_W$.
To reduce it, next-to-next-to leading (NNLL)
corrections to the Wilson coefficients were recently
calculated by Bobeth et al. \cite{Bobeth:2000mk}.
This required a two-loop matching calculation
of the effective theory to the full
SM theory, followed by a renormalization group evolution of the
Wilson coefficients, using up to three-loop anomalous dimensions
\cite{Bobeth:2000mk,Chetyrkin}.
Including these NNLL corrections to the Wilson coefficients, the
matching scale dependence is indeed removed to a large extent.

As pointed out in ref. \cite{Bobeth:2000mk}, this partially NNLL result
suffers from a relatively large ($\sim \pm 13\%$)
renormalization scale ($\mu_b$) dependence ($\mu_b \sim \Order(m_b)$) which,
interestingly enough,
is even larger than that of the pure NLL result.
Recently we showed in a letter \cite{AAGW:letter} that the NNLL
corrections to the matrix elements of the effective Hamiltonian
drastically reduce the renormalization scale dependence.
The aim of the current
paper is to present a detailed description of the rather involved
calculations and to extend the phenomenological part.
We will discuss in particular the methods which allowed us to tackle
with the most involved part, viz. the calculation of the
$\Order(\alpha_s)$ two-loop virtual corrections
to the matrix elements of the operators $O_1$ and $O_2$.
We also comment on
the $\Order(\alpha_s)$ one-loop corrections to
$O_7$--$O_{10}$. Furthermore, we include those bremsstrahlung contributions
which are needed
to cancel infrared and collinear singularities
in the virtual corrections. As shown already in \cite{AAGW:letter},
the new contributions reduce the renormalization
scale dependence
{}from $\sim \pm 13\%$ to $\sim \pm 6.5\%$.

The paper is organized as follows: In chapter \ref{section:eff} we review the
theoretical framework. Our results for the
virtual $\Order(\alpha_s)$ corrections to the matrix elements
of  the operators $O_1$ and $O_2$ are presented  in chapter III,
whereas the corresponding corrections to the matrix elements of $O_{7}$,
$O_8$, $O_9$ and $O_{10}$ are given in chapter IV.
Chapter V is devoted to the bremsstrahlung corrections. The combined corrections
(virtual and
bremsstrahlung) to $b \to s \ell^+ \ell^-$ are discussed in chapter VI. Finally, in
chapter VII, we analyze the invariant
mass distribution of the lepton pair in the range $0.05 \le \s \le 0.25$.

%
%
\section{Effective Hamiltonian}
\label{section:eff}
The appropriate framework for studying QCD corrections to rare $B$-decays in a
systematic way is the
effective Hamiltonian technique. For the specific decay channels $b\rightarrow
s\ell^+\ell^-$ ($\ell=\mu$, $e$),
the effective Hamiltonian is derived by integrating out the heavy degrees of
freedom. In the context of the standard model,
these are the $t$-quark, the $W$-boson and the $Z^0$-boson. Due to the unitarity
of the CKM matrix, the CKM structure factorizes when
neglecting the combination $V_{us}^* V_{ub}$. The effective Hamiltonian then
reads
\begin{equation} \label{Heff}
    {\cal H}_{\eff} =  - \frac{4 G_F}{\sqrt{2}} V_{ts}^* V_{tb} \sum_{i=1}^{10}
C_i(\mu) O_i(\mu)\, .
\end{equation}
Following ref.\cite{Bobeth:2000mk}, we choose the operator basis as follows:
\begin{equation} \label{oper}
\begin{array}{rclrcl}
    O_1    & = & (\bar{s}_L \gamma_\mu T^a c_L) (\bar{c}_L \gamma^\mu T^a b_L)
\, ,&
    \vspace{0.4cm}
    O_2    & = & (\bar{s}_L \gamma_\mu c_L) (\bar{c}_L \gamma^\mu b_L) \, ,\\
    \vspace{0.2cm}
    O_3    & = &  (\bar{s}_L \gamma_{\mu}  b_{L }) \sum_q (\bar{q}\gamma^{\mu}
q) \, ,&
    \vspace{0.2cm}
    O_4    & = &  (\bar{s}_{L}\gamma_{\mu} T^a b_{L }) \sum_q
(\bar{q}\gamma^{\mu} T^a q) \, ,\\
    \vspace{0.2cm}
    O_5    & = & (\bar{s}_L\gamma_\mu \gamma_\nu \gamma_\rho b_L)
                    \sum_q(\bar{q} \gamma^\mu \gamma^\nu \gamma^\rho q) \, ,&
    \vspace{0.2cm}
    O_6   & = & (\bar{s}_L\gamma_\mu \gamma_\nu \gamma_\rho T^a b_L)
                    \sum_q(\bar{q} \gamma^\mu \gamma^\nu \gamma^\rho T^a q) \,
,\\
    \vspace{0.2cm}
    O_7    & = & \frac{e}{g_s^2} m_b (\bar{s}_{L} \sigma^{\mu\nu} b_{R})
F_{\mu\nu} \, ,&
    \vspace{0.2cm}
    O_8    & = & \frac{1}{g_s} m_b (\bar{s}_{L} \sigma^{\mu\nu} T^a b_{R})
G_{\mu\nu}^a \, ,\\
    \vspace{0.2cm}
    O_9    & = & \frac{e^2}{g_s^2}(\bar{s}_L\gamma_{\mu} b_L)
\sum_l(\bar{l}\gamma^{\mu}l) \, ,&
    O_{10} & = & \frac{e^2}{g_s^2}(\bar{s}_L\gamma_{\mu} b_L)
\sum_l(\bar{l}\gamma^{\mu} \gamma_{5}l) \, ,
\end{array}
\end{equation}
where the subscripts $L$ and $R$ refer to left- and right- handed components of
the fermion fields.

The factors $1/g_s^2$ in the definition of the operators $O_7$, $O_9$ and
$O_{10}$, as well as the factor
$1/g_s$ present in $O_8$ have been chosen by Misiak \cite{Misiak:1993bc} in
order to simplify the organization of the calculation: With
these definitions, the one-loop anomalous dimensions (needed for a leading
logarithmic (LL) calculation) of the
operators $O_i$ are all proportional to $g_s^2$, while two-loop anomalous
dimensions (needed for a next-to-leading
logarithmic (NLL) calculation) are proportional to $g_s^4$, etc..

After this important remark we now outline the principal steps which lead to a
LL, NLL, NNLL prediction for the decay
amplitude for $b \to s \ell^+ \ell^-$:
\begin{enumerate}
\item
    A matching calculation between the full SM theory and the effective theory
has to be performed
    in order to determine the Wilson coefficients $C_i$ at the high scale
$\mu_W\sim m_W,m_t$. At this scale, the
    coefficients can be worked out in fixed order perturbation theory, i.e. they
can be expanded in $g_s^2$:
    \begin{equation}
        C_i(\mu_W) = C_i^{(0)}(\mu_W)
        + \frac{g_s^2}{16\pi^2} C_i^{(1)}(\mu_W)
        + \frac{g_s^4}{(16\pi^2)^2} C_i^{(2)}(\mu_W) + \Order(g_s^6) \, .
    \end{equation}
    At LL order, only $C_i^{(0)}$ are needed, at NLL order also $C_i^{(1)}$,
etc.. While the coefficient $C_7^{(2)}$, which is
    needed for a NNLL analysis, is known for quite some time
    \cite{matching}, $C_{9}^{(2)}$ and  $C_{10}^{(2)}$
    have been calculated only recently \cite{Bobeth:2000mk}
    (see also \cite{Buchalla:1999ba}).
\item
    The renormalization group equation (RGE) has to be solved in order to get
the Wilson coefficients at the low scale
    $\mu_b \sim m_b$. For this RGE step the anomalous dimension matrix to the
relevant order in $g_s$ is required, as
    described above. After these two steps one can decompose the Wilson
coefficients $C_i(\mu_b)$ into a LL, NLL
    and NNLL part according to
    \begin{equation}
        \label{wilsondecomplow}
        C_i(\mu_b) = C_i^{(0)}(\mu_b) + \frac{g_s^2(\mu_b)}{16\pi^2}
C_i^{(1)}(\mu_b)
        + \frac{g_s^4(\mu_b)}{(16\pi^2)^2} C_i^{(2)}(\mu_b) +
     \Order(g_s^6) \, .
    \end{equation}
\item
    In order to get the decay amplitude, the matrix elements $\bra s \ell^+
\ell^-|O_i(\mu_b)|b \ket$ have to be
    calculated. At LL precision, only the operator $O_9$ contributes, as this
operator is the only one which at the same
    time has a Wilson coefficient starting at lowest order and an explicit
$1/g_s^2$ factor in the definition. Hence, at
    NLL precision, QCD corrections (virtual and bremsstrahlung) to the matrix
element of $O_9$ are needed. They have
    been calculated a few years ago \cite{Misiak:1993bc,Buras:1995dj}. At NLL
precision, also the other operators start
    contributing, viz. $O_7(\mu_b)$ and $O_{10}(\mu_b)$ contribute at tree-level
and the four-quark operators
    $O_1,...,O_6$ at one-loop level. Accordingly, QCD corrections to the latter
matrix elements are needed for a NNLL
    prediction of the decay amplitude.
\end{enumerate}

The formally leading term $\sim (1/g_s^2) C_9^{(0)}(\mu_b)$ to the amplitude for
$b \to s \ell^+ \ell^-$ is smaller than
the NLL term $\sim (1/g_s^2) [g_s^2/(16 \pi^2)] \, C_9^{(1)}(\mu_b)$
\cite{Grinstein:1989}. We adapt our systematics to
the numerical situation and treat the sum of these two terms as a NLL
contribution. This is, admittedly some abuse of
language, because the decay amplitude then starts out with a term which is
called NLL.

As pointed out in step 3), $\Order(\alpha_s)$ QCD corrections to the matrix
elements $\bra s \ell^+ \ell^-|O_i(\mu_b)|b \ket$
have to be calculated in order to obtain the NNLL prediction for the decay
amplitude. In the present paper we
systematically evaluate virtual corrections of order $\alpha_s$ to the
matrix elements of $O_1$, $O_2$, $O_7$,
$O_8$, $O_9$ and $O_{10}$.
As the Wilson coefficients of the gluonic penguin operators $O_3,...,O_6$ are
much smaller than
those of $O_1$ and $O_2$, we neglect QCD corrections to their matrix elements.
As discussed in more detail later, we also
include those bremsstrahlung diagrams which are needed to cancel infrared and
collinear singularities from the virtual
contributions. The complete bremsstrahlung corrections, i.e. all the finite
parts, will be given elsewhere \cite{AAGW:brems}.
We anticipate that the QCD corrections calculated in the present
paper substantially reduce the scale
dependence of the NLL result.
%
%
\section{ Virtual  $\Order(\alpha_s)$ corrections to the current-current \\
                    operators $O_1$ and $O_2$}
\label{section:virtO1O2}
In this chapter we present a detailed calculation of the virtual
$\Order(\alpha_s)$ corrections to the matrix elements of the
current-current operators $O_1$ and $O_2$. Using the naive dimensional
regularization scheme (NDR) in $d=4-2\epsilon$
dimensions, both, ultraviolet and infrared singularities show up as
$1/\epsilon^n$ poles ($n=1,2$). The ultraviolet
singularities cancel after including the counterterms. Collinear singularities
are regularized by retaining a finite strange
quark mass $m_s$. They are cancelled together with the infrared singularities at
the level of the decay width, taking
the bremsstrahlung process $b\rightarrow s\ell^+\ell^-g$ into account.
Gauge invariance implies that the QCD corrected matrix elements of the operators
$O_i$ can be written as
\begin{equation}
    \label{formdef}
    \bra s\ell^+\ell^-|O_i|b\ket =
    \hat{F}_i^{(9)} \bra O_9 \ket_{\text{tree}} +
    \hat{F}_i^{(7)} \bra O_7 \ket_{\text{tree}} \, ,
\end{equation}
where  $\bra O_9 \ket_{\text{tree}}$ and  $\bra O_7 \ket_{\text{tree}}$
are the tree-level matrix elements of $O_9$ and $O_7$, respectively.
Equivalently, we may write
\begin{equation}
    \label{formdeftilde}
    \bra s\ell^+\ell^-|O_i|b\ket = - \frac{\alpha_s}{4\pi} \,
   \left[  F_i^{(9)} \bra \widetilde{O}_9 \ket_{\text{tree}} +
    F_i^{(7)} \bra \widetilde{O}_7 \ket_{\text{tree}} \right] \, ,
\end{equation}
where the operators $\widetilde{O}_7$ and $\widetilde{O}_9$
are defined as
\begin{equation}
\widetilde{O}_7 = \frac{\alpha_s}{4 \pi} \, O_7 \, ; \quad
\widetilde{O}_9 = \frac{\alpha_s}{4 \pi} \, O_9 \, .
\end{equation}
We present the final results for the QCD
corrected matrix elements in the form of eq. (\ref{formdeftilde}).
%
%
\subsection{Regularized $\Order(\alpha_s)$ contribution of $O_1$ and $O_2$}
\begin{figure}[t]
    \begin{center}
    \leavevmode
    \includegraphics[height=6cm]{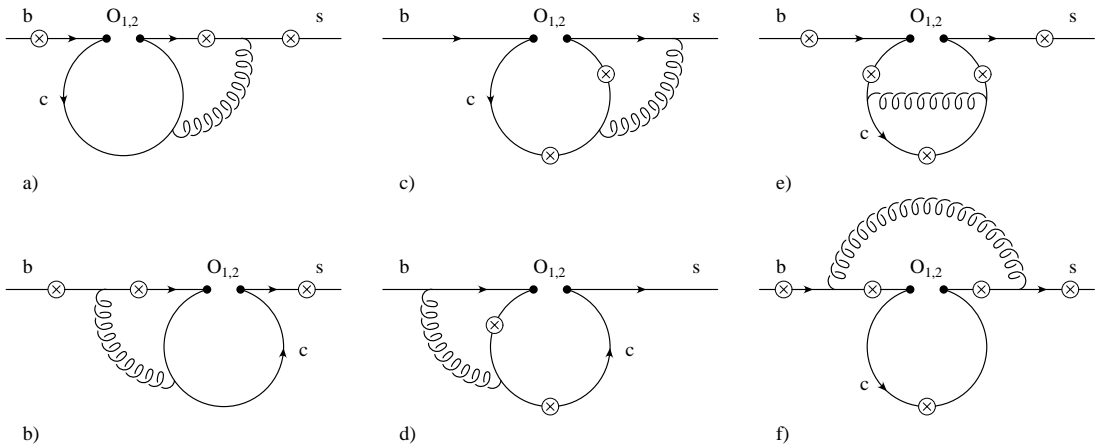}
    \vspace{2ex}
    \caption{Complete list of two-loop Feynman diagrams for
     $b \to s \gamma^*$
associated with the operators $O_1$
    and $O_2$. The fermions ($b$, $s$ and $c$ quarks) are represented by solid
lines, whereas the curly lines represent
    gluons. The circle-crosses denote the possible locations where the virtual
photon (which then splits into a lepton pair) is emitted.}
    \label{fig:1}
    \end{center}
\end{figure}

The full set of the diagrams contributing to the matrix elements
\begin{equation}
M_i=\bra s\ell^+\ell^-|O_i|b \ket \quad (i=1,2)
\end{equation}
at $\Order(\alpha_s)$ is shown in Fig. \ref{fig:1}.
%
As indicated in this figure, the diagrams associated with $O_1$ and
$O_2$ are topologically identical. They differ only by the color
structure. While the matrix elements of the operator $O_2$ all
involve the color structure
\begin{equation}
\nonumber
\sum_a T^a T^a = C_F {\bf 1};\quad
C_F=\frac{N_c^2-1}{2N_c},
\end{equation}
there are two possible color structures for the corresponding diagrams
of $O_1$, viz.
\begin{equation}\nonumber
\tau_1 = \sum_{a,b} T^a T^b T^a T^b
\quad\text{and}
\quad \tau_2 = \sum_{a,b} T^a T^b T^b T^a.
\end{equation}
The structure $\tau_1$ appears in diagrams \ref{fig:1}a)-d) and
$\tau_2$ in diagrams \ref{fig:1}e) and \ref{fig:1}f). Using the relation
\[
    \sum_a \, T^a_{\alpha\beta} T^a_{\gamma\delta} = -\frac{1}{2 N_c}\,
\delta_{\alpha\beta} \delta_{\gamma\delta} +
        \frac{1}{2}\, \delta_{\alpha\delta} \delta_{\beta\gamma} ,
\]
we find that $\tau_1=C_{\tau_1}{\bf 1}$ and
$\tau_2=C_{\tau_2}{\bf 1}$ with
\[
    C_{\tau_1} = - \frac{N_c^2-1}{4 N_c^2} \quad\text{and}\quad
    C_{\tau_2} =  \frac{\left(N_c^2-1\right)^2}{4 N_c^2} \, .
\]
Inserting $N_c=3$, the color factors are $C_F=\frac{4}{3},
C_{\tau_1}=-\frac{2}{9}$ and $C_{\tau_2}=\frac{16}{9}$.
The contributions from $O_1$ are obtained by multiplying those
{}from $O_2$ by the appropriate factors, i.e. by
$C_{\tau_1} / C_F = -1/6$ and $C_{\tau_2} / C_F = 4/3$, respectively.
In the following descriptions of the individual diagrams
we therefore restrict ourselves to those associated with the
operator $O_2$.

In the current paper we use the $\overline{\text{MS}}$
renormalization scheme which is technically
implemented by introducing the renormalization scale in
the form $\overline{\mu}^{\,2} = \mu^2 \exp(\gamma_E)/(4\pi)$,
followed by  minimal subtraction. The precise
definition of the evanescent operators, which is necessary to
fully specify the renormalization scheme, will be given later.
The remainder of this section is divided into 8 subsections.
Subsections 1-6 deal with the diagrams \ref{fig:1}a)-d)
which are calculated by means of Mellin-Barnes techniques
\cite{Smirnov_book}.
Subsection 7 is devoted to the diagrams  \ref{fig:1}e) which
are evaluated by using the heavy mass expansion procedure
\cite{Smirnov}.
Among the diagrams
\ref{fig:1}f) only the one where the virtual photon is emitted
{}from the charm quark line is non-zero. As it factorizes into
two one-loop diagrams, its calculation is straightforward and
does not require to be discussed in detail. It is, however, worth
mentioning already at this point that it is convenient to omit
this diagram in the discussion of the matrix elements of $O_1$
and $O_2$ and to take it into account together with the virtual
corrections to $O_9$. Finally, in subsection 8, we give the
results for the dimensionally regularized matrix elements
$\bra s\ell^+\ell^-|O_i|b\ket \,(i=1,2)$.
\subsubsection{The building blocks $I_\beta$ and $J_{\alpha\beta}$}
\begin{figure}
\begin{center}
    \includegraphics[height = 4cm]{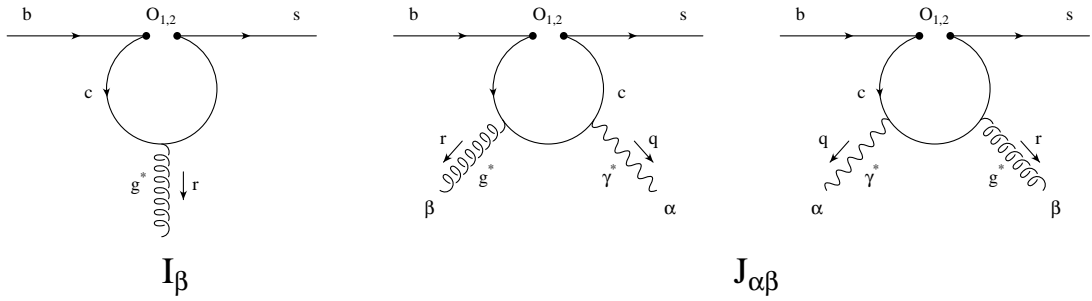}
\end{center}
\caption{The building blocks $I_{\beta}$ and $J_{\alpha\beta}$ which are used
for the calculation of the two-loop diagrams
\ref{fig:1}a)-d). The curly and wavy lines represent gluons and photons,
respectively.\label{fig:2}}
\end{figure}

For the calculation of diagrams \ref{fig:1}a) - d) it is advisable to evaluate
the building blocks $I_\beta$ and
$J_{\alpha\beta}$ first. The corresponding diagrams are depicted in Fig.
\ref{fig:2}.
%
%
After performing a straightforward Feynman parameterization followed by the
integration over the loop momentum,
the analytic expression for the building block $I_{\beta}$ reads
\begin{multline}
\label{build1}
    I_\beta = - \frac{g_s}{4 \pi^2} \, \Gamma(\epsilon) \, \mu^{2 \epsilon} \,
e^{\gamma_E \epsilon} \, (1-\epsilon) \,
    e^{i \pi \epsilon} \, \left(r_{\beta} r\hspace{-5pt}/ - r^2 \gamma_\beta
    \right) \, L \, \frac{\lambda}{2} \times \\
    \int_0^1 \! dx \, \left[x(1-x)\right]^{1-\epsilon} \,
    \left[ r^2 - \frac{m_c^2}{x(1-x)} + i \delta \right]^{-\epsilon} \, ,
\end{multline}
where $r$ is the momentum of the virtual gluon emitted from the $c$-quark loop.
The term $i\delta$ is the
"$i \epsilon$-prescription". In the full two-loop diagrams, the free index
$\beta$ will be contracted with the
corresponding gluon propagator. Note, that $I_\beta$ is gauge invariant in the
sense that $r^\beta I_\beta = 0$.

The building block $J_{\alpha\beta}$ is somewhat more complicated.
Using the notation introduced by Simma and Wyler \cite{SW90}, it reads
\begin{multline}
\label{bb2}
    J_{\alpha\beta} = \frac{eg_sQ_u}{16\pi^2}
    \left[
        E(\alpha,\beta,r) \Delta i_5 +
        E(\alpha,\beta,q) \Delta i_6 -
        E(\beta,r,q)\frac{r_{\alpha}}{\qr} \Delta i_{23}
    \right.\\
    \left.
        - E(\alpha,r,q)\frac{r_{\beta}}{\qr} \Delta i_{25}
        - E(\alpha,r,q)\frac{q_{\beta}}{\qr} \Delta i_{26}
        - E(\beta,r,q)\frac{q_{\alpha}}{\qr}\Delta i_{27}
    \right]
    \, L \, \frac{\lambda}{2} \, ,
\end{multline}
where $q$ and $r$ denote the momenta of the (virtual) photon and gluon,
respectively. The indices $\alpha$ and $\beta$
will be contracted with the propagators of the photon and the gluon,
respectively.
The matrix $E(\alpha,\beta,r)$ is defined as
\begin{equation}
    E(\alpha,\beta,r) = \frac{1}{2} (\gamma_{\alpha}\gamma_{\beta}\rsl -
\rsl\gamma_{\beta}\gamma_{\alpha})
\end{equation}
and the dimensionally regularized quantities $\Delta i_k$
occurring in eq. (\ref{bb2}) read
\begin{eqnarray}
    \Delta i_5  &=& 4\, B^+ \int_S\! dx\,dy \left[ 4 (\qr)\, x\, y \,(1-x) \epsilon
    + r^2 \,x\, (1-x) (1-2\,x) \epsilon
    \right. \nonumber \\ && \hspace{1cm} \left.
    + q^2\, y (2 - 2\,y + 2\,x\,y - x) \epsilon + (1-3x)C \right] \,
    C^{-1-\epsilon} \, , \nonumber \\
    \Delta i_6 &=& 4\, B^+ \int_S\! dx\,dy \left[ -4 (\qr)\, x\, y \,(1-y) \epsilon
    - q^2\, y\, (1-y) (1-2\,y) \epsilon
    \right. \nonumber \\ && \hspace{1cm} \left.
    -r^2\, x\, (2 - 2\,x + 2\,x\,y - y) \epsilon - (1-3\,y)C  \right] \,
    C^{-1-\epsilon} \, , \nonumber \\
    \Delta i_{23} &=& - \Delta i_{26} = 8\, B^+ (\qr) \int_S\! dx\,dy\,
        x\, y \,\epsilon \, C^{-1-\epsilon} \, ,
    \nonumber \\
    \Delta i_{25} &=& -8\, B^+ (\qr) \int_S\! dx\, dy\,
        x\, (1-x) \,\epsilon \,  C^{-1-\epsilon} \, ,
    \nonumber \\
    \Delta i_{27} &=& 8\, B^+ (\qr) \int_S\! dx\,dy\,
    y\, (1-y) \,\epsilon \, C^{-1-\epsilon} \, ,
\end{eqnarray}
where $B^+ = (1+\epsilon) \Gamma(\epsilon)\, e^{\gamma_E \epsilon}
\mu^{2\epsilon}$ and $C$ is given by
\begin{equation*}
    C = m_c^2 - 2\, x\, y (\qr) - r^2\, x\, (1-x) - q^2\, y\, (1-y).
\end{equation*}
The integration over the Feynman parameters $x$ and $y$
 is restricted to the simplex $S$, i.e. $y\in[0,1-x]$, $x\in[0,1]$.
Due to Ward identities, the quantities $\Delta i_k$ are not independent of one
another. Namely,
\[
    q^\alpha J_{\alpha\beta} = 0 \quad \text{and} \quad r^\beta J_{\alpha\beta}
= 0
\]
imply that $\Delta i_5$ and $\Delta i_6$ can be expressed as
\begin{equation}
    \Delta i_5 = \Delta i_{23} + \frac{q^2}{\qr}\Delta i_{27} \, ; \quad
    \Delta i_6 = \frac{r^2}{\qr}\Delta i_{25} + \Delta i_{26}.
\end{equation}

\subsubsection{General remarks}
After inserting the above expressions for the building blocks $I_\beta$ and
$J_{\alpha\beta}$ into diagrams
\ref{fig:1}a), b) and \ref{fig:1}c), d), respectively, and introducing
additional Feynman parameters, we can easily perform the
integration over the second loop momentum. The remaining Feynman parameter
integrals are, however, non-trivial.
In refs. \cite{matel} and \cite{Greub:2001sy}, where the analogous
corrections to the processes $b \to s \gamma$
and $b \to s g$ were studied,  the strategy used to evaluate
these integrals is the following:
\begin{itemize}
\item
    The denominators are represented as complex Mellin-Barnes integrals (see
below and refs.
     \cite{matel,Greub:2001sy}).
\item
    After interchanging the order of integration and appropriate variable
    transformations, the Feynman parameter integrals reduce
    to Euler $\beta$- and $\Gamma$- functions.

\item
    Finally, by Cauchy's theorem the remaining complex integral over the
Mellin variable
can be written as a sum over residues taken
at certain poles of $\beta$- and $\Gamma$- functions.
This leads in a natural
way to an expansion in the small ratio
    $z = m_c^2/m_b^2$.
\end{itemize}
However, this procedure cannot  be applied directly
in the present case: While the processes $b \to s \gamma$
and $b \to s g$ are characterized by the two mass scales $m_b$ and $m_c$, a
third mass scale, viz. $q^2$,  the
invariant mass squared of the lepton pair enters the
process $b \to s \ell^+ \ell^-$.
For values of $q^2$ satisfying
\[
    \frac{q^2}{m_b^2} < 1 \quad \text{and} \quad \frac{q^2}{4 m_c^2} < 1,
\]
most of the diagrams allow a naive Taylor series expansion in $q^2$ and the
dependence of  the charm quark mass
can again be calculated by means of  Mellin-Barnes representations.
This method does not work, however, for the diagram in Fig. \ref{fig:1}a)
where
the photon is emitted from
the internal $s$-quark line. Instead, we apply a Mellin-Barnes representation
twice, as we discuss in detail in subsection 4.
Using these methods, we get the results for diagrams \ref{fig:1}a)--d)
 as an expansion in
$\s = q^2/m_b^2$, $z = m_c^2/m_b^2$ and $\s/(4 z)$ as well as
$\ln(\s)$ and $\ln(z)$. This implies that our results are meaningful only for
small values of $\s$. Fortunately,
this is exactly the range of main theoretical and experimental interest
in the phenomenology of the process $b \to s \ell^+ \ell^-$.

\subsubsection{Calculation of diagram \ref{fig:1}b)}
\begin{figure}[htb]
\begin{center}
    \includegraphics[height=4cm]{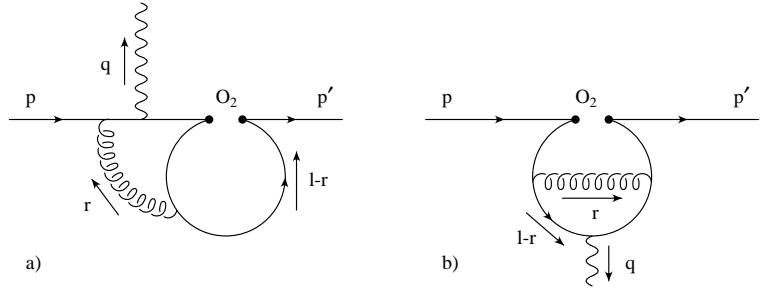}
    \vspace*{0.7cm}\caption{\label{fig:3}
    a) Momentum flow in diagram \ref{fig:1}b), where the virtual photon
       is emitted from the $b$-quark line (see subsection 3);
    b) Momentum flow in the vertex correction diagram in Fig. \ref{fig:1}e)
       (see subsection 7).}
\end{center}
\end{figure}

We describe the basic steps of our calculation of the diagram in Fig.
\ref{fig:1}b) where the photon is emitted from the
internal $b$-quark line. Our notations for the momenta are set up in Fig.
\ref{fig:3}a).
Inserting the building block $I_{\beta}$ yields the following analytic
expression for this diagram:
\begin{multline}
\label{m2c1}
    M_2[1b] = \frac{i\,e\,Q_d\,g_s^2}{4 \pi^2} \, C_F \,
    \Gamma(\epsilon) e^{2 \gamma_E \epsilon} \mu^{4 \epsilon}
    (1-\epsilon) e^{i \pi \epsilon} (4\pi)^{-\epsilon}
    \int_0^1 \! \, dx \, \frac{[x(1-x)]^{1-\epsilon}}{[r^2 - m_c^2/(x(1-x)) +
    i \delta]^\epsilon}
    \\
    \int\! \frac{d^dr}{(2\pi)^d} \, \bar{u}(p') \left(r_\beta r\hspace{-5pt}/ -
r^2 \gamma_\beta \right) \, L \,
    \frac{\pssl + \rsl+m_b }{(p'+r)^2-m_b^2} \, \gamma_{\alpha}\,
    \frac{\psl +\rsl+m_b }{(p+r)^2-m_b^2} \, \gamma^\beta \, u(p) \cdot
\frac{1}{r^2} \,.
\end{multline}
Applying a Feynman parameterization according to
\begin{equation}
    \label{feynman}
    \frac{1}{D_1 D_2 D_3 D_4^\epsilon} = \frac{\Gamma(3+\epsilon)}
    {\Gamma(\epsilon)} \,
    \int_S \frac{du\, dv\, dy\, \, y^{\epsilon-1} \, }{
    \big[u D_1 + v D_2 + (1-u-v-y) D_3 + y D_4 \big]^{3+\epsilon}}
    \, ,
\end{equation}
with
\begin{align}
    D_1 & = (p'+r)^2-m_b^2, & D_2 & = (p+r)^2-m_b^2, \\
    D_3 & = r^2,            & D_4 & = r^2-m_c^2/[x(1-x)],
\end{align}
and performing
the integral over the loop momentum $r$, we obtain
 \begin{multline}
\label{M2b}
    M_2[1b] =
    -\frac{e\, Q_d\, g_s^2}{64\pi^4}(1-\epsilon) C_F \, \Gamma(2\epsilon) e^{2
\gamma_E \epsilon} \mu^{4\epsilon} \times \\
    \int\limits^{1}_{0} \!\!dx \,[x(1-x)]^{1-\epsilon}
\int_S dv \, du \, dy \,
y^{\epsilon-1}\,   \bar{u}(p')\!
    \left[
    \frac{P_1}{\Delta_b^{1+2\epsilon}} + \frac{P_2}{\Delta_b^{2\epsilon}} +
\frac{P_3\Delta_b}{\Delta_b^{2\epsilon}}
    \right] \,
    u(p) \, ,
\end{multline}
 where the Feynman parameters $u$, $v$ and $y$ run over the simplex $S$,
i.e $u,v,y >0$ and $u+v+y \leq 1$.  $P_1$, $P_2$ and $P_3$ are polynomials in
the
Feynman parameters and the quantity $\Delta_b$ reads
\[ \Delta_b = m_b^2 (u+u \, v + v^2) - q^2 \, u \, v +
\frac{m_c^2 \, y}{x(1-x)} \, .\] 
For $q^2 \le m_b^2$ it is positive in the integration region.
Therefore, one is allowed to do a naive Taylor series
expansion of the integrand in $q^2$.
In order to simplify the resulting Feynman parameter integrals, it is
convenient to first transform
the integration variables $x$, $y$, $u$ and $v$ according to
\begin{equation*}
    u  \to \frac{(1-v')(-1+v'+u')}{v'}, \quad
    v  \to \frac{(1-v')(1-u')}{v'}, \quad
    x  \to x', \quad
    y \to y'v'.
\end{equation*}
The integration region of the new variables is given by
$u' \in [(1-v'),1]$ and $v',x',y' \in [0,1]$.
 Taking the corresponding Jacobian into account and omitting
primes in order to simplify the notation, we find
\begin{multline}
    M_2[1b] =
    -\frac{e\, Q_d\, g_s^2}{64\pi^4}(1-\epsilon) C_F \, \Gamma(2\epsilon) e^{2
\gamma_E \epsilon} \mu^{4\epsilon} \times \\
    \int\limits^{1}_{0} \!\!dx \,[x(1-x)]^{1-\epsilon} \int\limits_0^1 \!\!dv
\int\limits_{1-v}^1 \!\!du
    \int\limits_0^1 \!\!dy \, (v\,y)^{\epsilon-1}\,  (1-v) \, \bar{u}(p')\!
    \left[
    \frac{Q_1}{\Delta_b^{1+2\epsilon}} + \frac{Q_2}{\Delta_b^{2\epsilon}} +
\frac{Q_3\Delta_b}{\Delta_b^{2\epsilon}}
    \right]
    u(p) \, ,
\end{multline}
where, in terms of the new variables, $\Delta_b$ reads
\[  \Delta_b = m_b^2 (1-v)u + q^2 \frac{(1-v)^2(1-v-u)(1-u)}{v^2} + m_c^2
\frac{v\,y}{x(1-x)}\,. \]
$Q_1$, $Q_2$ and $Q_3$ are rational functions in the new Feynman parameters.
After performing the Taylor series expansion in $q^2$, the
remaining
integrals are of the form
\begin{equation}
    \label{int}
    \int\limits_0^1\! dx\,dv\,dy \int\limits_{1-v}^{1} du\,[x(1-x)]^{1-
     \epsilon}(v\,y)^{\epsilon-1}(1-v)
    \frac{1}{v^m} \frac{P(x,y,u,v)}{\Delta_{b,0}^{~n+2\epsilon}} \, ,
\end{equation}
where $P(x,y,u,v)$ is a polynomial in $x$, $y$, $u$, $v$; $\Delta_{b,0} =
\Delta_b(q^2=0)$; $n$ and $m$ are
non-negative integers. We further follow the strategy used in
\cite{matel,Greub:2001sy} and represent the
denominators $\Delta_{b,0}^\lambda$ as
Mellin-Barnes integrals. The Mellin-Barnes representation
for $(K^2-M^2)^{-\lambda}$ reads ($\lambda >0$)
\begin{equation}
    \label{Mellin}
    \frac{1}{(K^2 - M^2)^\lambda} = \frac{1}{(K^2)^\lambda} \,
    \frac{1}{\Gamma(\lambda)} \, \frac{1}{2 \pi i} \, \int_{\gamma} \!ds\,
    \left(-\frac{M^2}{K^2}\right)^s \Gamma(-s) \, \Gamma(\lambda+s) \, .
\end{equation}
The integration path $\gamma$ runs parallel to the imaginary axis and intersects
the real
axis somewhere between $-\lambda$ and 0. The Mellin-Barnes representation of
$\Delta_{b,0}^\lambda$ is obtained by making the identifications
\[
    K^2 \leftrightarrow m_b^2\,u(1-v)\quad\text{and}\quad M^2 \leftrightarrow -
m_c^2 \,y\, v / [x(1-x)].
\]
Interchanging the order of integration, it is now an easy task to perform the
Feynman parameter integrals since the most
complicated ones are of the form
\begin{equation}
\label{beta}
    \int\limits_0^1 \!da\, a^{p(s)} (1-a)^{q(s)}  =
\beta \big( p(s)+1,q(s)+1 \big) \, .
\end{equation}
The integration path $\gamma$ has to be chosen in such a way that the Feynman
parameter
integrals exist for values of $s \in \gamma$. By inspection of
the explicit expressions, one finds that this is the case if the path $\gamma$
is chosen such that $\rm{Re}(s) > - \epsilon$.
(Note that in this paper $\epsilon$ is always a positive number).
To perform the integration over the Mellin parameter $s$, we close the
integration path in the right half-plane and use the residue
theorem to identify the integral with the sum over the residues of the poles
located at
\begin{eqnarray}
\label{poles}
    s &=& 0,~ 1,~ 2,~ 3,~  ........ \nonumber \\
    s &=& 1-\epsilon,~ 2-\epsilon,~ 3-\epsilon,~ ....... \nonumber \\
    s &=& 1-2\epsilon,~ 2-2\epsilon,~ 3-2\epsilon,~ ....... \nonumber,\\
    s &=& 1/2-2\epsilon,~ 3/2-2\epsilon,~ 5/2-2\epsilon,~ ....... \quad
\end{eqnarray}
 In view of the factor $(m_c^2/m_b^2)^s$ stemming from the Mellin-Barnes formula
(\ref{Mellin}),
 the evaluation of  the residues at the pole positions listed in eq.
(\ref{poles}) corresponds directly to an expansion in $z=m_c^2/m_b^2$.
 Note, that closing the integration contour in the right half-plane
yields an overall minus sign due to the clockwise orientation of the integration
path. After expanding in $\epsilon$, we get
the form factors of $M_2[1b]$ (see eq. (\ref{formdeftilde}))
as an expansion of the form
\begin{equation}
\label{res2}
    F_2^{(7,9)}[1b] = \sum_{i,l,m} c^{(7,9)}_{2,ilm} \,
\s^i \, z^l \, \ln^m(z) \, ,
\end{equation}
where $i$ and $m$ are non-negative integers and $l$ is a natural multiple of
$\frac{1}{2}$
(see eq. (\ref{poles})).
Furthermore, the power $m$ of $\ln(z)$ is bounded by four, independent of the
values of $i$ and $l$. This becomes clear if
we consider the structure of the poles. There are three poles in $s$ located
near any natural number $k$, viz. at $s=k$,
$s=k-\epsilon$ and $s=k-2\epsilon$. Taking the residue at one of them yields a
term proportional to $1/\epsilon^2$ from the
other two poles. In addition, there can be an explicit $1/\epsilon^2$
term from
the integration over the two loop momenta.
Therefore, the most singular term can be of order $1/\epsilon^4$ and, after
expanding in $\epsilon$, the highest possible
power of $\ln(z)$ is four.

\subsubsection{Calculation of diagram \ref{fig:1}a)}
To calculate the diagram in Fig. \ref{fig:1}a) where the photon is emitted
{}from
the internal $s$-quark, we proceed in a similar
way as in the previous subsection, i.e., we insert the building block
$I_\beta$, introduce three additional
Feynman parameters and integrate over the loop momentum $r$.
The characteristic denominator $\Delta_a$ is of the form
\[
    \Delta_a = \left( A\,m_b^2 + B\,q^2 + C\,m_c^2+i\delta \right)
\]
and occurs with powers $2 \epsilon$ or $1+2\epsilon$.
The coefficients $A$, $B$ and $C$ are functions of the Feynman parameters.
After suitable transformations, they  read
\[
    A = u\, v (1-v); \quad B = u\, v^2 (1-u); \quad C = -\frac{y(1-v)}{x(1-x)} \, ,
\]
with $u,~x,~y,~v \in [0,1]$.
{}From this we conclude that the result of this diagram
is not analytic in $q^2$.
We are therefore not allowed to Taylor expand the integrand.
Instead, we
apply the Mellin-Barnes representation twice and write
\begin{equation}
\label{eq:doubleMB}
    \frac{1}{\Delta_a^\lambda} = \frac{1}{\left(B\,q^2\right)^\lambda}
    \int_\gamma ds \int_{\gamma'} ds'\,
    \frac{\Gamma(s+\lambda) \, \Gamma(-s') \, \Gamma(s'-s) \,
e^{i\pi s'}}{(2\pi i)^2 \, \Gamma(\lambda)}
    \left[\frac{A\,m_b^2}{B\,q^2}\right]^s \left[-
\frac{C\,m_c^2}{A\,m_b^2}\right]^{s'} \, .
\end{equation}
The integration paths $\gamma$ and $\gamma'$ are again parallel to the imaginary
axis and
$-\lambda < \Re(s) < \Re(s') < 0$.
 $\lambda$ takes one of the two values $2\epsilon$ and $ 1+2\epsilon$.
We have written eq. (\ref{eq:doubleMB}) in such a way that
non-integer powers appear only  for positive numbers, i.e. we made use of the
formula
\[
    (x\pm i\delta)^\alpha = e^{\pm i\pi\alpha}(-x\mp i\delta)^\alpha \, .
\]
As in the previous subsection, the exact positions of
the integration paths $\gamma$ and $\gamma'$ are
dictated by the condition
that the Feynman parameter integrals exist
for values of $s$ and $s'$ lying thereon.
For $\lambda = 2 \epsilon$, we find that these integrals exist if
\[
    -\epsilon < \Re(s) < \Re(s') < 0 .
\]
Closing the integration contour for the $s$- and $s'$-integration in the left
and right half-plane, respectively, and
applying the residue theorem results in an expansion in $\s$ and $z$. As
$\rm{Re}(s') > \rm{Re}(s)$,
the term $\Gamma(s'-s)$ in eq. (\ref{eq:doubleMB}) does not generate any poles.
For $\lambda = 2 \epsilon$, the poles which have to be taken into account  are
located at
\begin{align}
    s' & = 1-\epsilon,~2-\epsilon,~3-\epsilon,~...... &
    s  & = -\epsilon,~-1-\epsilon,~-2-\epsilon,~...... \nonumber \\
    s' & = 1-2\epsilon,~2-2\epsilon,~3-2\epsilon,~...... &
    s  & = -2\epsilon,~-1-2\epsilon,~-2-2\epsilon,~...... \nonumber \\
    s' & = 0,~1,~2,~...... \nonumber
\end{align}
For $\lambda =1+ 2 \epsilon$, we
find that the Feynman parameter integrals exist if
\[
    -\epsilon < \Re(s')  < 0  \quad \rm{and} \quad -1-\epsilon < \Re(s) < -2
\epsilon \, .
\]
This condition implies that the poles
at $s=-\epsilon,-2\epsilon$ in the above list must not be taken into account
when applying the residue theorem.

The final result for
the form factors (eq.(\ref{formdeftilde})) of this diagram is of the form
\begin{equation}
  F_2^{(7,9)}[1a] = \sum_{i,j,l,m} c^{(7,9)}_{i,j,l,m}
\, \s^i \, \ln^j(\s) \, z^l \, \ln^m(z) \, ,
\end{equation}
where $i,~j,~l$ and $m$ all are non-negative integers. The remaining four
diagrams in Fig. \ref{fig:1}a) and b) exhibit no further
difficulties.
\subsubsection{Calculation of diagrams \ref{fig:1}c)}
Inserting the building block $J_{\alpha\beta}$ allows us to calculate
directly the sum of the two
diagrams shown in Fig. \ref{fig:1}c). After performing the second
loop integral, one obtains
\begin{multline}
\label{M2c}
    M_2[1c] = \frac{e Q_u g_s^2 C_F }{256 \pi^4}(1+\epsilon) \,
    \Gamma(2\epsilon) e^{2 \gamma_E \epsilon} \mu^{4 \epsilon} e^{2i\pi\epsilon}
\\
    \int dx\,dy\,du\,dv\, \frac{v^\epsilon (1-u)^{1+\epsilon}(1-x)}{[x(1-
x)]^{1+\epsilon}} \,
    \bar{u}(p')\left[\frac{P_1}{\Delta_c^{1+2\epsilon}} +
    \frac{P_2}{\Delta_c^{2\epsilon}} +
\frac{P_3\Delta_c}{\Delta_c^{2\epsilon}}\right]u(p) \, ,
\end{multline}
where $P_1$, $P_2$ and $P_3$ are polynomials in the Feynman parameters,
which  all run in the interval [0,1].
$\Delta_c$ reads (using $v'=v(1-u)$)
\[
    \Delta_c = m_b^2\, u\, v'\, y - q^2\, y\, v' \left(u + y\, v'\right)  -
\frac{v'}{x(1-x)} \left[ m_c^2-q^2 \, y(1-x)
     \left(1-y(1-x)\right) \right] \, .
\]
Note that we do not expand in $q^2$ at this stage of the calculation. Instead,
we use the Mellin-Barnes representation
(\ref{Mellin}) with  the identification
\[
    K^2 \leftrightarrow m_b^2\, u\, v'\, y \quad\text{and}\quad
    M^2 \leftrightarrow  q^2\, y\, v' \left(u + y\, v' \right)  + \frac{v'}{x(1-x)}
\left[ m_c^2-q^2 \, y(1-x)
     \left(1-y(1-x)\right) \right] \, .
\]
This representation does a good job, since $\left(-M^2/K^2\right)^s$ turns out
to be analytic
in $q^2$ for $\s < 4z$, as in this range $M^2/K^2$ is positive for all
values of the Feynman parameters.
We therefore do the Taylor expansion with respect to $q^2$ only at this level.
Evaluating the Feynman parameter integrals as well as the Mellin-Barnes
integral,
we find the result as an expansion in $z$ and $\hat{s}/(4z)$ which
can be cast into the general form
\begin{equation}
\label{genformc}
    F_2^{(7,9)}[1c] = \sum_{i,l,m} c_{2,ilm}^{(7,9)}
    \, \hat{s}^i \, z^l \, \ln^m(z),
\end{equation}
where
$i$ and $m$ are non-negative integers and $l=-i,~-i + \frac{1}{2},~-i +
1,~....$.
\subsubsection{Calculation of diagrams \ref{fig:1}d)}
After inserting the building block $J_{\alpha\beta}$
and  performing the second loop integral,
the sum of the diagrams in
Fig. \ref{fig:1}d) yields
\begin{multline}
\label{M2d}
    M_2[1d] = \frac{e Q_u g_s^2 C_F }{256 \pi^4}(1+\epsilon) \,
    \Gamma(2\epsilon) e^{2 \gamma_E \epsilon} \mu^{4 \epsilon}  \\
    \int\limits_S dx\,dy\,\int\limits_S du\,dv\, \frac{v^\epsilon}{[x(1-
x)]^{1+\epsilon}} \,
    \bar{u}(p')\left[\frac{P_1}{\Delta_d^{1+2\epsilon}} +
    \frac{P_2}{\Delta_d^{2\epsilon}} +
\frac{P_3\Delta_d}{\Delta_d^{2\epsilon}}\right]u(p) \, ,
\end{multline}
where $P_1$, $P_2$ and $P_3$ are polynomials in the Feynman
parameters $x$, $y$, $u$ and $v$.
The parameters $(x,y)$ and $(u,v)$ run in their respective simplex.
The quantity $\Delta_d$ reads
\[
    \Delta_d = m_b^2\, u \left( u+ \frac{y\,v}{1-x}\right) + q^2\, y\,
v \left[\frac{y\,v}{(1-x)^2}+\frac{u}{1-x}
    -\frac{(1-y)}{x(1-x)}\right] + \frac{m_c^2\,v}{x(1-x)}  \, .
\]
Next, we use the Mellin-Barnes representation
(\ref{Mellin}) with  the identification
\[
    K^2 \leftrightarrow m_b^2\, u \left( u+ \frac{y\, v}{1-
x}\right) \, ; \quad
    M^2 \leftrightarrow  q^2\, y\, v \left[\frac{y\, v}{(1-
x)^2}+\frac{u}{1-x}
    -\frac{(1-y)}{x(1-x)}\right] + \frac{m_c^2\, v}{x(1-x)} \, .
\]
Again, $\left(-M^2/K^2\right)^s$ is analytic in $q^2$ for $\s < 4z$,
what allows us to
perform a Taylor series expansion with respect to $q^2$.
In order to perform the integrations over the Feynman parameters, we make
suitable substitutions, e.g.
\begin{eqnarray}
x \rightarrow x',\quad y \rightarrow \frac{(1-x')(y'-(1-v'))}{v'},\quad v
\rightarrow u'v',\quad
u \rightarrow u'(1-v').
\end{eqnarray}
The new variables $x', u', v'$ run in the interval $[0,1]$, while $y'$ varies in
$[1-v',1]$.
Evaluating the integrals over the Feynman and Mellin
parameters,
we find the result as an expansion in $z$ and $\hat{s}/(4z)$ which
can be cast into the general form
\begin{equation}
\label{genformd}
    F_2^{(7,9)}[1d] = \sum_{i,l,m} c_{2,ilm}^{(7,9)}
    \, \hat{s}^i \, z^l \, \ln^m(z) \, .
\end{equation}
$i$ and $m$ are non-negative integers and $l=-i,~-i + \frac{1}{2},~-i +
1,~....$.
\subsubsection{Calculation of diagram \ref{fig:1}e)}
We consider one of the diagrams in Fig. \ref{fig:1}e) in some detail and
redraw it in Fig. \ref{fig:3}b).
The matrix element is proportional to $1/\Delta_e$, where
\begin{eqnarray}
\Delta_e= [(l-r)^2-m_c^2] \, [(l-q-r)^2-m_c^2] \, [(l-q)^2-m_c^2] \,
          [l^2-m_c^2] \, r^2 \, .
\end{eqnarray}
$q$ is the four-momentum of the off-shell photon, while $l$ and $r$
denote loop
momenta. As $q^2 < 4 m_c^2$ in our application,
we use the heavy mass expansion (HME) technique
\cite{Smirnov} to evaluate this diagram.
In the present case, as the gluon is massless, the HME boils down to a naive
Taylor series expansion
of the diagram (before loop integrations) in the four-momentum $q$.
Expanding $1/\Delta_e$ in $q$, we obtain
\begin{equation}
\frac{1}{\Delta_e} =\sum \limits_{n,m,i,j,k} C_e(n,m,i,j,k)\frac{(q^2)^i
(q\cdot r)^j(q\cdot l)^k}{r^2 \, [l^2-m_c^2]^n \, [(l-r)^2-m_c^2]^m} \quad .
\end{equation}
Using the Feynman parameterization
\begin{eqnarray}
\frac{1}{[l^2-m_c^2]^n \, [(l-r)^2-m_c^2]^m}
=\frac{\Gamma(n+m)}{\Gamma(n)\Gamma(m)}\int \limits_0^1
\frac{v^{m-1}(1-v)^{n-1}}{[l^2-2v(l\cdot r)-m_c^2+vr^2]^{n+m}} \, dv \, ,
\end{eqnarray}
we can perform the integration over the loop momentum $l$.
The  integral over  the loop momentum $r$ can be done using the
parameterization
 \begin{equation}
 \frac{1}{r^2} \left (\frac{1}{r^2-\frac{m_c^2}{v (1-v)}}\right
)^p=\frac{\Gamma(1+p)}{\Gamma(p)}\int
 \limits_0^1\frac{u^{p-1}}{\left (r^2-\frac{u \, m_c^2}{v (1-v)}\right)^{p+1}}
\, du \, .
 \end{equation}
 The remaining integrals over the Feynman parameters
$u$ and $v$  all have the form of
eq. (\ref{beta}) and can be performed easily.
The other two diagrams in Fig. \ref{fig:1}e) 
where the virtual photon is emitted from the charm quark
can be evaluated in a similar way. The diagrams where the photon
is radiated from the $b$-quark or the $s$-quark vanish.

As the results for the sum of all the diagrams in 
Fig. \ref{fig:1}e) are compact,
we explicitly give their contribution
 to the form factors $F_{a}^{(j)}$ ($a=1,2$; $j=7,9$). We obtain
$F_a^{(7)}[1e]=0$,
 $F_1^{(9)}[1e]=\frac{4}{3} \, F_2^{(9)}[1e]$ and
\begin{eqnarray}
F_2^{(9)}[1e] &=&  \left( \frac{\mu}{m_c} \right)^{4 \epsilon} \,
\frac{1}{\epsilon} \,
 \left[ \frac{8}{3} + \frac{128}{45} \, \left( \frac{\s}{4z} \right)
+ \frac{256}{105} \left( \frac{\s}{4z} \right)^2+ \frac{2048}{945} \left(
\frac{\s}{4z} \right)^3  \right]   \nonumber \\
&& - \left[ \frac{124}{27} + \frac{12416}{3645} \, \left( \frac{\s}{4z} \right)
+ \frac{11072}{42525} \left( \frac{\s}{4z} \right)^2- \frac{4971776}{4465125}
\left( \frac{\s}{4z} \right)^3  \right] \, .
\end{eqnarray}
\subsubsection{Unrenormalized form factors of $O_1$ and $O_2$}
We stress that the diagram  \ref{fig:1}f) where the virtual photon is emitted
{}from the charm quark line
is the only one in Fig. \ref{fig:1} which suffers from infrared and collinear
singularities.
As this diagram can easily be combined with diagram \ref{fig:789}b) associated
with the operator $O_9$,
we take it into account only in section \ref{virtcorro9}
where the virtual corrections to $O_9$ are discussed.

The unrenormalized form factors $F^{(7,9)}_{a}$ of $\bra s \ell^+ \ell^-|
O_a|b \ket$
$(a=1,2)$, corresponding to diagrams \ref{fig:1}a)--\ref{fig:1}e),
are obtained in the form
\[
    F^{(7,9)}_{a} = \sum_{i,j,l,m} c_{a,ijlm}^{(7,9)} \, \s^i \, \ln^j(\s)
\, z^l \, \ln^m(z),
\]
where $i,~j$ and $m$ are non-negative integers and $l = -i,~-i + \frac{1}{2},~-i
+ 1,~....$. We keep the terms with $i$
and $l$ up to 3, after checking that higher order terms are small for $0.05 \le
\s \le 0.25$, the range considered in this
paper. As we will give the full results for the counterterm contributions to the
form factors in section
\ref{subsec:countertermso2} and the
 renormalized form factors in section \ref{subsec:renformfactors} and in
appendix \ref{appendix:aaa},
 it is not necessary to explicitly present the somewhat
 lengthy expressions for the unrenormalized form factors. But, in order to
demonstrate the cancellation of ultraviolet
 singularities in the next section, we list the divergent parts of the
unrenormalized form factors:
 $F_1^{(7)}$, $F_1^{(9)}$, $F_2^{(7)}$ and $F_2^{(9)}$:
\begin{eqnarray}
\label{fidiv}
    F^{(9)}_{2,\text{div}} & = &
    \frac{128}{81\, \epsilon^2}
    - \frac{4}{25515\, \epsilon}
    \,\left( 1890 + 1260\,i\pi + 5040\,L_\mu- 1260\,L_s + 252\,\s + 27\,\s^2 + 4\,\s^3  \right)
    \nonumber \\  \nonumber
&&
    + \frac{8}{2835\, \epsilon}\,
    \left (420 + 2520\, L_\mu  - 1260\,L_z+
    2016 \left(\frac{\hat{s}}{4z}\right) +
    1296 \left(\frac{\hat{s}}{4z}\right)^2 +
    1024 \left(\frac{\hat{s}}{4z}\right)^3
    \right)
    \, , \\ \nonumber
    F^{(7)}_{2,\text{div}} & = &
    \frac{92}{81\, \epsilon} \, , \\ \nonumber
    F_{1,\text{div}}^{(9)}&^= &
    -\frac{64}{243\, \epsilon^2} +
    \frac{2}{76545\, \epsilon}
    \left(1890 + 1260\, i\pi + 5040\, L_{\mu} - 1260\, L_s +
    252\, \hat{s} + 27\, \hat{s}^2 + 4\, \hat{s}^3\right) \\ \nonumber
    &&\hspace{-0.5cm}
    -\frac{4}{8505 \, \epsilon}\left( - 8085 + 2520\, L_{\mu} - 1260\, L_z
    - 7056  \left(\frac{\hat{s}}{4z}\right)-6480\left(\frac{\hat{s}}{4z}\right)^2-
    5888\left(\frac{\hat{s}}{4z}\right)^3\right) \, , \\
    F_{1,\text{div}}^{(7)} & = & -\frac{46}{243\, \epsilon} \, ,
\end{eqnarray}
where $L_s = \ln(\s)$, $L_z = \ln(z)$,
$L_\mu = \ln \left( \frac{\mu}{m_b} \right)$ and $z=m_c^2/m_b^2$.
%
%
\subsection{$\Order(\alpha_s)$ counterterms to $O_1$ and $O_2$}
\label{subsec:countertermso2}
So far, we have calculated the two-loop matrix elements $\bra s \ell^+ \ell^-
|C_i \, O_i|b \ket$ ($i=1,2$).
As the operators mix under renormalization, there are additional contributions
proportional to $C_i$.
These counterterms arise from the matrix elements of the operators
\begin{equation}
    \sum_{j=1}^{12} \delta Z_{ij}  O_j  \, , \quad i=1,2 ,
\end{equation}
where the operators $O_1$--$O_{10}$ are given in eq. (\ref{oper}). $O_{11}$ and
$O_{12}$
are evanescent operators, i.e., operators which vanish in $d=4$ dimensions.
In principle, there is some freedom in the choice of the evanescent operators.
However,
as we want to combine our matrix elements with the  Wilson coefficients
calculated by Bobeth et al.
 \cite{Bobeth:2000mk},
we must use the same definitions:
\begin{align}
    O_{11} & = \left( \sb_L \gamma_\mu \gamma_\nu \gamma_\sigma T^a c_L \right)
             \left( \cb_L \gamma^\mu \gamma^\nu \gamma^\sigma T^a b_L \right) -
16 \, O_1 \, ,\\
    O_{12} & = \left( \sb_L \gamma_\mu \gamma_\nu \gamma_\sigma c_L \right)
             \left( \cb_L \gamma^\mu \gamma^\nu \gamma^\sigma b_L \right) - 16
\, O_2 \, .
\end{align}
The operator renormalization constants $Z_{ij} = \delta_{ij} + \delta Z_{ij}$
are of the form
\begin{equation}
    \delta Z_{ij} = \frac{\alpha_s}{4 \pi} \left( a_{ij}^{01} +
\frac{1}{\epsilon} a_{ij}^{11}\right) +
        \frac{\alpha_s^2}{(4 \pi)^2}
        \left( a_{ij}^{02} + \frac{1}{\epsilon} a_{ij}^{12} +
\frac{1}{\epsilon^2} a_{ij}^{22}\right) +
        \Order(\alpha_s^3).
\end{equation}
Most of the coefficients $a_{ij}^{lm}$ needed for our calculation are given in
ref. \cite{Bobeth:2000mk}. As some are new
(or not explicitly given in \cite{Bobeth:2000mk}), we list those for $i=1,~2$
and $j=1,...,~12$:
\begin{equation}
    \hat{a}^{11}=
    \left(
    \begin{array}{cccccccccccc}
        -2& \frac{4}{3}&0& -\frac{1}{9} &0&0 & 0 &0& -
\frac{16}{27}&0&\frac{5}{12}& \frac{2}{9} \\
        \\
         6& 0&0 & \frac{2}{3} &0&0&0 &0& -\frac{4}{9}& 0& 1& 0
    \end{array}
    \right),
    \begin{array}{lll}
        a^{12}_{17} = -\frac{58}{243}\,, \hspace{0cm} &
        a^{12}_{19} = -\frac{64}{729}\,, \hspace{0cm} &
        a^{22}_{19} = \frac{1168}{243}\,, \\ \\
        a^{12}_{27} = \frac{116}{81}\,, \hspace{0.cm} &
        a^{12}_{29} = \frac{776}{243}\,, \hspace{0.cm} &
        a^{22}_{29} = \frac{148}{81}\,.
    \end{array}
\end{equation}
We denote the counterterm contributions to $b \to s \ell^+ \ell^-$ which are due
to the mixing of
$O_1$ or $O_2$ into four-quark operators by $F_{i \to \rm{4
quark}}^{\rm{ct}(7)}$ and $F_{i \to \rm{4 quark}}^{\rm{ct}(9)}$.
They can be extracted from the equation
\begin{equation}
\label{fi4quark}
\sum_{j} \left( \frac{\alpha_s}{4\pi}\right) \, \frac{1}{\epsilon} \,
a_{ij}^{11} \bra s \ell^+ \ell^-|O_j|b \ket_{\text{1-loop}}
= - \left( \frac{\alpha_s}{4\pi}\right) \, \left[
 F_{i \to \rm{4 quark}}^{\rm{ct}(7)} \bra \widetilde{O}_7\ket_{\text{tree}} +
 F_{i \to \rm{4 quark}}^{\rm{ct}(9)} \bra \widetilde{O}_9\ket_{\text{tree}}
\right] \, ,
\end{equation}
where $j$ runs over the four-quark operators. As certain entries of
$\hat{a}^{11}$ are zero, only the one-loop matrix
elements of $O_1$, $O_2$,  $O_4$,  $O_{11}$ and $O_{12}$ are needed.
In order to keep the presentation transparent, we relegate their explicit form
to appendix
\ref{appendix:oneloop}.

The counterterms which are related  to the mixing of $O_i$ ($i=1,2$)  into $O_9$
can be split into two classes:
The first class consists of the one-loop mixing $O_i \to O_9$, followed by
taking the one-loop corrected matrix element of $O_9$.
It is obvious  that this class contributes to the renormalization of diagram
\ref{fig:1}f). As we decided to treat diagram \ref{fig:1}f)
only in section \ref{virtcorro9}
 (when discussing  virtual corrections to $O_9$), we proceed in the
same way with the counterterm just mentioned.
There is, however, a second class of counterterm contributions due to $O_i \to
O_9$ mixing. These contributions are generated by
 two-loop mixing of $O_2$ into $O_9$ as well as by one-loop mixing and
 one-loop renormalization of the $g_s$ factor in the definition of
the operator $O_9$.
We denote the corresponding contribution to the counterterm form
factors by
$F_{i  \to 9}^{\rm{ct}(7)}$ and $F_{i \to 9}^{\rm{ct}(9)}$. We obtain
 \begin{equation}
 F_{i  \to 9}^{\rm{ct}(9)} = -\left(\frac{a_{i9}^{22}}{\epsilon^2} +
\frac{a_{i9}^{12}}{\epsilon} \right) -
 \frac{a_{i9}^{11} \, \beta_0}{\epsilon^2} \, ; \quad
  F_{i  \to 9}^{\rm{ct}(7)} = 0 \, ,
  \end{equation}
where we made use of the renormalization constant $Z_{g_s}$  given by
\begin{equation}
\label{eq:Zg}
    Z_{g_s} = 1 - \frac{\alpha_s}{4\pi} \, \frac{\beta_0}{2} \,
\frac{1}{\epsilon} \, ;  \quad \beta_0 = 11 - \frac{2}{3} N_f \, ; \quad N_f = 5
\, .
\end{equation}

Besides the contribution from operator mixing, there are ordinary QCD
counterterms. The renormalization of the charm
quark mass is taken into account by replacing $m_c$ through $Z_{m_c} \cdot m_c$
in the one-loop matrix elements
of $O_1$ and $O_2$ (see appendix \ref{appendix:oneloop}).
We denote the corresponding contribution to the counterterm form factors by
$F_{i, m_c \text{ren}}^{\rm{ct}(7)}$ and $F_{i, m_c \text{ren}}^{\rm{ct}(9)}$.
We obtain
 \begin{eqnarray}\label{fimcren}\nonumber
&& F_{1, m_c \text{ren}}^{\rm{ct}(7)} =  F_{2, m_c \text{ren}}^{\rm{ct}(7)} =0
\, ; \quad F_{1, m_c \text{ren}}^{\rm{ct}(9)} =
 \frac{4}{3} F_{2, m_c \text{ren}}^{\rm{ct}(9)} \\
 \nonumber
F_{2, m_c \text{ren}}^{\rm{ct}(9)}
&=& \left[-\frac{32}{945\epsilon}\left(105+84  \left(\frac{\hat{s}}{4z}\right)+
72  \left(\frac{\hat{s}}{4z}\right)^2+64
\left(\frac{\hat{s}}{4z}\right)^3\right)\right.\\
\nonumber
&-&\left.\frac{32}{2835}\left(105+1260 \ln{\frac{\mu}{m_c}}+
\left(\frac{\hat{s}}{4z}\right)\left(336+1008 \ln{\frac{\mu}{m_c}}\right)
+\left(\frac{\hat{s}}{4z}\right)^2\left(396+864 \ln{\frac{\mu}{m_c}}
\right)\right.\right.\\
&+&\left.\left. \left(\frac{\hat{s}}{4z}\right)^3\left(416+768
\ln{\frac{\mu}{m_c}}\right)\right)\right] \, ,
\end{eqnarray}
where we have used the pole mass definition of $m_c$ which is characterized by the
renormalization constant
\begin{equation}
\label{eq:Zm}
    Z_m = 1 -\frac{\alpha_s}{4\pi} \frac{4}{3} \left( \frac{3}{\epsilon} + 6 \ln
\left(\frac{\mu}{m}\right) + 4 \right) .
\end{equation}
If one wishes to express the results for $F_{i, m_c \text{ren}}^{\rm{ct}(9)}$ in
terms of the $\overline{\rm{MS}}$
definition of the charm quark mass, the expressions in eqs. (\ref{fimcren}) get
changed according to
\begin{equation}
F_{i, m_c \text{ren}}^{\rm{ct}(9)} \to F_{i, m_c \text{ren}}^{\rm{ct}(9)} +
\Delta F_{i, m_c \text{ren}}^{\rm{ct}(9)} \, ,
\end{equation}
where $\Delta F_{i, m_c \text{ren}}^{\rm{ct}(9)}$ reads
\begin{eqnarray}\label{deltafimcren}\nonumber
&& \Delta F_{1, m_c \text{ren}}^{\rm{ct}(9)} =  \frac{4}{3} \Delta F_{2, m_c
\text{ren}}^{\rm{ct}(9)} \\
\Delta F_{2, m_c \text{ren}}^{\rm{ct}(9)}
&=& \frac{64}{945} \, \left(105+84  \left(\frac{\hat{s}}{4z}\right)+
72  \left(\frac{\hat{s}}{4z}\right)^2+64
\left(\frac{\hat{s}}{4z}\right)^3\right) \,
\left(  \ln \frac{\mu}{m_c} +\frac{2}{3} \right) \, .
\end{eqnarray}
We stress at this point that we always use the pole mass definition in the
following, i.e., eqs. (\ref{fimcren})
for  $F_{i, m_c \text{ren}}^{\rm{ct}(j)}$.

The total counterterms $F_{i}^{\rm{ct}(j)}$ ($i=1,2$; $j=7,9$) which renormalize
diagrams
\ref{fig:1}a)--\ref{fig:1}e) are given by
\begin{equation}
 F_{i}^{\rm{ct}(j)} = F_{i \to \rm{4 quark}}^{\rm{ct}(j)} + F_{i  \to
9}^{\rm{ct}(j)} +  F_{i, m_c \text{ren}}^{\rm{ct}(j)} \, .
\end{equation}
Explicitly, they read
\begin{eqnarray}
\nonumber
F_{2}^{\rm{ct}(9)}&=&-F_{2, \,\text{div}}^{(9)}-\frac{4}{25515}\left[
5740 + 2520\, \pi^2 - 840\, i \pi\right.\\
\nonumber
&+&\left.
840\, L_{\mu}\left(19 - 3\, i\pi - 54\, L_z + 48\, L_{\mu}\right)
+ 3780\,  L_z( - 2 + 3\, L_z)\right.\\
\nonumber
&+&\left.
420 \,L_s (3\, i \pi + 2 + 6\, L_{\mu}) - 630\, L_s^2
+252\,\hat{s}(1-2\, L_{\mu})-54\, L_{\mu}\,\hat{s}^2
-2\,\hat{s}^3 (1+4\, L_{\mu})\right.\\
\nonumber
&+&\left.
6048 \left(\frac{\hat{s}}{4z}\right)\left(18\, L_{\mu}-9\, L_z-1\right)
+7776 \left(\frac{\hat{s}}{4z}\right)^2(10\, L_{\mu}-5\, L_z + 3)\right.\\
&+&\left.
1536 \left(\frac{\hat{s}}{4z}\right)^3 (42\, L_{\mu}-21\, L_z+19)\right]
 \, ,  \nonumber \\
\nonumber
F_{2}^{\rm{ct}(7)} &=& -F_{2, \,\text{div}}^{(7)}+\frac{2}{2835}
\left(840\, L_{\mu}+70\, \hat{s}+7\, \hat{s}^2+\hat{s}^3 \right) \, , \\
\nonumber
F_{1}^{\rm{ct}(9)}&=&-F_{1, \,\text{div}}^{(9)}+
\frac{2}{76545}\left[ -62300 - 840\, i \pi + 2520\, \pi^2 \right.\\
\nonumber &+&\left.
840\, L_{\mu} (-3\, i\pi - 54\, L_z  + 48\, L_{\mu} - 791)
+3780\, L_z (3\, L_z + 88)\right.\\
\nonumber&+&\left.
420\, L_s ( 3\, i \pi + 2 + 6\, L_{\mu})
-630\, L_s^2+\hat{s}(252 - 504\, L_{\mu})-54\, \hat{s}^2\, L_{\mu}
-2\, \hat{s}^3 ( 1 + 4\, L_{\mu} )\right.\\
\nonumber &-&\left.
6048 \left(\frac{\hat{s}}{4z}\right) (28 + 90\, L_{\mu} - 45\, L_z)
-7776 \left(\frac{\hat{s}}{4z}\right)^2 (27 + 62\, L_{\mu} - 31\, L_z)\right.\\
&-&\left.
768 \left(\frac{\hat{s}}{4z}\right)^3 ( 295 + 564\, L_{\mu} - 282\, L_z)\right] \,
, \nonumber \\
F_{1}^{\rm{ct}(7)} &=& -F_{1, \,\text{div}}^{(7)}-\frac{1}{8505}
\left(840\, L_{\mu}+70\, \hat{s}+7\, \hat{s}^2+\hat{s}^3 \right) \, .
\end{eqnarray}
The divergent parts of these counterterms are, up to a sign, identical to
those of the unrenormalized matrix elements given in eq. (\ref{fidiv}),
which proves the cancellation of ultraviolet singularities.

As mentioned before, we will take diagram \ref{fig:1}f) into account only in
section \ref{virtcorro9}.
The same holds for the
counterterms associated with the $b$- and $s$-quark wave function
renormalization and, as mentioned
earlier in this subsection, the $\Order(\alpha_s)$ correction
to the matrix element of $\delta Z_{i9} O_9$. The sum of these contributions
is
\[
   \delta \bar Z_\psi \bra O_i \ket_{\text{1-loop}} + \frac{\alpha_s}{4\pi} \,
\frac{a_{i9}^{11}}{\epsilon} \,
        \left[ \delta \bar Z_\psi \bra O_9 \ket_{\text{tree}} + \bra O_9
\ket_{\text{1-loop}} \right]  \, , \quad
    \delta \bar Z_\psi = \sqrt{ Z_\psi(m_b) Z_\psi(m_s)} - 1,
\]
and provides the counterterm that renormalizes diagram \ref{fig:1}f). We use
on-shell renormalization for the
external $b$- and $s$-quark. In this scheme the field strength renormalization
constants are
given by
\begin{equation}
\label{quarkren}
    Z_\psi(m) = 1 - \frac{\alpha_s}{4\pi} \, \frac{4}{3} \,
\left(\frac{\mu}{m}\right)^{2\epsilon}
        \left( \frac{1}{\epsilon} + \frac{2}{\epsilon_\IR} + 4 \right) .
\end{equation}
So far, we have discussed the counterterms which
renormalize the $\Order(\alpha_s)$
corrected matrix elements
 $\bra s \ell^+ \ell^-|O_i|b \ket$ ($i=1,2$).
The corresponding one-loop matrix
elements (of order $\Order(\alpha_s^0)$)
 are renormalized by adding the counterterms
\[\frac{\alpha_s}{4\pi} \, \frac{a_{i9}^{11}}{\epsilon} \,
\bra O_9 \ket_\tree \, . \]
%
\subsection{Renormalized form factors of $O_1$ and $O_2$}
\label{subsec:renformfactors}
We decompose the renormalized matrix elements of $O_i$ ($i=1,2$) as
\begin{equation}
\label{o12decomp}
    \bra s \ell^+ \ell^-|C_i^{(0)} O_i | b \ket = C_i^{(0)} \left( -
\frac{\alpha_s}{4 \pi} \right)
    \left[ F_i^{(9)} \bra \widetilde{O}_9 \ket_{\text{tree}} + F_i^{(7)} \bra
\widetilde{O}_7 \ket_{\text{tree}} \right],
\end{equation}
where $\widetilde{O}_9=\frac{\alpha_s}{4 \pi} \, O_9$ and
$\widetilde{O}_7=\frac{\alpha_s}{4 \pi} \, O_7$.
The form
factors $F_i^{(9)}$ and $F_i^{(7)}$, expanded up to $\s^3$ and $z^3$, of the
renormalized sum of diagrams
\ref{fig:1}a)-e) read ($L_c = \ln m_c/m_b =\ln \hat{m}_c =L_z/2$)
\begin{multline}
    F_1^{(9)} =
    \left( -\frac{1424}{729} + \frac{16}{243} \,i\pi +
    \frac{64}{27} \,L_c \right) L_\mu - \frac{16}{243} \, L_\mu \, L_s +
    \left( \frac{16}{1215} - \frac{32}{135} \, z^{-1} \right) L_\mu \,\s
    \\
    + \left( \frac{4}{2835} - \frac{8}{315} \, z^{-2} \right) L_\mu \,\s^2 +
    \left( \frac{16}{76545} - \frac{32}{8505} \, z^{-3}\right) L_\mu \,\s^3
    - \frac{256}{243} \, L_\mu^2 + f_1^{(9)} \, ,
\end{multline}
\begin{multline}
    F_2^{(9)} = \left( \frac{256}{243} - \frac{32}{81} \,i\pi -
    \frac{128}{9} \, L_c \right) L_\mu +
    \frac{32}{81} \, L_\mu \, L_s +
    \left( -\frac{32}{405} + \frac{64}{45} \,
    z^{-1} \right) L_\mu \, \s
    \\
    + \left( -\frac{8}{945} + \frac{16}{105} \,
    z^{-2} \right) L_\mu \, \s^2 +
    \left( -\frac{32}{25515} + \frac{64}{2835} \,
    z^{-3} \right) L_\mu \, \s^3
    + \frac{512}{81} \, L_\mu^2
    + f_2^{(9)} \, ,
\end{multline}
\begin{equation}
    F_1^{(7)} = -\frac{208}{243} \, L_\mu + f_1^{(7)} ,\quad \quad
    F_2^{(7)} = \frac{416}{81}   \, L_\mu + f_2^{(7)} \, .
\end{equation}
The analytic results for $f_1^{(9)}$, $f_1^{(7)}$, $f_2^{(9)}$,
and $f_2^{(7)}$ are rather lengthy. We decompose
them as follows:
\begin{equation}
\label{f1decomp}
    f_a^{(b)} = \sum_{i,j,l,m} \, \kappa^{(b)}_{a,ijlm} \,
\s^i \, L^j_s \, z^l \,
L_c^m +
        \sum_{i,j} \rho^{(b)}_{a,ij} \, \s^i \, L_s^j.
\end{equation}
The quantities $\rho^{(b)}_{a,ij}$ collect the half-integer powers of
$z=m_c^2/m_b^2=\hat{m}_c^2$. This way, the summation indices
in the above equation run over integers only. We list the coefficients
$\kappa^{(b)}_{a,ijlm}$ and $\rho^{(b)}_{a,ij}$ in
appendix \ref{appendix:aaa}.

If we give the charm quark mass
dependence in numerical form,
the formulas become simpler.
For this purpose, we write the functions
$f_a^{(b)}$ as
\begin{equation}
\label{f1decompsimple}
f_a^{(b)} = \sum_{i,j} \, k_a^{(b)}(i,j) \, \s^i \, L_s^j
\quad (a=1,2; \, b = 7,9; \, i=0,...,3; \, j = 0,1) \, .
\end{equation}
The numerical values for the quantities $k_a^{(b)}(i,j)$
are given in Tab. \ref{table1} and \ref{table2}
for $\hat{m}_c = 0.25$, 0.29, 0.33.
For numerical values corresponding to
$\hat{m}_c = 0.27$, 0.29, 0.31 we refer to Tab. I and Tab. II
in the letter version \cite{AAGW:letter}.
\begin{table}[htb]
\begin{center}
\begin{tabular}{| c | c | c | c |}
 \hline & $\hat{m}_c=0.25$ & $\hat{m}_c=0.29$ &  $\hat{m}_c=0.33$ \\ \hline \hline
$k_1^{(9)}(0,0)$  & $ -12.715+0.094699\,i $  & $ -11.973+0.16371\,i $  & $ -11.355+0.19217\,i $  \\
$k_1^{(9)}(0,1)$  & $ -0.078830-0.074138\,i $  & $ -0.081271-0.059691\,i $  & $ -0.079426-0.043950\,i $  \\
$k_1^{(9)}(1,0)$  & $ -38.742-0.67862\,i $  & $ -28.432-0.25044\,i $  & $ -21.648-0.063493\,i $  \\
$k_1^{(9)}(1,1)$  & $ -0.039301-0.00017258\,i $  & $ -0.040243+0.016442\,i $  & $ -0.029733+0.031803\,i $  \\
$k_1^{(9)}(2,0)$  & $ -103.83-2.5388\,i $  & $ -57.114-0.86486\,i $  & $ -33.788-0.24902\,i $  \\
$k_1^{(9)}(2,1)$  & $ -0.044702+0.0026283\,i $  & $ -0.035191+0.027909\,i $  & $ -0.0020505+0.040170\,i $  \\
$k_1^{(9)}(3,0)$  & $ -313.75-8.4554\,i $  & $ -128.80-2.5243\,i $  & $ -59.105-0.72977\,i $  \\
$k_1^{(9)}(3,1)$  & $ -0.051133+0.022753\,i $  & $ -0.017587+0.050639\,i $  & $ 0.052779+0.038212\,i $  \\
\hline
$k_1^{(7)}(0,0)$  & $ -0.76730-0.11418\,i$ & $ -0.68192-0.074998\,i$ & $ -0.59736-0.044915\,i$ \\
$k_1^{(7)}(0,1)$  & $ 0$ & $ 0$ & $ 0$ \\
$k_1^{(7)}(1,0)$  & $ -0.28480-0.18278\,i$ & $ -0.23935-0.12289\,i$ & $ -0.19850-0.081587\,i$ \\
$k_1^{(7)}(1,1)$  & $ -0.0032808+0.020827\,i$ & $ 0.0027424+0.019676\,i$ & $ 0.0074152+0.016527\,i$ \\
$k_1^{(7)}(2,0)$  & $ 0.056108-0.23357\,i$ & $ -0.0018555-0.17500\,i$ & $ -0.039209-0.12242\,i$ \\
$k_1^{(7)}(2,1)$  & $ 0.016370+0.020913\,i$ & $ 0.022864+0.011456\,i$ & $ 0.022282+0.00062522\,i$ \\
$k_1^{(7)}(3,0)$  & $ 0.62438-0.027438\,i$ & $ 0.28248-0.12783\,i$ & $ 0.085946-0.11020\,i$ \\
$k_1^{(7)}(3,1)$  & $ 0.030536+0.0091424\,i$ & $ 0.029027-0.0082265\,i$ & $ 0.012166-0.019772\,i$ \\
\hline
\end{tabular}
\end{center}
\caption{Coefficients in the decomposition of
$f_1^{(9)}$ and $f_1^{(7)}$ for   three values of $\hat{m}_c$.
See eq. (\ref{f1decompsimple}).}
\label{table1}
\end{table}
\begin{table}[htb]
\begin{center}
\begin{tabular}{| c | c | c | c |}
 \hline & $\hat{m}_c=0.25$ & $\hat{m}_c=0.29$ &  $\hat{m}_c=0.33$ \\ \hline \hline
$k_2^{(9)}(0,0)$  & $ 9.5042-0.56819\,i $  & $ 6.6338-0.98225\,i $  & $ 4.3035-1.1530\,i $  \\
$k_2^{(9)}(0,1)$  & $ 0.47298+0.44483\,i $  & $ 0.48763+0.35815\,i $  & $ 0.47656+0.26370\,i $  \\
$k_2^{(9)}(1,0)$  & $ 7.4238+4.0717\,i $  & $ 3.3585+1.5026\,i $  & $ 0.73780+0.38096\,i $  \\
$k_2^{(9)}(1,1)$  & $ 0.23581+0.0010355\,i $  & $ 0.24146-0.098649\,i $  & $ 0.17840-0.19082\,i $  \\
$k_2^{(9)}(2,0)$  & $ 0.33806+15.233\,i $  & $ -1.1906+5.1892\,i $  & $ -2.3570+1.4941\,i $  \\
$k_2^{(9)}(2,1)$  & $ 0.26821-0.015770\,i $  & $ 0.21115-0.16745\,i $  & $ 0.012303-0.24102\,i $  \\
$k_2^{(9)}(3,0)$  & $ -42.085+50.732\,i $  & $ -17.120+15.146\,i $  & $ -9.2008+4.3786\,i $  \\
$k_2^{(9)}(3,1)$  & $ 0.30680-0.13652\,i $  & $ 0.10552-0.30383\,i $  & $ -0.31667-0.22927\,i $  \\
\hline
$k_2^{(7)}(0,0)$  & $ 4.6038+0.68510\,i $  & $ 4.0915+0.44999\,i $  & $ 3.5842+0.26949\,i $  \\
$k_2^{(7)}(0,1)$  & $ 0 $  & $ 0 $  & $ 0 $  \\
$k_2^{(7)}(1,0)$  & $ 1.7088+1.0967\,i $  & $ 1.4361+0.73732\,i $  & $ 1.1910+0.48952\,i $  \\
$k_2^{(7)}(1,1)$  & $ 0.019685-0.12496\,i $  & $ -0.016454-0.11806\,i $  & $ -0.044491-0.099160\,i $  \\
$k_2^{(7)}(2,0)$  & $ -0.33665+1.4014\,i $  & $ 0.011133+1.0500\,i $  & $ 0.23525+0.73452\,i $  \\
$k_2^{(7)}(2,1)$  & $ -0.098219-0.12548\,i $  & $ -0.13718-0.068733\,i $  & $ -0.13369-0.0037513\,i $  \\
$k_2^{(7)}(3,0)$  & $ -3.7463+0.16463\,i $  & $ -1.6949+0.76698\,i $  & $ -0.51568+0.66118\,i $  \\
$k_2^{(7)}(3,1)$  & $ -0.18321-0.054854\,i $  & $ -0.17416+0.049359\,i $  & $ -0.072997+0.11863\,i $  \\
\hline
\end{tabular}
\end{center}
\caption{Coefficients in the decomposition of $f_2^{(9)}$ and $f_2^{(7)}$ for   three values of $\hat{m}_c$. See eq. (\ref{f1decompsimple}).}
\label{table2}
\end{table}
%
%
\section{Virtual corrections to the matrix elements of \\ the operators
$O_7$, $O_8$, $O_9$ and $O_{10}$}
\label{section:virto789}
%
%
\subsection{Virtual corrections to the matrix element of $O_9$ and $O_{10}$}
\label{virtcorro9}
\begin{figure}[t]
    \begin{center}
    \leavevmode
    \includegraphics[height=6cm]{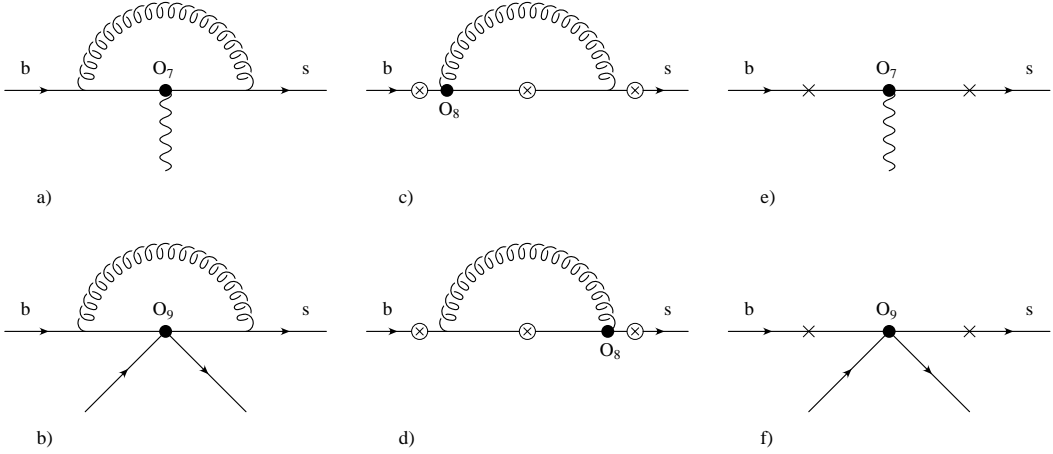}
    \vspace{2ex}
    \caption{Some Feynman diagrams for
    $b \to s \gamma^*$ or $b \to s \ell^+ \ell^-$
    associated with the operators $O_7$, $O_8$ and $O_9$.
    The circle-crosses denote the possible locations where
    the virtual photon
    is emitted, while the crosses mark the possible locations
    for gluon bremsstrahlung. See text.}
    \label{fig:789}
    \end{center}
\end{figure}
As the hadronic parts of the operators $O_9$ and $O_{10}$ are
identical, the QCD corrected matrix element of $O_{10}$
can easily be obtained from the one of $O_9$. We therefore
present only the calculation for
$ \bra s \ell^+ \ell^-|O_9|b \ket $ in some
detail. The virtual corrections to this matrix element
consist of the vertex correction
shown in Fig. \ref{fig:789}b) and of the quark self-energy contributions.
The result can be written as
\begin{equation}
    \label{o9decomp}
    \bra s \ell^+ \ell^-|C_9 O_9 | b \ket =
    \widetilde{C}_9^{(0)} \left( -\frac{\alpha_s}{4 \pi} \right)
    \left[ F_9^{(9)} \bra \widetilde{O}_9 \ket_{\text{tree}} +
        F_9^{(7)} \bra \widetilde{O}_7 \ket_{\text{tree}} \right],
\end{equation}
with $\widetilde{O}_9=\frac{\alpha_s}{4 \pi} \, O_9$ and
$\widetilde{C}_9^{(0)}=\frac{4 \pi}{\alpha_s}
\left(C_9^{(0)}+\frac{\alpha_s}{4 \pi}
C_9^{(1)}\right)$.

We evaluate diagram \ref{fig:789}b) keeping the strange quark mass $m_s$
as a regulator of collinear singularities.
The unrenormalized contributions of diagram \ref{fig:789}b)
to the form factors $F_9^{(7)}$ and $F_9^{(9)}$ read
\begin{eqnarray}
\label{F79}
\nonumber
    F_9^{(9)}[4b] &=&  - \frac{\mubeps}{\epsilon}\frac{4}{3} +
    \frac{\mubeps}{\epsilon_\IR} \frac{8}{3}
    \left( \s + \frac{1}{2}\, \s^2 + \frac{1}{3}\, \s^3 + \frac{1}{2}\ln(r)
\right)
    \\
&&    + \frac{8}{3} \ln(r) - \frac{2}{3} \ln^2(r)
    + \frac{16}{3} + \frac{20}{3}\, \s + \frac{16}{3}\, \s^2  +
\frac{116}{27}\, \s^3 \, , \nonumber \\
 F_9^{(7)}[4b] &=&  -\frac{2}{3}\s
\left(1 + \frac{1}{2}\, \s + \frac{1}{3}\, \s^2 \right)\, ,
\end{eqnarray}
where we kept all terms up to $\s^3$.
$\epsilon_\IR$ and $r=(m_s^2/m_b^2)$ regularize the infrared
and collinear singularities in eq. (\ref{F79}).

The $b$- and $s$-quark self-energy contributions are obtained
by multiplying the tree level matrix element
of $O_9$ by the quark field renormalization factor
$\delta \bar Z_\psi = \sqrt{ Z_\psi(m_b) Z_\psi(m_s)} - 1$,
where the explicit form for $Z_\psi(m)$ (in the on-shell scheme)
is given in eq. (\ref{quarkren}).

Adding the self-energy contributions and the
vertex correction, we get the ultraviolet finite results
\begin{eqnarray}
   \label{eq:FO9}
   && F_9^{(9)} = \frac{16}{3} + \frac{20}{3}\, \s +
        \frac{16}{3}\, \s^2  + \frac{116}{27}\, \s^3 + f_{\rm{inf}}\, , \\
   \label{eq:finf}
   && f_{\rm{inf}}=\frac{\mubeps}{\epsilon_\IR} \, \frac{8}{3} \, (1 + \s +
\frac{1}{2}\, \s^2 + \frac{1}{3}\, \s^3)
    + \frac{4}{3} \frac{\mubeps}{\epsilon_\IR} \, \ln(r) + \frac{2}{3} \ln(r)
- \frac{2}{3} \ln^2(r) \, , \\
&&  F_9^{(7)}  =  -\frac{2}{3}\s
\left(1 + \frac{1}{2}\, \s + \frac{1}{3}\, \s^2 \right)\, .
\end{eqnarray}

At this place, it is convenient to incorporate diagram \ref{fig:1}f)
together with its counterterms discussed in section
\ref{subsec:countertermso2}.

It is easy to see that the two loops in diagram
\ref{fig:1}f) factorize into two one-loop
contributions. The charm loop has the Lorentz structure of $O_9$
and can therefore be absorbed into a modified Wilson coefficient:
The renormalized diagram \ref{fig:1}f) is properly included
by modifying $\widetilde{C}^{(0)}_9$ in eq. (\ref{o9decomp}) as
follows:
\begin{equation}
    \label{c9replacement}
    \widetilde{C}^{(0)}_9 \longrightarrow
    \widetilde{C}^{(0,\text{mod})}_9=
    \widetilde{C}^{(0)}_9 + \left(C_2^{(0)} +
    \frac{4}{3} C_1^{(0)} \right)
\, H_0 \, ,
\end{equation}
where the charm-loop function $H_0$
reads (in expanded form)
\begin{equation}
\label{h0fun}
    H_0 = \frac{1}{2835} \left[
    -1260 + 2520 \ln \!\left(\frac{\mu}{m_c}\right) + 1008  \left(\frac{\hat{s}}{4z}\right) +
    432  \left(\frac{\hat{s}}{4z}\right)^2 +
    256  \left(\frac{\hat{s}}{4z}\right)^3 \right] \, .
\end{equation}
In the context of virtual corrections
also the $\Order(\epsilon)$-part of this loop function is needed.
We neglect it here
since it will drop out in combination with gluon bremsstrahlung.
Note that $H_0=h(z,\s)+8/9 \, \ln(\mu/m_b)$,
with $h$ defined in \cite{Buras:1995dj,Bobeth:2000mk}.

%
%
\subsection{Virtual corrections to the matrix element of $O_7$}
We now turn to the virtual corrections to the matrix element of the operator
$O_7$, consisting of the vertex-
(see Fig. \ref{fig:789}a)) and self-energy corrections. The ultraviolet
singularities of the sum of these diagrams are
cancelled when adding the counterterm amplitude
\begin{equation}
\label{countero7}
    C_7\left[Z_{77}Z_{m_b}/Z_{g_s}^2-1\right]\OSevenTree \, ,
\quad\text{where}\quad
    Z_{77} = 1 - \frac{\alpha_s}{4\pi} \frac{7}{3\epsilon} \, .
\end{equation}
The expressions for $Z_{m_b}$ and $Z_{g_s}$ are given in eqs.
(\ref{eq:Zm}) and (\ref{eq:Zg}), respectively.
The renormalized result for the contribution proportional to $C_7$ can be
written as
\begin{equation}
    \label{o7decomp}
    \bra s \ell^+ \ell^-|C_7 O_7 | b \ket =
    \widetilde{C}_7^{(0)} \left( -\frac{\alpha_s}{4 \pi} \right)
    \left[ F_7^{(9)} \bra \widetilde{O}_9 \ket_{\text{tree}} +
        F_7^{(7)} \bra \widetilde{O}_7 \ket_{\text{tree}} \right],
\end{equation}
with $\widetilde{O}_7=\frac{\alpha_s}{4 \pi} \, O_7$ and
$\widetilde{C}_7^{(0)}=C_7^{(1)}$.
The expanded form factors $F_7^{(9)}$ and $F_7^{(7)}$ read
\begin{eqnarray}
    F_7^{(9)} &=& -\frac{16}{3}\left(1 + \frac{1}{2}\, \s +
\frac{1}{3}\, \s^2 + \frac{1}{4}\, \s^3 \right)\, , \\
\label{eq:FO7}
    F_7^{(7)} &=& \frac{32}{3} \, L_\mu + \frac{32}{3} + 8\, \s + 6\, \s^2 +
\frac{128}{27}\, \s^3 + f_{\rm{inf}}\, ,
\end{eqnarray}
where the infrared- and collinear singular part $f_{\rm{inf}}$
is identical to the one of $F_9^{(9)}$ in eq. (\ref{eq:finf}).
Note that the on-shell value for the renormalization factor $Z_{m_b}$
was used in eq. (\ref{countero7}).
Therefore, when using the expression for $F_7^{(7,9)}$ in the form given
above, the
pole mass for $m_b$ has to be used at lowest order.
%
%
\subsection{Virtual corrections to the matrix element of $O_8$}
Finally, we present our results for the corrections to the matrix elements of
$O_8$. The corresponding diagrams are shown
in Fig. \ref{fig:789}c) and d). Including the counterterm
\[
    C_8 \, \delta Z_{87} \, \OSevenTree\, ,
\quad\text{where}\quad \delta Z_{87} =
- \frac{\alpha_s}{4\pi} \frac{16}{9\epsilon} \, ,
\]
yields the ultraviolet and infrared finite result
\begin{equation}
    \label{o8decomp}
    \bra s \ell^+ \ell^-|C_8 O_8 | b \ket =
    \widetilde{C}_8^{(0)} \left( -\frac{\alpha_s}{4 \pi} \right)
    \left[ F_8^{(9)} \bra \widetilde{O}_9 \ket_{\text{tree}} +
        F_8^{(7)} \bra \widetilde{O}_7 \ket_{\text{tree}} \right],
\end{equation}
with $\widetilde{C}_8^{(0)}=C_8^{(1)}$.
The expanded form factors $F_8^{(9)}$ and
$F_8^{(7)}$ read
\begin{eqnarray}
    F_8^{(9)} &=&
    \frac{104}{9} - \frac{32}{27}\, \pi^2 +
    \left(\frac{1184}{27} - \frac{40}{9}\, \pi^2\right) \s +
    \left(\frac{14212}{135} - \frac{32}{3}\, \pi^2 \right)\s^2
    \nonumber \\
 &&   + \left(\frac{193444}{945} - \frac{560}{27}\, \pi^2 \right) \s^3 +
    \frac{16}{9}\, L_s \left( 1 + \s + \s^2 +  \s^3 \right)\, ,
\end{eqnarray}
\begin{eqnarray}
    F_8^{(7)} &=&
    - \frac {32}{9}\, L_\mu +
    \frac{8}{27}\, \pi^2 - \frac{44}{9} - \frac{8}{9}\, i \pi +
    \left(\frac{4}{3}\, \pi^2 - \frac{40}{3} \right) \s
    + \left(\frac{32}{9}\, \pi^2 - \frac{316}{9} \right) \s^2
    \nonumber \\
 &&   + \left(\frac{200}{27}\, \pi^2 - \frac{658}{9} \right) \s^3 -
    \frac{8}{9}\, L_s  \left( \s + \s^2 + \s^3 \right)\, .
\end{eqnarray}
%
%
%
\section{Bremsstrahlung corrections}
\label{section:brems}
First of all, we remark
that in the present paper only those bremsstrahlung diagrams are taken
into account which are needed to cancel
the infrared and collinear singularities appearing in the
virtual corrections. All the
other bremsstrahlung contributions
(which are finite), will be given elsewhere \cite{AAGW:brems}.

It is known \cite{Buras:1995dj,Misiak:1993bc} that the contribution to the
inclusive decay width coming from
the interference between the tree-level and the one-loop matrix elements of
$O_9$ (Fig. \ref{fig:789}b)) and  from the
corresponding bremsstrahlung corrections (Fig. \ref{fig:789}f)) can be written
in the form
\begin{eqnarray}
\label{gamma99}
    \frac{d\Gamma_{99}}{d\s} &=&  \frac{d\Gamma_{99}^{\rm{virt}}}{d\s} +
\frac{d\Gamma_{99}^{\rm{brems}}}{d\s} \nonumber \\
   \frac{d\Gamma_{99}}{d\s} &=&
    \left( \frac{\alpha_{em}}{4\pi} \right)^2
    \frac{G_F^2\, m_{b,pole}^5 \left|V_{ts}^*V_{tb}\right|^2} {48\pi^3}(1-\s)^2
    \left( 1 + 2\s \right) \left[ 2 \left| \widetilde C_9^{(0)} \right|^2
\frac{\alpha_s}{\pi}\, \omega_9(\s)\right] \, ,
\end{eqnarray}
where $\widetilde C_9^{(0)}=\frac {4 \pi}{\alpha_s} \left (C_9^{(0)}+\frac
{\alpha_s}{4 \pi} C_9^{(1)} \right )$. This
procedure corresponds to encapsulating the virtual and bremsstrahlung
corrections in the tree-level calculation by
replacing $\bra O_9 \ket_\tree$ through
$\left( 1 + \frac{\alpha_s}{\pi}\,\omega_9(\s) \right)\bra O_9 \ket_\tree$.
The function $\omega_9(\s) \equiv \omega(\s)$,
which contains all information on
virtual and bremsstrahlung corrections,
can be found in \cite{Misiak:1993bc,Buras:1995dj} and is given by
\begin{eqnarray}
\label{omega9}
    \omega_9(\s) &=& -\frac{4}{3} \mbox{Li}(\s) -
\frac{2}{3} \ln(1-\s) \ln(\s) -
\frac{2}{9}\pi^2
    - \frac{5+4\s}{3(1+2\s)} \ln(1-\s) \nonumber \\
&&
    - \frac{2\s(1+\s)(1-2\s)}{3(1-\s)^2(1+2\s)} \ln(\s)
    + \frac{5+9\s-6\s^2}{6(1-\s)(1+2\s)} \, .
\end{eqnarray}
Replacing $\widetilde{C}_9^{(0)}$ by
$\widetilde{C}_9^{(0,\text{mod})}$ (see eq. (\ref{c9replacement})) in
eq. (\ref{gamma99}), diagram \ref{fig:1}f) and the corresponding
bremsstrahlung corrections are automatically included.

For the combination of the interference terms between the tree-level and the
one-loop matrix element of $O_7$
(Fig. \ref{fig:789}a)) and the corresponding bremsstrahlung corrections (Fig.
\ref{fig:789}e)) we make the ansatz
\begin{eqnarray}
    \frac{d\Gamma_{77}}{d\s} &=& \frac{d\Gamma_{77}^\virt}{d\s} +
\frac{d\Gamma_{77}^\brems}{d\s} \nonumber \\
    \frac{d\Gamma_{77}}{d\s} &=&
    \left(\frac{\alpha_{em}}{4\pi}\right)^2
    \frac{G_F^2\, m_{b,pole}^5 \left|V_{ts}^*V_{tb}\right|^2} {48\pi^3}(1-\s)^2
   \, 4 \, (1+2/\s) \left[ 2\left| \widetilde C_7^{(0)} \right|^2
\frac{\alpha_s}{\pi}\, \omega_7(\s) \right] \, ,
\end{eqnarray}
where $\widetilde C_7^{(0)}=C_7^{(1)}$. This time, the
encapsulation of virtual
and bremsstrahlung corrections is provided
by the replacement
$\bra O_7 \ket_\tree \rightarrow
\left( 1 + \frac{\alpha_s}{\pi}\, \omega_7(\s)
\right)\bra O_7 \ket_\tree$.
In order to simplify the calculation of $\omega_7(\s)$,
we make the important observation that the form factors $F^{(7)}_7$ and
$F^{(9)}_9$ have the same infrared divergent part
$f_{\rm{inf}}$
(eq. (\ref{eq:FO7}) and (\ref{eq:FO9})), whereas $F^{(9)}_7$ and
$F^{(7)}_9$ are finite. Taking into account that
in $d$ dimensions the decay width $d\Gamma(b \to s \ell^+ \ell^-)/d\s$
corresponding to the matrix element
\begin{equation}
M(b \to s \ell^+ \ell^- ) = \bra s \ell^+ \ell^-|
\widetilde{C}_7^{(0)} \, \widetilde{O}_7^{(0)} +
\widetilde{C}_9^{(0)} \, \widetilde{O}_9^{(0)} +
\widetilde{C}_{10}^{(0)} \, \widetilde{O}_{10}^{(0)} |b \ket_{\text{tree}}
\end{equation}
is given by
\begin{eqnarray}
&&\nonumber \frac{d\Gamma(b\to X_s \ell^+\ell^-)}{d\s} =
    \left(\frac{\alpha_{em}}{4\pi}\right)^2 \,
    \frac{G_F^2 m_{b,pole}^5\left|V_{ts}^*V_{tb}\right|^2}{48 \pi^3} \,
    (1-\s)^2 \, (1+ {\cal O}(d-4))
    \times \\
    &&
    \left ( \left (1+(d-2)\hat s\right)
    \left (\left |\widetilde C_9^{(0)}\right |^2+
    \left |\widetilde C_{10}^{(0)}\right |^2 \right )
    + 4(1+(d-2)/\s)\left
    |\widetilde C_7^{(0)}\right |^2+
    4(d-1) \mbox{Re}\left (\widetilde C_7^{(0)}
    \widetilde C_9^{(0)*}\right ) \right ) \, ,
\end{eqnarray}
one concludes that the combination
\begin{equation}
    \Delta\Gamma^\virt  =  \frac{\left| \widetilde C_9^{(0)} \right|^{-
2}}{1+(d-2)\s} \frac{d\Gamma^\virt_{99}}{d\s}
    - \frac{\left| \widetilde C_7^{(0)} \right|^{-2}}{4(1+(d-2)/ \s)}
\frac{d\Gamma^\virt_{77}}{d\s} \, ,
\end{equation}
is free of infrared and collinear
singularities. Defining analogously
\begin{equation}
\Delta\Gamma^\brems  =  \frac{\left| \widetilde C_9^{(0)}
\right|^{-2}}{1+(d-2)\s} \frac{d\Gamma^\brems_{99}}{d\s}
    - \frac{\left| \widetilde C_7^{(0)} \right|^{-2}}{4(1+(d-2)/ \s)}
\frac{d\Gamma^\brems_{77}}{d\s}
\end{equation}
and using the identity
\begin{eqnarray}
\label{omega7det}
\frac{\left| \widetilde C_9^{(0)} \right|^{-2}}{1+(d-
2)\s}\frac{d\Gamma_{99}}{d\s}-
\frac{\left| \widetilde C_7^{(0)} \right|^{-2}}{4(1+(d-2)/
\s)}\frac{d\Gamma_{77}}{d\s}=\Delta\Gamma^\virt
+\Delta\Gamma^\brems \, ,
\end{eqnarray}
one concludes that also $\Delta\Gamma^\brems$ is finite. This is
because $d\Gamma_{99}/d\s$ and $d\Gamma_{77}/d\s$ are finite
due to the Kinoshita-Lee-Nauenberg theorem and because
$\Delta\Gamma^\virt$ is finite as mentioned above. The calculation
of $\Delta\Gamma^\brems$ is straightforward, as the integrand,
expanded in $\epsilon$, leads to unproblematic integrals. Using
the explicit results for $\Delta\Gamma^\virt$,
$\Delta\Gamma^\brems$ and $\omega_9(\s)$, one can readily extract
$\omega_7(\s)$ from eq. (\ref{omega7det}):
\begin{eqnarray}
    \omega_7(\s) &=&
    - \frac{8}{3}\,\ln \left( \frac{\mu}{m_b} \right)
    - \frac{4}{3}\, \mbox{Li}(\s)
    - \frac{2}{9}\, \pi^2
    - \frac{2}{3}\, \ln(\s) \ln(1-\s)
    - \frac{1}{3}\, \frac{8 + \s}{2 + \s} \ln (1-\s) \nonumber \\ &&
    - \frac{2}{3}\, \frac {\s \left( 2 - 2\,\s - \s^2 \right)}{
       (1-\s)^2 (2+\s)} \ln(\s)
    - \frac{1}{18}\, \frac {16 - 11\,\s - 17\,\s^2}{(2+\s)(1-\s)} \, .
\end{eqnarray}
The reasoning for the interference terms between the tree-level matrix element
of $O_7$ and the one-loop matrix element of
$O_9$ and vice versa is analogous: We may combine this contribution with the
corresponding bremsstrahlung terms coming from
the interference of diagrams \ref{fig:789}e) and \ref{fig:789}f) making the
ansatz
\begin{eqnarray}
    \frac{d\Gamma_{79}}{d\s} &=& \frac{d\Gamma_{79}^\virt}{d\s} +
\frac{d\Gamma_{79}^\brems}{d\s} \nonumber \\
\frac{d\Gamma_{79}}{d\s} &=&
    \left(\frac{\alpha_{em}}{4\pi}\right)^2
    \frac{G_F^2\, m_{b,pole}^5 \left|V_{ts}^*V_{tb}\right|^2}{48\pi^3}(1-\s)^2
    \, 12 \, \left[ 2 \, \Re \left(
     \widetilde{C}_7^{(0)} \widetilde{C}_9^{(0)*} \right)
 \, \frac{\alpha_s}{\pi} \,\omega_{79}(\s)\, \right]\, .
\end{eqnarray}
The corresponding encapsulation is realized by the replacement
$\bra O_{7,9} \ket_\tree \rightarrow \left( 1 +
\frac{\alpha_s}{\pi}\omega_{79}(\s) \right)\bra O_{7,9} \ket_\tree$.
This time, we make use of the fact that the quantities
\begin{eqnarray}
    \Delta\Gamma^\virt_{\text{mixed}} & = & \frac{\left| \widetilde C_9^{(0)}
\right|^{-2}}{1+(d-2)\s}
        \frac{d\Gamma^\virt_{99}}{d\s}
    - \frac{ \Re\left[ \widetilde{C}_7^{(0)} \widetilde{C}_9^{(0)*}\right]^{-
1}}{4(d-1)}
        \frac{d\Gamma^\virt_{79}}{d\s} \quad\quad\quad\text{and} \\
    \Delta\Gamma^\brems_{\text{mixed}} & = & \frac{\left| \widetilde C_9^{(0)}
\right|^{-2}}{1+(d-2)\s}
        \frac{d\Gamma^\brems_{99}}{d\s}
    - \frac{ \Re\left[ \widetilde{C}_7^{(0)} \widetilde{C}_9^{(0)*}\right]^{-
1}}{4(d-1)}
        \frac{d\Gamma^\brems_{79}}{d\s}
\end{eqnarray}
are finite. For the function $\omega_{79}(\s)$ we obtain
\begin{eqnarray}
    \omega_{79}(\s) &=&
    - \frac{4}{3}\,\ln \left (\frac {\mu}{m_b}\right)
    - \frac{4}{3}\, \mbox{Li}(\s) -
    \frac{2}{9}\,\pi^2 -
    \frac{2}{3}\,\ln(\s) \ln(1 - \s)
    - \frac{1}{9}\,\frac{2 + 7\s}{\s} \ln (1-\s) \nonumber \\ &&
    - \frac{2}{9}\, \frac{\s \left(3 - 2\,\s \right)}
    {\left( 1 - \s \right)^{2}} \ln(\s)
    + \frac{1}{18}\, \frac{5- 9\,\s}{1-\s} \, .
\end{eqnarray}
Note that the procedure described here does work only if one of the functions
$\omega_7(\s)$, $\omega_9(\s)$ or
$\omega_{79}(\s)$ is known already.

Finally, we remark that the combined virtual- and bremsstrahlung
corrections to the operator $O_{10}$ (which has the same hadronic structure
as $O_9$) is described by the function $\omega_9(\s)$, too:
\begin{eqnarray}
\label{gamma1010}
    \frac{d\Gamma_{10,10}}{d\s} &=&  \frac{d\Gamma_{10,10}^{\rm{virt}}}{d\s} +
\frac{d\Gamma_{10,10}^{\rm{brems}}}{d\s} \nonumber \\
   \frac{d\Gamma_{10,10}}{d\s} &=&
    \left( \frac{\alpha_{em}}{4\pi} \right)^2
    \frac{G_F^2\, m_{b,pole}^5 \left|V_{ts}^*V_{tb}\right|^2} {48\pi^3}(1-\s)^2
    \left( 1 + 2\s \right) \left[ 2 \left( \widetilde C_{10}^{(0)} \right)^2
\frac{\alpha_s}{\pi}\, \omega_9(\s)\right] \, ,
\end{eqnarray}
where $\widetilde C_{10}^{(0)}=C_{10}^{(1)}$.
%
%
\section{Corrections to the decay width for $b \to X_s \ell^+ \ell^-$}
\label{section:decaywidth}
In this chapter we combine  the virtual corrections calculated in chapters
\ref{section:virtO1O2}, \ref{section:virto789} and the bremsstrahlung
contributions discussed in chapter \ref{section:brems} and study their
influence on the decay width
$d\Gamma(b \to X_s \ell^+ \ell^-)/d\s$. In the literature (see e.g.
\cite{Bobeth:2000mk}), this decay width is usually
written as
\begin{multline}
\label{rarewidth}
    \frac{d\Gamma(b\to X_s \ell^+\ell^-)}{d\s} =
    \left(\frac{\alpha_{em}}{4\pi}\right)^2
    \frac{G_F^2 m_{b,pole}^5\left|V_{ts}^*V_{tb}\right|^2}
    {48\pi^3}(1-\s)^2  \, \times
    \\
    \left ( \left (1+2\hat s\right)
    \left (\left |\widetilde C_9^{\eff}\right |^2+
    \left |\widetilde C_{10}^{\eff}\right |^2 \right )
    + 4(1+2/\s)\left
    |\widetilde C_7^{\eff}\right |^2+
    12 \, \mbox{Re}\left (\widetilde C_7^{\eff}
    \widetilde C_9^{\eff*}\right ) \right ) \, ,
\end{multline}
where the contributions calculated so far have been absorbed into the effective
Wilson coefficients
$\widetilde{C}_7^{\eff}$,
$\widetilde{C}_9^{\eff}$ and $\widetilde{C}_{10}^{\eff}$.
It turns out that also the new contributions calculated in the
present paper can
be absorbed into these coefficients.
Following as closely as possible the 'parameterization'
given recently by Bobeth
et al. \cite{Bobeth:2000mk}, we write
\begin{eqnarray}
    \label{effcoeff}
    \nonumber
    \widetilde C_9^{\eff} & = &
    \left (1+\frac{\alpha_s(\mu)}{\pi}
    \omega_9 (\hat{s})\right )
    \big( A_9 + T_9 \, h(z,\s) + U_9 \, h(1,\s) + W_9 \, h(0,\s) \big)
    \\
    && \quad\quad\quad - \frac{\alpha_{s}(\mu)}{4\pi}\left(C_1^{(0)} F_1^{(9)} +
    C_2^{(0)} F_2^{(9)} +
    A_8^{(0)} F_8^{(9)}\right)\\
    \nonumber
    \widetilde C_7^{\eff} & = & \left (1+\frac{\alpha_s(\mu)}{\pi}
    \omega_7 (\s)\right )A_7
    -\frac{\alpha_s(\mu)}{4\pi}\left(C_1^{(0)} F_1^{(7)}+
    C_2^{(0)} F_2^{(7)}+A_8^{(0)} F_8^{(7)}\right)\\
    \nonumber
    \widetilde C_{10}^{\eff} & = & \left( 1 + \frac{\alpha_s(\mu)}{\pi}
    \omega_9(\s) \right) A_{10} \, ,
\end{eqnarray}
where the expressions for $h(z,\s)$ and $\omega_9(\s)$
(see eqs. (\ref{h0fun}) and (\ref{omega9}))
were already available in the literature
\cite{Misiak:1993bc,Buras:1995dj,Bobeth:2000mk}.
The quantities $\omega_7(\s)$ and $F_{1,2,8}^{(7,9)}$, on the other hand,
have been calculated in the present paper.
We take the numerical values for $A_7$, $A_9$, $A_{10}$, $T_9$, $U_9$, and
$W_9$ from \cite{Bobeth:2000mk},
while $C_1^{(0)}$, $C_2^{(0)}$ and $A_8^{(0)}=\widetilde{C}_8^{(0,\eff)}$
can be found in \cite{Greub:2001sy}.
For completeness we list them  in Tab. \ref{table3}.
\begin{table}[htb]
\begin{center}
\begin{tabular}{| c | c | c | c |}
    &$\mu=2.5$ GeV& $\mu=5$ GeV  & $\mu=10$ GeV\\ \hline
    $\alpha_s$ & $0.267$&  $0.215$ & $  0.180$\\ \hline
    $C_1^{(0)}$ & $-0.697$ & $-0.487$ & $-0.326$\\ \hline
    $C_2^{(0)}$ & $1.046 $& $ 1.024$ &  $ 1.011$\\ \hline
    $\big(A_7^{(0)},~A_7^{(1)}\big)$ & $ (-0.360,~0.031)$ & $(-0.321,~0.019)$
& $ (-0.287,~0.008)$\\ \hline
    $A_8^{(0)}$ &$ -0.164 $& $ -0.148$& $ -0.134$\\ \hline
    $\big(A_9^{(0)},~A_9^{(1)}\big)$ & $ (4.241,~-0.170) $& $(4.129,~0.013)$ & $
(4.131,~0.155)$ \\ \hline
    $\big(T_9^{((0))},~T_9^{(1)}\big)$ & $ (0.115,~0.278) $& $ (0.374,~0.251)$ &
$ (0.576,~0.231)$ \\ \hline
    $\big(U_9^{(0)},~U_9^{(1)}\big)$ & $ (0.045,~0.023)$ & $ (0.032,~0.016)$ & $
(0.022,~0.011)$ \\ \hline
    $\big(W_{9}^{(0)},~W_{9}^{(1)}\big)$ & $(0.044,~0.016)$ & $ (0.032,~0.012)$
& $ (0.022,~0.009)$ \\ \hline
    $\big(A_{10}^{(0)},~A_{10}^{(1)}\big)$ & $ (-4.372,~0.135)$ &  $(-
4.372,~0.135)$ &  $ (-4.372,~0.135)$ \\
\end{tabular}
\end{center}
\caption{
    Coefficients appearing in eq. (\ref{effcoeff}) for $\mu = 2.5$ GeV, $\mu =5$
GeV and $\mu = 10$ GeV.
    For $\alpha_s(\mu)$ (in the $\overline{\mbox{MS}}$ scheme) we used the two-
loop expression with 5 flavors and
    $\alpha_s(m_Z)=0.119$. The entries correspond to the pole top quark mass
$m_t= 174 $ GeV. The superscript (0) refers
    to lowest order quantities while the superscript (1) denotes
the correction terms of order $\alpha_s$.}
\label{table3}
\end{table}

\begin{figure}[htb]
\begin{center}
    \includegraphics[width=13.0cm, bb=18 177 579 582]{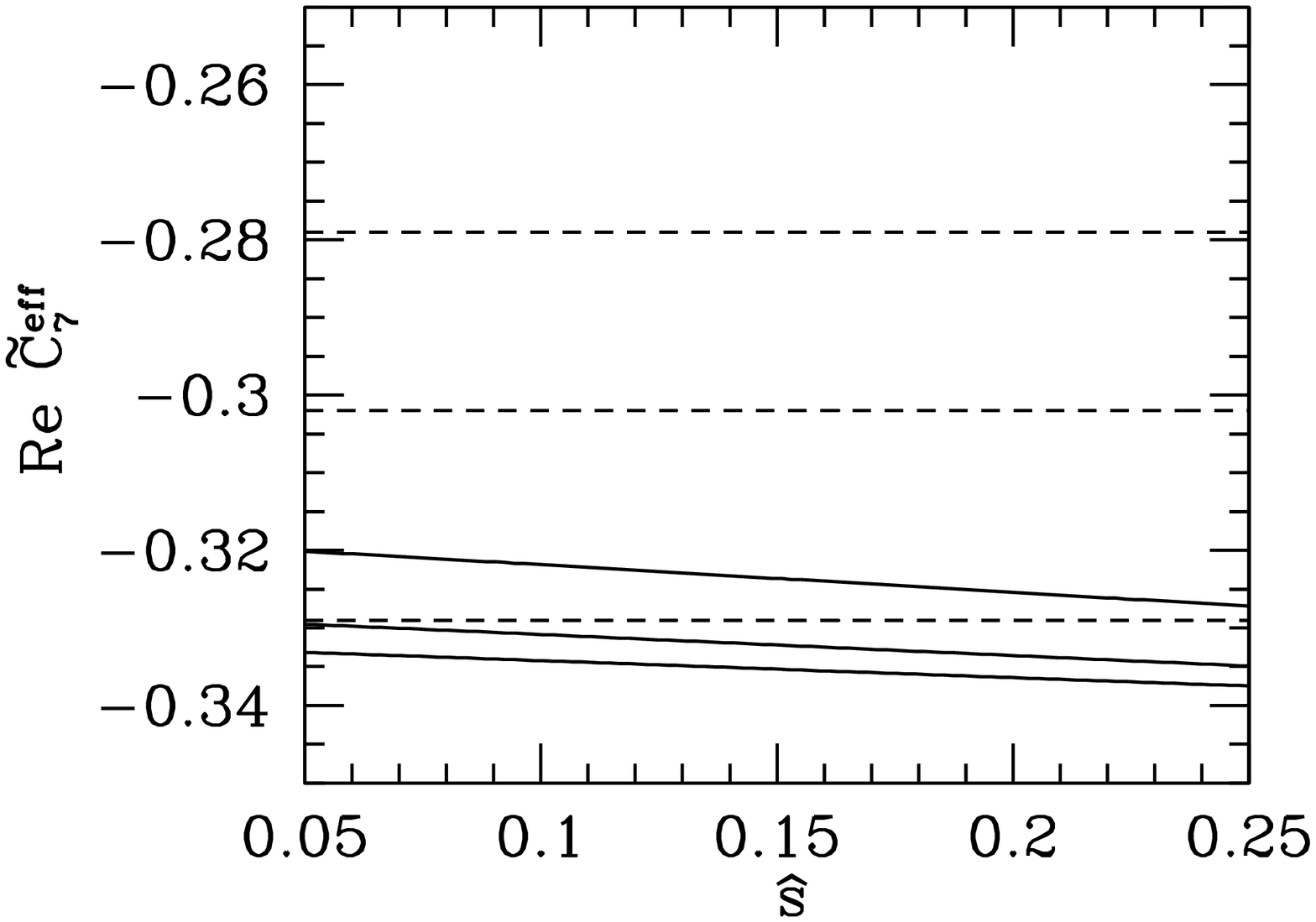}
    \vspace{0.5cm}
    \caption[]{The three solid curves illustrate the $\mu$ dependence of
    $\mbox{Re} \, \widetilde C_7^{\eff}(\s)$ when the new
    corrections are included.
    The dashed curves are obtained when switching off these corrections.
    We set $\hat{m}_c=0.29$. See text.}
\label{fig:c7eff}
\end{center}
\end{figure}
In Fig. \ref{fig:c7eff} we illustrate the renormalization scale
dependence of  $\mbox{Re} \, \widetilde C_7^{\eff}(\s)$. The dashed curves
are obtained by neglecting the corrections calculated in this paper,
i.e., $\omega_7(\s)$, $F_1^{(7)}$, $F_2^{(7)}$ and $F_8^{(7)}$
are put equal to
zero in eq. (\ref{effcoeff}).
The three curves correspond to the values of the renormalization scale
$\mu=2.5$ GeV (lowest), $\mu=5$ GeV (middle)
and $\mu=10$ GeV (uppermost). The solid curves are obtained
by taking into account the new corrections. In this case, the lowest, middle
and uppermost curve correspond to
$\mu=10$ GeV, 5 GeV and 2.5 GeV, respectively.
We conclude that the new corrections significantly reduce the
renormalization scale dependence of $\mbox{Re} \, \widetilde C_7^{\eff}(\s)$.

\begin{figure}[htb]
\begin{center}
    \includegraphics[width=13.0cm,bb=18 177 579 582]{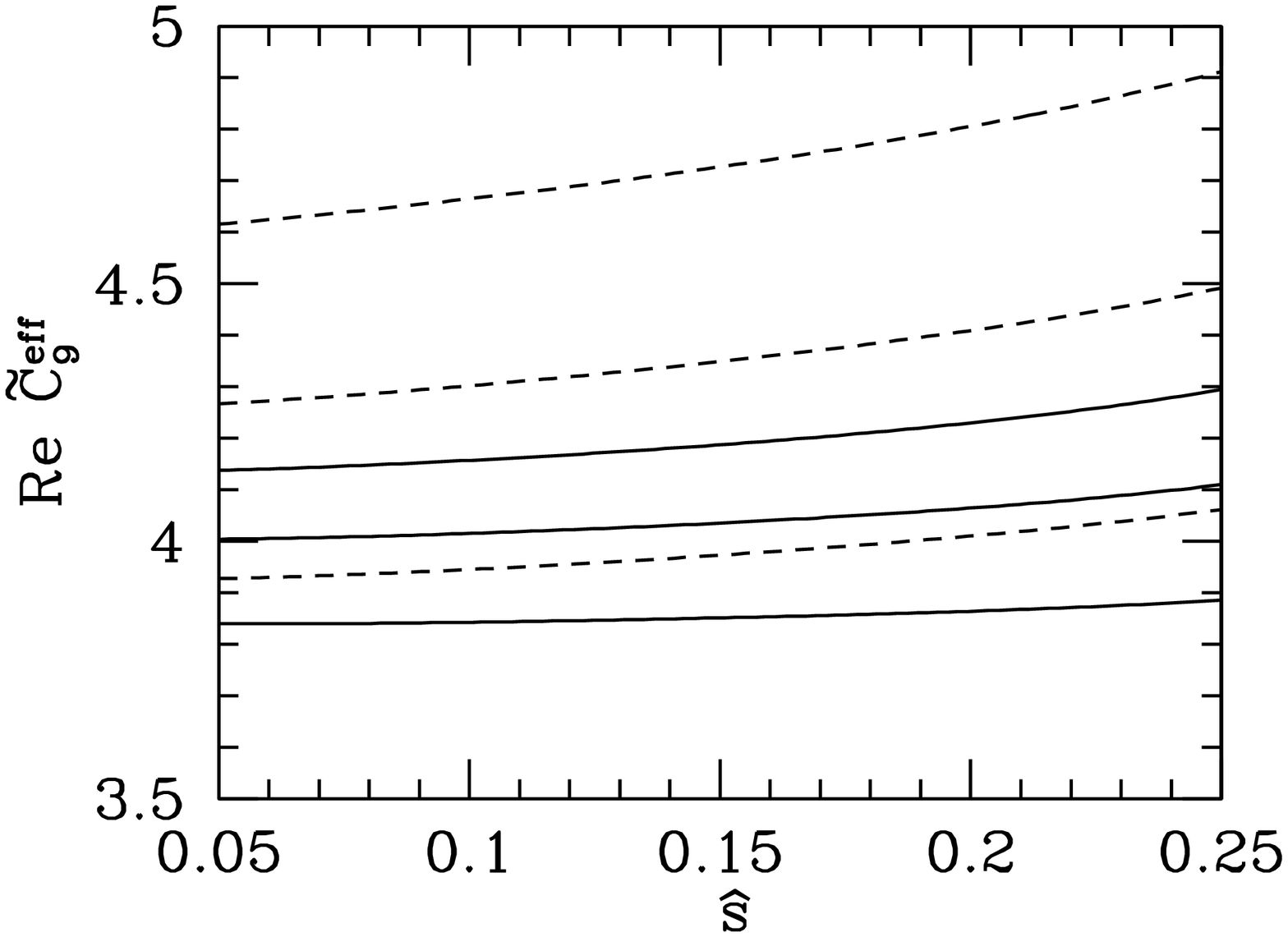}
    \vspace{0.5cm}
    \caption[]{The three solid curves illustrate the $\mu$ dependence of
    $\mbox{Re} \, \widetilde C_9^{\eff}(\s)$
    when the new corrections are included.
    The dashed curves are obtained when switching off these corrections.
    We set $\hat{m}_c=0.29$. See text.}
\label{fig:c9eff}
\end{center}
\end{figure}
Fig. \ref{fig:c9eff} shows  the renormalization scale
dependence of  $\mbox{Re} \, \widetilde C_9^{\eff}(\s)$.
Again, the dashed curves
are obtained by neglecting the new corrections in eq. (\ref{effcoeff}),
i.e., $F_1^{(9)}$, $F_2^{(9)}$ and $F_8^{(9)}$ are put to
zero. We stress that $\omega_9(\s)$ is retained, as this function has
been known before.
The three curves correspond to the values of the renormalization scale
$\mu=2.5$ GeV (lowest), $\mu=5$ GeV (middle)
and $\mu=10$ GeV (uppermost). The solid curves take the new corrections
into account. Now, the lowest, middle
and uppermost curve correspond to $\mu=2.5$ GeV, 5 GeV and 10 GeV,
respectively.
We conclude that the new corrections significantly reduce the
renormalization scale dependence of
$\mbox{Re} \, \widetilde C_9^{\eff}(\s)$, too.

When calculating the decay width (\ref{rarewidth}),
we retain only terms linear in $\alpha_s$
(and thus in  $\omega_7$, $\omega_9$) in the expressions for
$|\widetilde C_7^{\eff}|^2$,
$|\widetilde C_9^{\eff}|^2$
and $|\widetilde C_{10}^{\eff}|^2$.
In the interference term $\Re \left (\widetilde C_7^{\eff} \widetilde
C_9^{\eff*}\right )$ too, we keep only  linear contributions
in $\alpha_s$. By construction, one has to make the replacements $\omega_9 \to
\omega_{79}$ and
$\omega_7 \to \omega_{79}$ in this term.

Our results include all the relevant virtual corrections and those
bremsstrahlung diagrams which generate infrared and collinear
singularities. There exist
additional bremsstrahlung terms coming e.g. from one-loop $O_1$ and $O_2$
diagrams in which both, the virtual photon and
the gluon, are emitted from the charm quark line. These contributions do not
induce additional renormalization scale
dependence as they are ultraviolet finite. Using our experience from $b \to s
\gamma$ and $b \to s g$, these contributions
are not expected to be large, but to give a definitive answer
concerning their size, they have to be calculated
\cite{AAGW:brems}.
%
%
\section{Numerical results for $R_{\text{quark}}(\s)$}
\label{section:numres}
The decay width in eq. (\ref{rarewidth}) has a large uncertainty due to the
factor
$m_{b,pole}^5$. Following common practice, we consider the ratio
\begin{equation}
    R_{\text{quark}}(\s)=\frac{1}{\Gamma(b \to X_c e\bar{\nu})}
\, \frac{d\Gamma(b\to X_s \ell^+ \ell^-)}{d\hat s} \, ,
\end{equation}
in which the factor $m_{b,pole}^5$ drops out. The explicit expression for the
semi-leptonic decay width $\Gamma(b \to X_c e \nu_e)$ reads
\begin{equation}
\label{widthsl}
\Gamma(b \to X_c e\bar{\nu}) = \frac{G_F^2 \, m_{b,pole}^5}{192 \pi^3}
\, |V_{cb}|^2 \, g \! \left( \frac{m_{c,pole}^2}{m_{b,pole}^2} \right) \,
K \! \left( \frac{m_c^2}{m_b^2} \right) \, ,
\end{equation}
where $g(z)=1-8 \,z +8 \, z^3 - z^4 -12 \, z^2 \, \ln(z) \, $
is the phase space factor, and
\begin{equation}
K(z) = 1 - \frac{2 \alpha_s(m_b)}{3\pi} \, \frac{f(z)}{g(z)}
\end{equation}
incorporates the next-to-leading QCD correction to the semi-leptonic
decay \cite{Cabibbo}. The function $f(z)$ has been given analytically
in ref. \cite{Nir}:
\begin{eqnarray}
\label{ffun}
&&f(z) = -(1-z^2) \, \left( \frac{25}{4} - \frac{239}{3} \, z +
\frac{25}{4} \, z^2 \right) + z \, \ln(z) \left( 20 + 90 \, z
-\frac{4}{3} \, z^2 + \frac{17}{3} \, z^3 \right)
\nonumber \\
&& + z^2 \, \ln^2(z) \, (36+z^2)
+ (1-z^2) \, \left(\frac{17}{3} -\frac{64}{3} \, z + \frac{17}{3} \, z^2
\right) \, \ln (1-z) \nonumber \\
&& -4 \, (1+30 \, z^2 + z^4) \, \ln(z) \ln(1-z)
-(1+16 \, z^2 +z^4)  \left( 6 \, \mbox{Li}(z) - \pi^2 \right)
\nonumber \\
&& -32 \, z^{3/2} (1+z) \left[\pi^2 - 4 \, \mbox{Li}(\sqrt{z})+
4 \, \mbox{Li}(-\sqrt{z}) - 2 \ln(z) \, \ln \left(
\frac{1-\sqrt{z}}{1+\sqrt{z}} \right) \right] \, .
\end{eqnarray}

\begin{figure}[htb]
\begin{center}
    \includegraphics[width=13.0cm, bb=10 205 592 540]{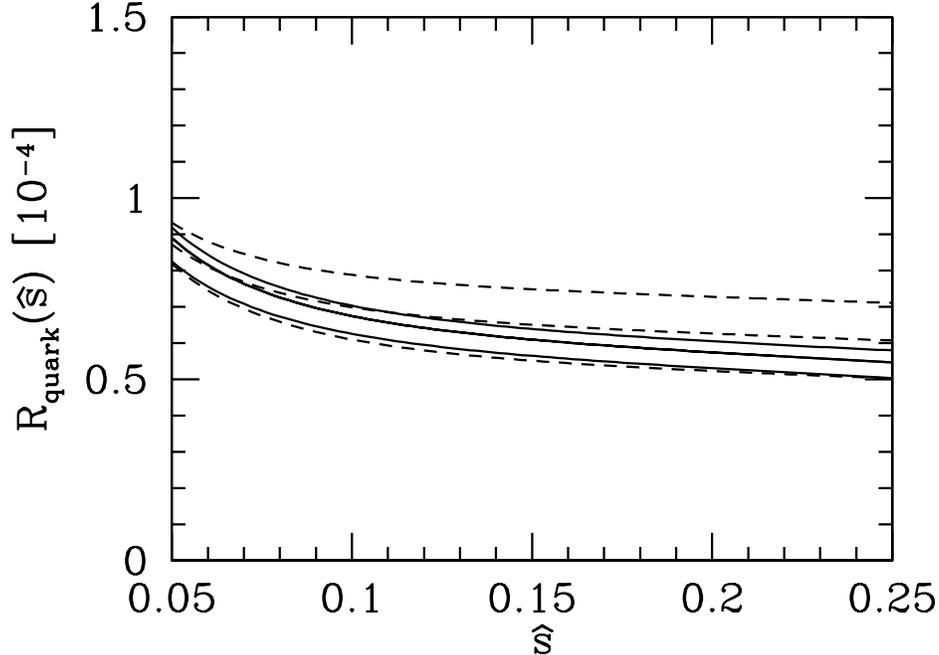}
    \vspace{0.5cm}
    \caption[]{    The three solid lines show the $\mu$ dependence
                   of  $R_{\text{quark}}(\s)$ when
                   including the corrections to the matrix elements
                   calculated in this paper; the dashed lines are obtained when
                   switching off these corrections. We set $\hat{m}_c=0.29$.
                   See text.}
\label{fig:mudependence}
\end{center}
\end{figure}
We now turn to the numerical results for $R_{\text{quark}}(\s)$ for
$0.05 \le \s \le 0.25$. In Fig. \ref{fig:mudependence} we investigate  the
dependence of
$R_{\text{quark}}(\s)$ on the renormalization scale $\mu$. The solid lines are
obtained by including the new NNLL
contributions, as explained in chapter \ref{section:decaywidth}. The
three solid curves correspond
to $\mu=2.5$ GeV (lowest line), $\mu=5$ GeV (middle line) and
$\mu=10$ GeV (uppermost line).
The three dashed curves (again $\mu=2.5$ GeV for the lowest,
$\mu=5$ GeV for the
middle and $\mu=10$ GeV for the uppermost line),
on the other hand, show the results without the new NNLL corrections,
i.e., they include the NLL
results combined with the NNLL corrections to the matching conditions
as obtained by Bobeth et al. \cite{Bobeth:2000mk}.
{}From this figure, we conclude that the renormalization scale dependence
gets reduced by more than a factor of 2. Only for low values of $\s$
($\s \sim 0.05$), where the NLL $\mu$-dependence is small already, the
reduction factor is smaller. For the integrated quantity we obtain
\begin{equation}
\label{Rint}
    R_{\text{quark}} = \int_{0.05}^{0.25} \, d\s \, R_{\text{quark}}(\s) =
    (1.25 \pm 0.08 (\mu)) \times 10^{-5} \, ,
\end{equation}
where the error is obtained by varying $\mu$ between 2.5 GeV and 10 GeV.
Before our corrections, the result was $R_{\text{quark}}=(1.36 \pm 0.18)\times
10^{-5}$ \cite{Bobeth:2000mk}. In other
words, the renormalization scale dependence got reduced from
$\sim \pm 13\%$ to $\sim \pm  6.5\%$.

\begin{figure}
\begin{center}
    \includegraphics[width=7.5cm, bb=18 177 579 582]{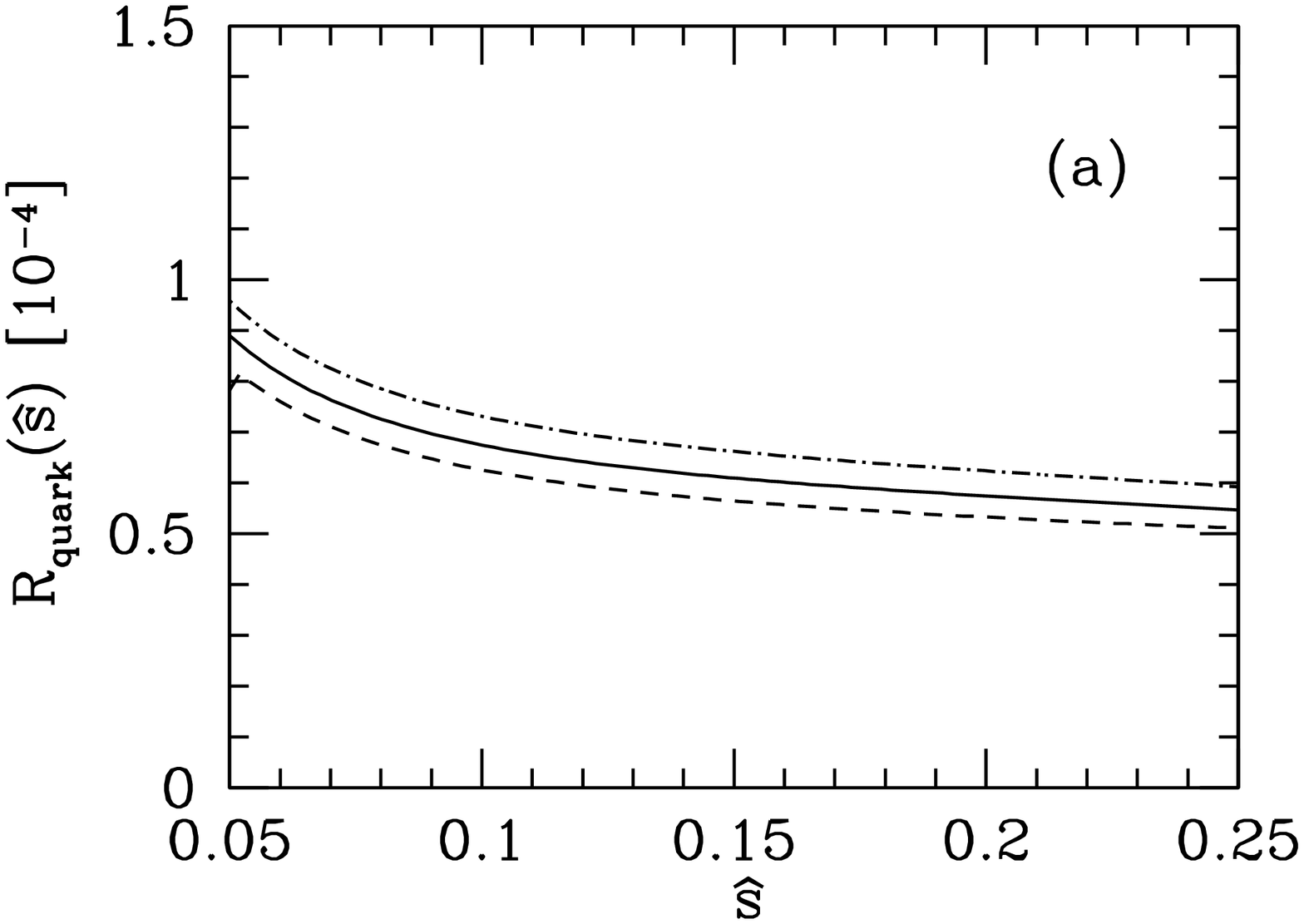}
    \includegraphics[width=7.5cm, bb=18 177 579 582]{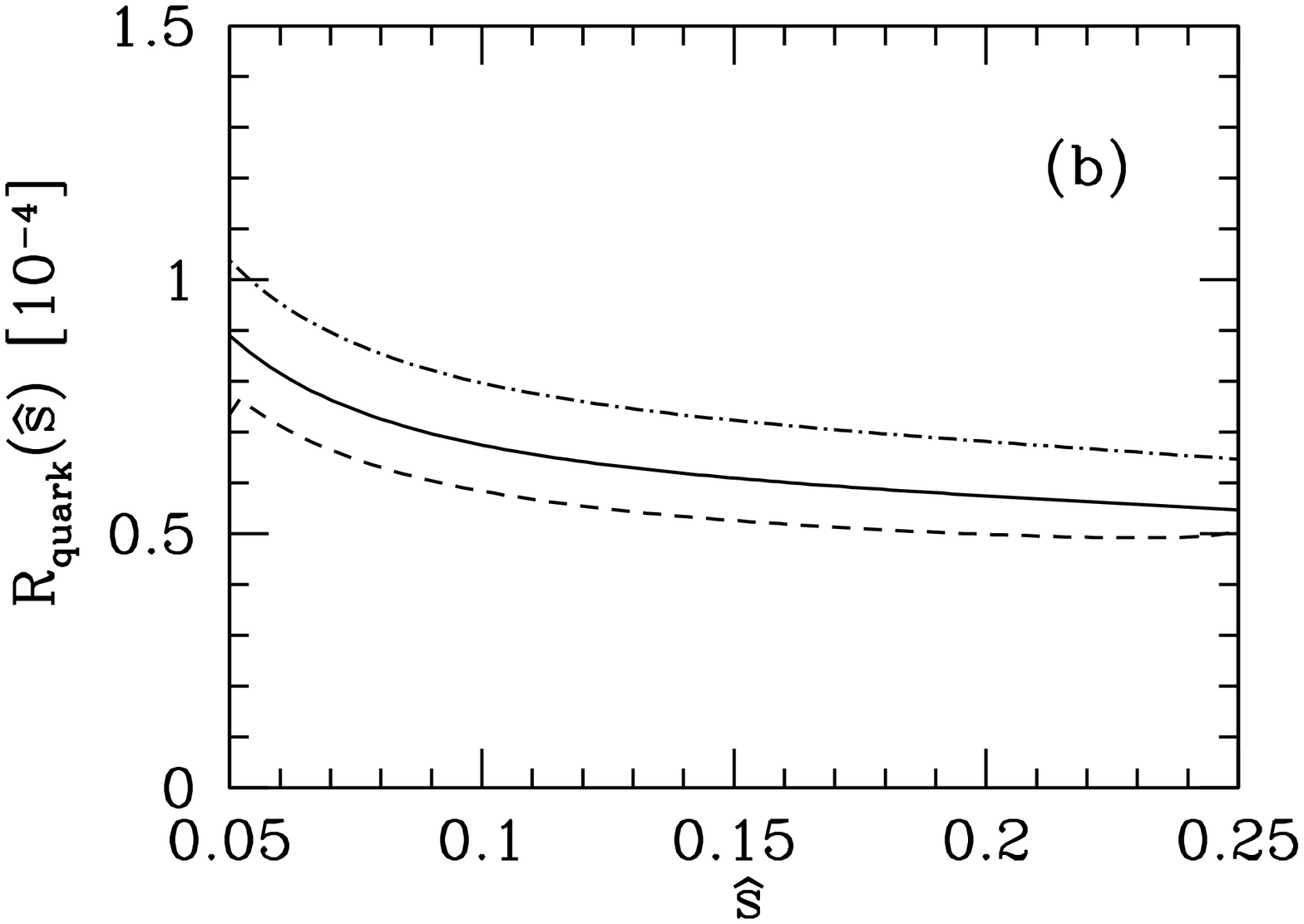}
    \vspace{0.5cm}
    \caption[]{a)  $R_{\text{quark}}(\s)$ for
                    $\hat{m}_c = 0.27$ (dashed line),
                    $\hat{m}_c = 0.29$ (solid line) and
                    $\hat{m}_c = 0.31$ (dash-dotted line) and $\mu$=5 GeV.
               b)  $R_{\text{quark}}(\s)$ for
                    $\hat{m}_c = 0.25$ (dashed line),
                    $\hat{m}_c = 0.29$ (solid line) and
                    $\hat{m}_c = 0.33$ (dash-dotted line) and $\mu$=5 GeV.
                    See text.}
\label{fig:mcdependence}
\end{center}
\end{figure}
Among the errors on  $R_{\text{quark}}(\s)$ which are due to the uncertainties
in the input parameters, the one induced by
$\hat{m}_c = m_c/m_b$ is known to be the largest. 
We repeat at this point that $m_c$ enters 
(unlike in $B \to X_s \gamma$) already  the one-loop diagrams associated
with $O_1$ and $O_2$. We did the renormalization of the charm quark mass in 
such a way that $m_c$ has the meaning of the pole mass in the one-loop
expressions. The meaning of $m_c$ in the corresponding two-loop matrix
elements, on the other hand, is not fixed (for a discussion of this issue
for $B \to X_s \gamma$, see ref. \cite{Gambino01}). 
As the running charm mass at a
scale of $\Order(m_b)$ is smaller than the pole mass, it numerically makes a 
difference whether one inserts a pole mass- or a running mass value for
$m_c$ in the two-loop contributions. In a thorough phenomenological
analysis this issue should certainly  be included when estimating the 
theoretical error. We decide, however, to postpone the quantitative 
discussion of this point and will take it up when also the finite
bremsstrahlung contibutions, which complete the NNLL calculation
of $R_{\text{quark}}(\s)$, are available \cite{AAGW:brems}.
For the time being,
we interpret $m_c$ to be the pole mass in the two-loop contributions. 
In Fig. \ref{fig:mcdependence}a)
we vary $\hat{m}_c$ between 0.27 and 0.31, while in Fig. 
\ref{fig:mcdependence}b)
the more conservative range $0.25 \le \hat{m}_c \le 0.33$ is considered.
Comparing Fig. \ref{fig:mudependence}
with Figs. \ref{fig:mcdependence}a) and b), we find
that at the NNLL level the uncertainty due to $\hat{m}_c$ is larger
 than the left-over
$\mu$-dependence, even for the less conservative range
of $\hat{m}_c$. For the
integrated quantity $R_{\text{quark}}$ we have an uncertainty of 
$\pm 7.6\%$ when
$\hat{m}_c$ is varied between 0.27 and 0.31.
Varying $\hat{m}_c$ in the more conservative range,
the corresponding uncertainty amounts to $\pm 15 \%$.

A more detailed numerical analysis for $R_{\text{quark}}(\s)$ and
$R_{\text{quark}}$, including the errors which are due to uncertainties
in other input parameters as well as non-perturbative effects,
will be given in ref. \cite{AAGW:brems}.


To conclude:
We have calculated virtual corrections of $\Order(\alpha_s)$ to the
matrix elements of $O_1$, $O_2$, $O_7$, $O_8$, $O_9$ and $O_{10}$.
We also took into account those bremsstrahlung corrections which cancel
the infrared and collinear singularities  in the virtual corrections.
The renormalization scale dependence of $ R_{\text{quark}}(\s) $
gets reduced by more than a factor of 2. The calculation of the remaining
bremsstrahlung contributions (which are expected to be rather small) and
a more detailed numerical analysis are in progress \cite{AAGW:brems}.

Acknowledgements: C.G. would like to thank the members of the
Yerevan Physics Institute for the kind hospitality extended to him
when this paper was finalized. We thank K. Bieri and P. Liniger for 
helpful discussions.

%
\newpage
\begin{appendix}
\section{One-loop matrix elements of the four quark operators}
\label{appendix:oneloop}
In order to fix the counterterms
$F_{i \to \rm{4 quark}}^{\rm{ct}(7,9)}$ ($i=1,2$) in eq. (\ref{fi4quark}),
we need the one-loop matrix elements
$\bra s \ell^+ \ell^-|O_j|b \ket_{\text{1-loop}}$ of the four-quark
operators $O_1$, $O_2$, $O_4$, $O_{11}$ and $O_{12}$. Due to the
$1/\epsilon$ factor in eq. (\ref{fi4quark}), they are needed up
to $\Order\left(\epsilon^1\right)$. The explicit results (in expanded form) read
\begin{eqnarray}
\nonumber
\OTwoOneLoop & =& \left( \frac{\mu}{m_c}\right)^{2\epsilon} \,
\left[\frac{4}{9\,\epsilon}
+\frac{4}{2835}\left(-315+252 \left(\frac{\hat{s}}{4z}\right)+108
\left(\frac{\hat{s}}{4z}\right)^2+64\left(\frac{\hat{s}}{4z}\right)^3
\right)\right.\\
&+&\left.
\frac{\epsilon}{2835}\left(105\,\pi^2-1008\left(\frac{\hat{s}}{4z}\right)+128
\left(\frac{\hat{s}}{4z}\right)^3\right)\right] \,
\bra \widetilde{O}_9\ket_{\text{tree}} \, ,
\end{eqnarray}
\begin{equation}
   \OOneOneLoop = \frac{4}{3}\, \OTwoOneLoop \, ,
\end{equation}
\begin{eqnarray}
\nonumber
\OFourOneLoop & = &
    -\left(\frac{\mu}{m_b}\right)^{2\epsilon}\left\{
    \left[\frac{4}{9}+
    \frac{\epsilon}{945}
    \left(70\, \hat{s} + 7\, \hat{s}^2 + \hat{s}^3\right)\right] \,
    \bra \widetilde{O}_7\ket_{\text{tree}} \right. \nonumber \\
    & + & \left[ \frac{16}{27\,\epsilon} +
    \frac{2}{8505} \left( - 420 + 1260\, i\pi - 1260\, L_s +
    252\, \hat{s} + 27\, \hat{s}^2 + 4\, \hat{s}^3 \right) \right.\nonumber \\
\nonumber
    & + & \frac{4\,\epsilon }{8505}
    \left(420\, i\pi + 910 - 630\, i\, L_s\,\pi -
    420\, L_s - 315\,\pi^2 \right. \nonumber \\
    & + & \left. \left. \left. 315\,L_s^2 - 126\,\hat{s}+ \hat{s}^3
    \right)\right] \,
    \bra \widetilde{O}_9\ket_{\text{tree}} \right\} \, ,
\end{eqnarray}
\begin{equation}
\OEleven =-\frac{64}{27} \,
\left(\frac{\mu}{m_c}\right )^{2\epsilon}
\left[
1+\frac {4\,\epsilon}{5}
    \left (
     {\frac {\hat{s}}{4z}}+{\frac {3}{7}}\,\left({\frac {{\hat{s}}}{{
4z}}}\right)^2+{\frac {16}{63}}\,\left({\frac {{\hat{s}}}{{4z}}}\right)^3
\right)
\right]  \,
\bra \widetilde{O}_9\ket_{\text{tree}} \, ,
\end{equation}
\begin{equation}
    \OTwelve = \frac{3}{4}\, \OEleven \, .
\end{equation}
\newpage
\section{Full $\s$ and $\zz$ dependence of the form factors $F^{(7,9)}_{1,2}$}
\label{appendix:aaa}
In this appendix we give the dependence of $f^{(b)}_a$ ($a=1,2;~b=7,9$)
(see eq. (\ref{f1decomp})) on $\s$ and $z$. We decompose them as follows:
\[
    f^{(b)}_a = \sum_{i,j,l,m} \kappa^{(b)}_{a,ijlm} \, \s^i \, L^j_s \,
z^l \, L_c^m +
                \sum_{i,j} \rho^{(b)}_{a,ij} \, \s^i \, L^j_s \, .
\]
The quantities $\rho^{(b)}_{a,ij}$ collect the half-integer powers of $z =
m_c^2/m_b^2 = \hat{m}_c^2$. This way, the summation
indices in the above equation run over integers only.
On the following pages, we list the numerical values of
$\kappa^{(b)}_{a,ijlm}$ and $\rho^{(b)}_{a,ij}$ for
\[
    i=0,...,~3; \quad j=0,~1; \quad l=-3,...,~3
\quad\text{and}\quad m=0,...,~4 \, .
\]
Coefficients not explicitly mentioned below vanish.
\\
\hrule
\noindent
Coefficients $\kappa^{(9)}_{1,ijlm}$ and $\rho^{(9)}_{1,ij}$ for the
decomposition of $f_1^{(9)}$
\begin{align}
\rho^{(9)}_{1,00} & = 3.8991\,\hat{m}_c^{3}
& \rho^{(9)}_{1,10} & = -23.3946\,\hat{m}_c\nonumber\\
\rho^{(9)}_{1,20} & = -140.368\,\hat{m}_c
& \rho^{(9)}_{1,30} & = 7.79821\,\hat{m}_c^{-1}-319.726\,\hat{m}_c\nonumber
\end{align}
$\kappa^{(9)}_{1,00lm} = 
\begin{scriptsize}
\left(\begin{tabular}{rcl rcl rcl rcl rcl}
~ & 0 &~ ~ & ~ & 0 &~ ~ & ~ & 0 &~ ~ & ~ & 0 &~ ~ & ~ & 0 &~ \\
~ & 0 &~ ~ & ~ & 0 &~ ~ & ~ & 0 &~ ~ & ~ & 0 &~ ~ & ~ & 0 &~ \\
~ & 0 &~ ~ & ~ & 0 &~ ~ & ~ & 0 &~ ~ & ~ & 0 &~ ~ & ~ & 0 &~ \\
$-4.61812$ &$+$ & $3.67166\,i$~ & $5.62963$ &$+$ & $1.86168\,i$~ & ~ & 0 &~ ~ & ~ & 0 &~ ~ & ~ & 0 &~ \\
$14.4621$ &$-$ & $16.2155\,i$~ & $9.59321$ &$-$ & $11.1701\,i$~ & $-1.18519$ &$-$ & $7.44674\,i$~ & \multicolumn{3}{c}{~$\,-0.790123$}~ & ~ & 0 &~ \\
$-16.0864$ &$+$ & $26.7517\,i$~ & $54.2439$ &$-$ & $14.8935\,i$~ & $-15.4074$ &$-$ & $29.787\,i$~ & \multicolumn{3}{c}{~$\,-3.95062$}~ & ~ & 0 &~ \\
$-14.73$ &$-$ & $23.6892\,i$~ & $-28.5761$ &$+$ & $34.7514\,i$~ & \multicolumn{3}{c}{~$\,20.1481$}~ & ~ & 0 &~ ~ & ~ & 0 &~ \\
\end{tabular}\right)
\end{scriptsize}
$
\\\\\\
$\kappa^{(9)}_{1,01lm} = 
\begin{scriptsize}
\left(\begin{tabular}{rcl rcl rcl rcl rcl}
~ & 0 &~ ~ & ~ & 0 &~ ~ & ~ & 0 &~ ~ & ~ & 0 &~ ~ & ~ & 0 &~ \\
~ & 0 &~ ~ & ~ & 0 &~ ~ & ~ & 0 &~ ~ & ~ & 0 &~ ~ & ~ & 0 &~ \\
~ & 0 &~ ~ & ~ & 0 &~ ~ & ~ & 0 &~ ~ & ~ & 0 &~ ~ & ~ & 0 &~ \\
$-0.0493827$ &$-$ & $0.103427\,i$~ & ~ & 0 &~ ~ & ~ & 0 &~ ~ & ~ & 0 &~ ~ & ~ & 0 &~ \\
\multicolumn{3}{c}{~$\,-0.592593$}~ & ~ & 0 &~ ~ & ~ & 0 &~ ~ & ~ & 0 &~ ~ & ~ & 0 &~ \\
$4.95977$ &$-$ & $1.86168\,i$~ & $-1.18519$ &$-$ & $7.44674\,i$~ & \multicolumn{3}{c}{~$\,-2.37037$}~ & ~ & 0 &~ ~ & ~ & 0 &~ \\
$-9.20287$ &$-$ & $1.65483\,i$~ & $-1.0535$ &$+$ & $9.92898\,i$~ & \multicolumn{3}{c}{~$\,3.16049$}~ & ~ & 0 &~ ~ & ~ & 0 &~ \\
\end{tabular}\right)
\end{scriptsize}
$
\\\\\\
$\kappa^{(9)}_{1,10lm} = 
\begin{scriptsize}
\left(\begin{tabular}{rcl rcl rcl rcl rcl}
~ & 0 &~ ~ & ~ & 0 &~ ~ & ~ & 0 &~ ~ & ~ & 0 &~ ~ & ~ & 0 &~ \\
~ & 0 &~ ~ & ~ & 0 &~ ~ & ~ & 0 &~ ~ & ~ & 0 &~ ~ & ~ & 0 &~ \\
$-2.48507$ &$-$ & $0.186168\,i$~ & ~ & 0 &~ ~ & ~ & 0 &~ ~ & ~ & 0 &~ ~ & ~ & 0 &~ \\
$4.47441$ &$-$ & $0.310281\,i$~ & $1.48148$ &$-$ & $1.86168\,i$~ & ~ & 0 &~ ~ & ~ & 0 &~ ~ & ~ & 0 &~ \\
$71.3855$ &$-$ & $30.7987\,i$~ & $8.47677$ &$-$ & $33.5103\,i$~ & $12.5389$ &$-$ & $7.44674\,i$~ & \multicolumn{3}{c}{~$\,-0.790123$}~ & \multicolumn{3}{c}{~$\,0.790123$}\\
$-18.1301$ &$+$ & $66.1439\,i$~ & $149.596$ &$-$ & $67.0206\,i$~ & $-49.1852$ &$-$ & $81.9141\,i$~ & \multicolumn{3}{c}{~$\,-11.0617$}~ & ~ & 0 &~ \\
$-72.89$ &$-$ & $63.7828\,i$~ & $-68.135$ &$+$ & $134.041\,i$~ & \multicolumn{3}{c}{~$\,63.6049$}~ & ~ & 0 &~ ~ & ~ & 0 &~ \\
\end{tabular}\right)
\end{scriptsize}
$
\\\\\\
$\kappa^{(9)}_{1,11lm} = 
\begin{scriptsize}
\left(\begin{tabular}{rcl rcl rcl rcl rcl}
~ & 0 &~ ~ & ~ & 0 &~ ~ & ~ & 0 &~ ~ & ~ & 0 &~ ~ & ~ & 0 &~ \\
~ & 0 &~ ~ & ~ & 0 &~ ~ & ~ & 0 &~ ~ & ~ & 0 &~ ~ & ~ & 0 &~ \\
~ & 0 &~ ~ & ~ & 0 &~ ~ & ~ & 0 &~ ~ & ~ & 0 &~ ~ & ~ & 0 &~ \\
~ & 0 &~ ~ & ~ & 0 &~ ~ & ~ & 0 &~ ~ & ~ & 0 &~ ~ & ~ & 0 &~ \\
$-2.66667$ &$-$ & $1.86168\,i$~ & \multicolumn{3}{c}{~$\,-1.18519$}~ & ~ & 0 &~ ~ & ~ & 0 &~ ~ & ~ & 0 &~ \\
$18.6539$ &$-$ & $7.44674\,i$~ & $-4.74074$ &$-$ & $29.787\,i$~ & \multicolumn{3}{c}{~$\,-9.48148$}~ & ~ & 0 &~ ~ & ~ & 0 &~ \\
$-41.6104$ &$-$ & $3.72337\,i$~ & $-2.37037$ &$+$ & $44.6804\,i$~ & \multicolumn{3}{c}{~$\,14.2222$}~ & ~ & 0 &~ ~ & ~ & 0 &~ \\
\end{tabular}\right)
\end{scriptsize}
$
\\\\\\
$\kappa^{(9)}_{1,20lm} = 
\begin{scriptsize}
\left(\begin{tabular}{rcl rcl rcl rcl rcl}
~ & 0 &~ ~ & ~ & 0 &~ ~ & ~ & 0 &~ ~ & ~ & 0 &~ ~ & ~ & 0 &~ \\
$-0.403158$ &$-$ & $0.0199466\,i$~ & ~ & 0 &~ ~ & ~ & 0 &~ ~ & ~ & 0 &~ ~ & ~ & 0 &~ \\
$-0.0613169$ &$+$ & $0.0620562\,i$~ & ~ & 0 &~ ~ & ~ & 0 &~ ~ & ~ & 0 &~ ~ & ~ & 0 &~ \\
$37.1282$ &$-$ & $1.36524\,i$~ & $22.0621$ &$-$ & $1.86168\,i$~ & \multicolumn{3}{c}{~$\,5.33333$}~ & \multicolumn{3}{c}{~$\,0.790123$}~ & ~ & 0 &~ \\
$212.74$ &$-$ & $52.2081\,i$~ & $-21.9215$ &$-$ & $52.1272\,i$~ & $57.1724$ &$-$ & $7.44674\,i$~ & \multicolumn{3}{c}{~$\,-2.37037$}~ & \multicolumn{3}{c}{~$\,2.37037$}\\
$-44.6829$ &$+$ & $108.713\,i$~ & $272.015$ &$-$ & $163.828\,i$~ & $-119.111$ &$-$ & $156.382\,i$~ & \multicolumn{3}{c}{~$\,-21.3333$}~ & ~ & 0 &~ \\
$-137.203$ &$-$ & $106.832\,i$~ & $-99.437$ &$+$ & $330.139\,i$~ & \multicolumn{3}{c}{~$\,168.889$}~ & ~ & 0 &~ ~ & ~ & 0 &~ \\
\end{tabular}\right)
\end{scriptsize}
$
\\\\\\
$\kappa^{(9)}_{1,21lm} = 
\begin{scriptsize}
\left(\begin{tabular}{rcl rcl rcl rcl rcl}
~ & 0 &~ ~ & ~ & 0 &~ ~ & ~ & 0 &~ ~ & ~ & 0 &~ ~ & ~ & 0 &~ \\
~ & 0 &~ ~ & ~ & 0 &~ ~ & ~ & 0 &~ ~ & ~ & 0 &~ ~ & ~ & 0 &~ \\
~ & 0 &~ ~ & ~ & 0 &~ ~ & ~ & 0 &~ ~ & ~ & 0 &~ ~ & ~ & 0 &~ \\
\multicolumn{3}{c}{~$\,0.0164609$}~ & ~ & 0 &~ ~ & ~ & 0 &~ ~ & ~ & 0 &~ ~ & ~ & 0 &~ \\
$-5.33333$ &$-$ & $3.72337\,i$~ & \multicolumn{3}{c}{~$\,-2.37037$}~ & ~ & 0 &~ ~ & ~ & 0 &~ ~ & ~ & 0 &~ \\
$40.786$ &$-$ & $22.3402\,i$~ & $-14.2222$ &$-$ & $67.0206\,i$~ & \multicolumn{3}{c}{~$\,-21.3333$}~ & ~ & 0 &~ ~ & ~ & 0 &~ \\
\multicolumn{3}{c}{~$\,-111.356$}~ & \multicolumn{3}{c}{~$\,119.148\,i$}~ & \multicolumn{3}{c}{~$\,37.9259$}~ & ~ & 0 &~ ~ & ~ & 0 &~ \\
\end{tabular}\right)
\end{scriptsize}
$
\\\\\\
$\kappa^{(9)}_{1,30lm} = 
\begin{scriptsize}
\left(\begin{tabular}{rcl rcl rcl rcl rcl}
$-0.0759415$ &$-$ & $0.00295505\,i$~ & ~ & 0 &~ ~ & ~ & 0 &~ ~ & ~ & 0 &~ ~ & ~ & 0 &~ \\
$-0.00480894$ &$+$ & $0.00369382\,i$~ & ~ & 0 &~ ~ & ~ & 0 &~ ~ & ~ & 0 &~ ~ & ~ & 0 &~ \\
$-1.81002$ &$+$ & $0.0871741\,i$~ & \multicolumn{3}{c}{~$\,-0.919459$}~ & \multicolumn{3}{c}{~$\,-0.197531$}~ & ~ & 0 &~ ~ & ~ & 0 &~ \\
$79.7475$ &$-$ & $1.72206\,i$~ & $57.3171$ &$-$ & $1.86168\,i$~ & \multicolumn{3}{c}{~$\,11.2593$}~ & \multicolumn{3}{c}{~$\,2.37037$}~ & ~ & 0 &~ \\
$425.579$ &$-$ & $76.6479\,i$~ & $-68.8016$ &$-$ & $69.5029\,i$~ & $129.357$ &$-$ & $7.44674\,i$~ & \multicolumn{3}{c}{~$\,-5.53086$}~ & \multicolumn{3}{c}{~$\,4.74074$}\\
$-87.8946$ &$+$ & $148.481\,i$~ & $417.612$ &$-$ & $311.522\,i$~ & $-227.16$ &$-$ & $253.189\,i$~ & \multicolumn{3}{c}{~$\,-34.7654$}~ & ~ & 0 &~ \\
$-279.268$ &$-$ & $135.118\,i$~ & $-146.853$ &$+$ & $652.831\,i$~ & \multicolumn{3}{c}{~$\,331.259$}~ & ~ & 0 &~ ~ & ~ & 0 &~ \\
\end{tabular}\right)
\end{scriptsize}
$
\\\\\\
$\kappa^{(9)}_{1,31lm} = 
\begin{scriptsize}
\left(\begin{tabular}{rcl rcl rcl rcl rcl}
~ & 0 &~ ~ & ~ & 0 &~ ~ & ~ & 0 &~ ~ & ~ & 0 &~ ~ & ~ & 0 &~ \\
~ & 0 &~ ~ & ~ & 0 &~ ~ & ~ & 0 &~ ~ & ~ & 0 &~ ~ & ~ & 0 &~ \\
~ & 0 &~ ~ & ~ & 0 &~ ~ & ~ & 0 &~ ~ & ~ & 0 &~ ~ & ~ & 0 &~ \\
\multicolumn{3}{c}{~$\,0.0219479$}~ & ~ & 0 &~ ~ & ~ & 0 &~ ~ & ~ & 0 &~ ~ & ~ & 0 &~ \\
$-8.2963$ &$-$ & $5.58505\,i$~ & \multicolumn{3}{c}{~$\,-3.55556$}~ & ~ & 0 &~ ~ & ~ & 0 &~ ~ & ~ & 0 &~ \\
$70.2698$ &$-$ & $49.6449\,i$~ & $-31.6049$ &$-$ & $119.148\,i$~ & \multicolumn{3}{c}{~$\,-37.9259$}~ & ~ & 0 &~ ~ & ~ & 0 &~ \\
$-231.893$ &$+$ & $18.6168\,i$~ & $11.8519$ &$+$ & $248.225\,i$~ & \multicolumn{3}{c}{~$\,79.0123$}~ & ~ & 0 &~ ~ & ~ & 0 &~ \\
\end{tabular}\right)
\end{scriptsize}
$
\\\\\\

\hrule
\noindent
Coefficients $\kappa^{(7)}_{1,ijlm}$ and $\rho^{(7)}_{1,ij}$ for the
decomposition of $f_1^{(7)}$
\begin{align}\rho^{(7)}_{1,00} & = 1.94955\,\hat{m}_c^{3}
& \rho^{(7)}_{1,10} & = 11.6973\,\hat{m}_c\nonumber\\\rho^{(7)}_{1,20} & = 70.1839\,\hat{m}_c
& \rho^{(7)}_{1,30} & = -3.8991\,\hat{m}_c^{-1}+159.863\,\hat{m}_c\nonumber\end{align}
$\kappa^{(7)}_{1,00lm} =
\begin{scriptsize}
\left(\begin{tabular}{rcl rcll rcl rcl rcl}
~ & 0 &~ ~ & ~ & 0 &~ ~ & ~ & 0 &~ ~ & ~ & 0 &~ ~ & ~ & 0 &~ \\
~ & 0 &~ ~ & ~ & 0 &~ ~ & ~ & 0 &~ ~ & ~ & 0 &~ ~ & ~ & 0 &~ \\
~ & 0 &~ ~ & ~ & 0 &~ ~ & ~ & 0 &~ ~ & ~ & 0 &~ ~ & ~ & 0 &~ \\
$-1.14266$ &$-$ & $0.517135\,i$~ & ~ & 0 &~ ~ & ~ & 0 &~ ~ & ~ & 0 &~ ~ & ~ & 0 &~ \\
$-2.20356$ &$+$ & $1.59186\,i$~ & $-5.21743$ &$+$ & $1.86168\,i$~ & $0.592593$ &$+$ & $3.72337\,i$~ & \multicolumn{3}{c}{~$\,0.395062$}~ & ~ & 0 &~ \\
$1.86366$ &$-$ & $3.06235\,i$~ & \multicolumn{3}{c}{~$\,-4.66347$}~ & \multicolumn{3}{c}{~$\,3.72337\,i$}~ & \multicolumn{3}{c}{~$\,0.395062$}~ & ~ & 0 &~ \\
$-1.21131$ &$+$ & $2.89595\,i$~ & $2.99588$ &$-$ & $2.48225\,i$~ & \multicolumn{3}{c}{~$\,-4.14815$}~ & ~ & 0 &~ ~ & ~ & 0 &~ \\
\end{tabular}\right)
\end{scriptsize}
$
\\\\\\
$\kappa^{(7)}_{1,01lm} = 0$
\\\\\\
$\kappa^{(7)}_{1,10lm} =
\begin{scriptsize}
\left(\begin{tabular}{rcl rcll rcl rcl rcl}
~ & 0 &~ ~ & ~ & 0 &~ ~ & ~ & 0 &~ ~ & ~ & 0 &~ ~ & ~ & 0 &~ \\
~ & 0 &~ ~ & ~ & 0 &~ ~ & ~ & 0 &~ ~ & ~ & 0 &~ ~ & ~ & 0 &~ \\
~ & 0 &~ ~ & ~ & 0 &~ ~ & ~ & 0 &~ ~ & ~ & 0 &~ ~ & ~ & 0 &~ \\
$-2.07503$ &$+$ & $1.39626\,i$~ & $-0.444444$ &$+$ & $0.930842\,i$~ & ~ & 0 &~ ~ & ~ & 0 &~ ~ & ~ & 0 &~ \\
$-25.9259$ &$+$ & $5.78065\,i$~ & $-3.40101$ &$+$ & $13.0318\,i$~ & $-4.4917$ &$+$ & $3.72337\,i$~ & \multicolumn{3}{c}{~$\,0.395062$}~ & \multicolumn{3}{c}{~$\,-0.395062$}\\
$11.4229$ &$-$ & $15.2375\,i$~ & $-34.0806$ &$+$ & $11.1701\,i$~ & $10.3704$ &$+$ & $18.6168\,i$~ & \multicolumn{3}{c}{~$\,2.37037$}~ & ~ & 0 &~ \\
$11.7509$ &$+$ & $15.6984\,i$~ & $18.9564$ &$-$ & $24.8225\,i$~ & \multicolumn{3}{c}{~$\,-14.6173$}~ & ~ & 0 &~ ~ & ~ & 0 &~ \\
\end{tabular}\right)
\end{scriptsize}
$
\\\\\\
$\kappa^{(7)}_{1,11lm} =
\begin{scriptsize}
\left(\begin{tabular}{rcl rcll rcl rcl rcl}
~ & 0 &~ ~ & ~ & 0 &~ ~ & ~ & 0 &~ ~ & ~ & 0 &~ ~ & ~ & 0 &~ \\
~ & 0 &~ ~ & ~ & 0 &~ ~ & ~ & 0 &~ ~ & ~ & 0 &~ ~ & ~ & 0 &~ \\
~ & 0 &~ ~ & ~ & 0 &~ ~ & ~ & 0 &~ ~ & ~ & 0 &~ ~ & ~ & 0 &~ \\
\multicolumn{3}{c}{~$\,-0.0164609$}~ & ~ & 0 &~ ~ & ~ & 0 &~ ~ & ~ & 0 &~ ~ & ~ & 0 &~ \\
$1.03704$ &$+$ & $0.930842\,i$~ & \multicolumn{3}{c}{~$\,0.592593$}~ & ~ & 0 &~ ~ & ~ & 0 &~ ~ & ~ & 0 &~ \\
\multicolumn{3}{c}{~$\,-4.66347$}~ & \multicolumn{3}{c}{~$\,7.44674\,i$}~ & \multicolumn{3}{c}{~$\,2.37037$}~ & ~ & 0 &~ ~ & ~ & 0 &~ \\
$6.73754$ &$+$ & $1.86168\,i$~ & $1.18519$ &$-$ & $7.44674\,i$~ & \multicolumn{3}{c}{~$\,-2.37037$}~ & ~ & 0 &~ ~ & ~ & 0 &~ \\
\end{tabular}\right)
\end{scriptsize}
$
\\\\\\
$\kappa^{(7)}_{1,20lm} =
\begin{scriptsize}
\left(\begin{tabular}{rcl rcll rcl rcl rcl}
~ & 0 &~ ~ & ~ & 0 &~ ~ & ~ & 0 &~ ~ & ~ & 0 &~ ~ & ~ & 0 &~ \\
~ & 0 &~ ~ & ~ & 0 &~ ~ & ~ & 0 &~ ~ & ~ & 0 &~ ~ & ~ & 0 &~ \\
\multicolumn{3}{c}{~$\,0.00555556$}~ & ~ & 0 &~ ~ & ~ & 0 &~ ~ & ~ & 0 &~ ~ & ~ & 0 &~ \\
$-19.4691$ &$+$ & $1.59019\,i$~ & $-11.6779$ &$+$ & $0.930842\,i$~ & \multicolumn{3}{c}{~$\,-2.96296$}~ & \multicolumn{3}{c}{~$\,-0.395062$}~ & ~ & 0 &~ \\
$-90.4953$ &$+$ & $14.7788\,i$~ & $14.9329$ &$+$ & $22.3402\,i$~ & $-24.438$ &$+$ & $3.72337\,i$~ & \multicolumn{3}{c}{~$\,1.18519$}~ & \multicolumn{3}{c}{~$\,-1.18519$}\\
$23.8816$ &$-$ & $32.8021\,i$~ & $-82.7915$ &$+$ & $39.0954\,i$~ & $32.2963$ &$+$ & $44.6804\,i$~ & \multicolumn{3}{c}{~$\,5.92593$}~ & ~ & 0 &~ \\
$38.1415$ &$+$ & $34.8683\,i$~ & $38.6436$ &$-$ & $80.673\,i$~ & \multicolumn{3}{c}{~$\,-41.5802$}~ & ~ & 0 &~ ~ & ~ & 0 &~ \\
\end{tabular}\right)
\end{scriptsize}
$
\\\\\\
$\kappa^{(7)}_{1,21lm} =
\begin{scriptsize}
\left(\begin{tabular}{rcl rcll rcl rcl rcl}
~ & 0 &~ ~ & ~ & 0 &~ ~ & ~ & 0 &~ ~ & ~ & 0 &~ ~ & ~ & 0 &~ \\
~ & 0 &~ ~ & ~ & 0 &~ ~ & ~ & 0 &~ ~ & ~ & 0 &~ ~ & ~ & 0 &~ \\
~ & 0 &~ ~ & ~ & 0 &~ ~ & ~ & 0 &~ ~ & ~ & 0 &~ ~ & ~ & 0 &~ \\
\multicolumn{3}{c}{~$\,-0.0164609$}~ & ~ & 0 &~ ~ & ~ & 0 &~ ~ & ~ & 0 &~ ~ & ~ & 0 &~ \\
$2.37037$ &$+$ & $1.86168\,i$~ & \multicolumn{3}{c}{~$\,1.18519$}~ & ~ & 0 &~ ~ & ~ & 0 &~ ~ & ~ & 0 &~ \\
$-13.9904$ &$+$ & $3.72337\,i$~ & $2.37037$ &$+$ & $22.3402\,i$~ & \multicolumn{3}{c}{~$\,7.11111$}~ & ~ & 0 &~ ~ & ~ & 0 &~ \\
$27.5428$ &$+$ & $3.72337\,i$~ & $2.37037$ &$-$ & $29.787\,i$~ & \multicolumn{3}{c}{~$\,-9.48148$}~ & ~ & 0 &~ ~ & ~ & 0 &~ \\
\end{tabular}\right)
\end{scriptsize}
$
\\\\\\
$\kappa^{(7)}_{1,30lm} =
\begin{scriptsize}
\left(\begin{tabular}{rcl rcll rcl rcl rcl}
~ & 0 &~ ~ & ~ & 0 &~ ~ & ~ & 0 &~ ~ & ~ & 0 &~ ~ & ~ & 0 &~ \\
$-0.00010778$ &$+$ & $0.00258567\,i$~ & ~ & 0 &~ ~ & ~ & 0 &~ ~ & ~ & 0 &~ ~ & ~ & 0 &~ \\
$0.946811$ &$-$ & $0.0258567\,i$~ & \multicolumn{3}{c}{~$\,0.488889$}~ & \multicolumn{3}{c}{~$\,0.0987654$}~ & ~ & 0 &~ ~ & ~ & 0 &~ \\
$-41.9952$ &$+$ & $1.63673\,i$~ & $-30.2091$ &$+$ & $0.930842\,i$~ & \multicolumn{3}{c}{~$\,-6.22222$}~ & \multicolumn{3}{c}{~$\,-1.18519$}~ & ~ & 0 &~ \\
$-189.354$ &$+$ & $25.8196\,i$~ & $42.6566$ &$+$ & $31.0281\,i$~ & $-57.765$ &$+$ & $3.72337\,i$~ & \multicolumn{3}{c}{~$\,2.76543$}~ & \multicolumn{3}{c}{~$\,-2.37037$}\\
$45.1784$ &$-$ & $52.4207\,i$~ & $-145.181$ &$+$ & $88.7403\,i$~ & $70.9136$ &$+$ & $81.9141\,i$~ & \multicolumn{3}{c}{~$\,11.0617$}~ & ~ & 0 &~ \\
$77.3602$ &$+$ & $54.2499\,i$~ & $58.4491$ &$-$ & $184.927\,i$~ & \multicolumn{3}{c}{~$\,-96.0988$}~ & ~ & 0 &~ ~ & ~ & 0 &~ \\
\end{tabular}\right)
\end{scriptsize}
$
\\\\\\
$\kappa^{(7)}_{1,31lm} =
\begin{scriptsize}
\left(\begin{tabular}{rcl rcll rcl rcl rcl}
~ & 0 &~ ~ & ~ & 0 &~ ~ & ~ & 0 &~ ~ & ~ & 0 &~ ~ & ~ & 0 &~ \\
~ & 0 &~ ~ & ~ & 0 &~ ~ & ~ & 0 &~ ~ & ~ & 0 &~ ~ & ~ & 0 &~ \\
~ & 0 &~ ~ & ~ & 0 &~ ~ & ~ & 0 &~ ~ & ~ & 0 &~ ~ & ~ & 0 &~ \\
\multicolumn{3}{c}{~$\,-0.0164609$}~ & ~ & 0 &~ ~ & ~ & 0 &~ ~ & ~ & 0 &~ ~ & ~ & 0 &~ \\
$3.85185$ &$+$ & $2.79253\,i$~ & \multicolumn{3}{c}{~$\,1.77778$}~ & ~ & 0 &~ ~ & ~ & 0 &~ ~ & ~ & 0 &~ \\
$-27.3882$ &$+$ & $13.0318\,i$~ & $8.2963$ &$+$ & $44.6804\,i$~ & \multicolumn{3}{c}{~$\,14.2222$}~ & ~ & 0 &~ ~ & ~ & 0 &~ \\
$69.4495$ &$+$ & $1.86168\,i$~ & $1.18519$ &$-$ & $74.4674\,i$~ & \multicolumn{3}{c}{~$\,-23.7037$}~ & ~ & 0 &~ ~ & ~ & 0 &~ \\
\end{tabular}\right)
\end{scriptsize}
$
\\\\\\

\hrule
\noindent
Coefficients $\kappa^{(9)}_{2,ijlm}$ and $\rho^{(9)}_{2,ij}$ for the
decomposition of $f_2^{(9)}$
\begin{align}\rho^{(9)}_{2,00} & = -23.3946\,\hat{m}_c^{3}
& \rho^{(9)}_{2,10} & = 140.368\,\hat{m}_c\nonumber\\\rho^{(9)}_{2,20} & = 842.206\,\hat{m}_c
& \rho^{(9)}_{2,30} & = -46.7892\,\hat{m}_c^{-1}+1918.36\,\hat{m}_c\nonumber\end{align}
$\kappa^{(9)}_{2,00lm} = 
\begin{scriptsize}
\left(\begin{tabular}{rcl rcl rcl rcl rcl}
~ & 0 &~ ~ & ~ & 0 &~ ~ & ~ & 0 &~ ~ & ~ & 0 &~ ~ & ~ & 0 &~ \\
~ & 0 &~ ~ & ~ & 0 &~ ~ & ~ & 0 &~ ~ & ~ & 0 &~ ~ & ~ & 0 &~ \\
~ & 0 &~ ~ & ~ & 0 &~ ~ & ~ & 0 &~ ~ & ~ & 0 &~ ~ & ~ & 0 &~ \\
$-24.2913$ &$-$ & $22.0299\,i$~ & $-23.1111$ &$-$ & $11.1701\,i$~ & ~ & 0 &~ ~ & ~ & 0 &~ ~ & ~ & 0 &~ \\
$-86.7723$ &$+$ & $97.2931\,i$~ & $-57.5593$ &$+$ & $67.0206\,i$~ & $7.11111$ &$+$ & $44.6804\,i$~ & \multicolumn{3}{c}{~$\,4.74074$}~ & ~ & 0 &~ \\
$96.5187$ &$-$ & $160.51\,i$~ & $-325.463$ &$+$ & $89.3609\,i$~ & $92.4444$ &$+$ & $178.722\,i$~ & \multicolumn{3}{c}{~$\,23.7037$}~ & ~ & 0 &~ \\
$88.3801$ &$+$ & $142.135\,i$~ & $171.457$ &$-$ & $208.509\,i$~ & \multicolumn{3}{c}{~$\,-120.889$}~ & ~ & 0 &~ ~ & ~ & 0 &~ \\
\end{tabular}\right)
\end{scriptsize}
$
\\\\\\
$\kappa^{(9)}_{2,01lm} = 
\begin{scriptsize}
\left(\begin{tabular}{rcl rcl rcl rcl rcl}
~ & 0 &~ ~ & ~ & 0 &~ ~ & ~ & 0 &~ ~ & ~ & 0 &~ ~ & ~ & 0 &~ \\
~ & 0 &~ ~ & ~ & 0 &~ ~ & ~ & 0 &~ ~ & ~ & 0 &~ ~ & ~ & 0 &~ \\
~ & 0 &~ ~ & ~ & 0 &~ ~ & ~ & 0 &~ ~ & ~ & 0 &~ ~ & ~ & 0 &~ \\
$0.296296$ &$+$ & $0.620562\,i$~ & ~ & 0 &~ ~ & ~ & 0 &~ ~ & ~ & 0 &~ ~ & ~ & 0 &~ \\
\multicolumn{3}{c}{~$\,3.55556$}~ & ~ & 0 &~ ~ & ~ & 0 &~ ~ & ~ & 0 &~ ~ & ~ & 0 &~ \\
$-29.7586$ &$+$ & $11.1701\,i$~ & $7.11111$ &$+$ & $44.6804\,i$~ & \multicolumn{3}{c}{~$\,14.2222$}~ & ~ & 0 &~ ~ & ~ & 0 &~ \\
$55.2172$ &$+$ & $9.92898\,i$~ & $6.32099$ &$-$ & $59.5739\,i$~ & \multicolumn{3}{c}{~$\,-18.963$}~ & ~ & 0 &~ ~ & ~ & 0 &~ \\
\end{tabular}\right)
\end{scriptsize}
$
\\\\\\
$\kappa^{(9)}_{2,10lm} = 
\begin{scriptsize}
\left(\begin{tabular}{rcl rcl rcl rcl rcl}
~ & 0 &~ ~ & ~ & 0 &~ ~ & ~ & 0 &~ ~ & ~ & 0 &~ ~ & ~ & 0 &~ \\
~ & 0 &~ ~ & ~ & 0 &~ ~ & ~ & 0 &~ ~ & ~ & 0 &~ ~ & ~ & 0 &~ \\
$0.8462$ &$+$ & $1.11701\,i$~ & ~ & 0 &~ ~ & ~ & 0 &~ ~ & ~ & 0 &~ ~ & ~ & 0 &~ \\
$-26.8464$ &$+$ & $1.86168\,i$~ & $-8.88889$ &$+$ & $11.1701\,i$~ & ~ & 0 &~ ~ & ~ & 0 &~ ~ & ~ & 0 &~ \\
$-428.313$ &$+$ & $184.792\,i$~ & $-50.8606$ &$+$ & $201.062\,i$~ & $-75.2337$ &$+$ & $44.6804\,i$~ & \multicolumn{3}{c}{~$\,4.74074$}~ & \multicolumn{3}{c}{~$\,-4.74074$}\\
$108.781$ &$-$ & $396.864\,i$~ & $-897.575$ &$+$ & $402.124\,i$~ & $295.111$ &$+$ & $491.485\,i$~ & \multicolumn{3}{c}{~$\,66.3704$}~ & ~ & 0 &~ \\
$437.34$ &$+$ & $382.697\,i$~ & $408.81$ &$-$ & $804.248\,i$~ & \multicolumn{3}{c}{~$\,-381.63$}~ & ~ & 0 &~ ~ & ~ & 0 &~ \\
\end{tabular}\right)
\end{scriptsize}
$
\\\\\\
$\kappa^{(9)}_{2,11lm} = 
\begin{scriptsize}
\left(\begin{tabular}{rcl rcl rcl rcl rcl}
~ & 0 &~ ~ & ~ & 0 &~ ~ & ~ & 0 &~ ~ & ~ & 0 &~ ~ & ~ & 0 &~ \\
~ & 0 &~ ~ & ~ & 0 &~ ~ & ~ & 0 &~ ~ & ~ & 0 &~ ~ & ~ & 0 &~ \\
~ & 0 &~ ~ & ~ & 0 &~ ~ & ~ & 0 &~ ~ & ~ & 0 &~ ~ & ~ & 0 &~ \\
~ & 0 &~ ~ & ~ & 0 &~ ~ & ~ & 0 &~ ~ & ~ & 0 &~ ~ & ~ & 0 &~ \\
$16.$ &$+$ & $11.1701\,i$~ & \multicolumn{3}{c}{~$\,7.11111$}~ & ~ & 0 &~ ~ & ~ & 0 &~ ~ & ~ & 0 &~ \\
$-111.923$ &$+$ & $44.6804\,i$~ & $28.4444$ &$+$ & $178.722\,i$~ & \multicolumn{3}{c}{~$\,56.8889$}~ & ~ & 0 &~ ~ & ~ & 0 &~ \\
$249.663$ &$+$ & $22.3402\,i$~ & $14.2222$ &$-$ & $268.083\,i$~ & \multicolumn{3}{c}{~$\,-85.3333$}~ & ~ & 0 &~ ~ & ~ & 0 &~ \\
\end{tabular}\right)
\end{scriptsize}
$
\\\\\\
$\kappa^{(9)}_{2,20lm} = 
\begin{scriptsize}
\left(\begin{tabular}{rcl rcl rcl rcl rcl}
~ & 0 &~ ~ & ~ & 0 &~ ~ & ~ & 0 &~ ~ & ~ & 0 &~ ~ & ~ & 0 &~ \\
$-0.0132191$ &$+$ & $0.11968\,i$~ & ~ & 0 &~ ~ & ~ & 0 &~ ~ & ~ & 0 &~ ~ & ~ & 0 &~ \\
$0.367901$ &$-$ & $0.372337\,i$~ & ~ & 0 &~ ~ & ~ & 0 &~ ~ & ~ & 0 &~ ~ & ~ & 0 &~ \\
$-222.769$ &$+$ & $8.19141\,i$~ & $-132.372$ &$+$ & $11.1701\,i$~ & \multicolumn{3}{c}{~$\,-32.$}~ & \multicolumn{3}{c}{~$\,-4.74074$}~ & ~ & 0 &~ \\
$-1276.44$ &$+$ & $313.249\,i$~ & $131.529$ &$+$ & $312.763\,i$~ & $-343.034$ &$+$ & $44.6804\,i$~ & \multicolumn{3}{c}{~$\,14.2222$}~ & \multicolumn{3}{c}{~$\,-14.2222$}\\
$268.098$ &$-$ & $652.279\,i$~ & $-1632.09$ &$+$ & $982.969\,i$~ & $714.667$ &$+$ & $938.289\,i$~ & \multicolumn{3}{c}{~$\,128.$}~ & ~ & 0 &~ \\
$823.218$ &$+$ & $640.989\,i$~ & $596.622$ &$-$ & $1980.83\,i$~ & \multicolumn{3}{c}{~$\,-1013.33$}~ & ~ & 0 &~ ~ & ~ & 0 &~ \\
\end{tabular}\right)
\end{scriptsize}
$
\\\\\\
$\kappa^{(9)}_{2,21lm} = 
\begin{scriptsize}
\left(\begin{tabular}{rcl rcl rcl rcl rcl}
~ & 0 &~ ~ & ~ & 0 &~ ~ & ~ & 0 &~ ~ & ~ & 0 &~ ~ & ~ & 0 &~ \\
~ & 0 &~ ~ & ~ & 0 &~ ~ & ~ & 0 &~ ~ & ~ & 0 &~ ~ & ~ & 0 &~ \\
~ & 0 &~ ~ & ~ & 0 &~ ~ & ~ & 0 &~ ~ & ~ & 0 &~ ~ & ~ & 0 &~ \\
\multicolumn{3}{c}{~$\,-0.0987654$}~ & ~ & 0 &~ ~ & ~ & 0 &~ ~ & ~ & 0 &~ ~ & ~ & 0 &~ \\
$32.$ &$+$ & $22.3402\,i$~ & \multicolumn{3}{c}{~$\,14.2222$}~ & ~ & 0 &~ ~ & ~ & 0 &~ ~ & ~ & 0 &~ \\
$-244.716$ &$+$ & $134.041\,i$~ & $85.3333$ &$+$ & $402.124\,i$~ & \multicolumn{3}{c}{~$\,128.$}~ & ~ & 0 &~ ~ & ~ & 0 &~ \\
\multicolumn{3}{c}{~$\,668.137$}~ & \multicolumn{3}{c}{~$\,-714.887\,i$}~ & \multicolumn{3}{c}{~$\,-227.556$}~ & ~ & 0 &~ ~ & ~ & 0 &~ \\
\end{tabular}\right)
\end{scriptsize}
$
\\\\\\
$\kappa^{(9)}_{2,30lm} = 
\begin{scriptsize}
\left(\begin{tabular}{rcl rcl rcl rcl rcl}
$-0.0142243$ &$+$ & $0.0177303\,i$~ & ~ & 0 &~ ~ & ~ & 0 &~ ~ & ~ & 0 &~ ~ & ~ & 0 &~ \\
$0.0288536$ &$-$ & $0.0221629\,i$~ & ~ & 0 &~ ~ & ~ & 0 &~ ~ & ~ & 0 &~ ~ & ~ & 0 &~ \\
$10.8601$ &$-$ & $0.523045\,i$~ & \multicolumn{3}{c}{~$\,5.51675$}~ & \multicolumn{3}{c}{~$\,1.18519$}~ & ~ & 0 &~ ~ & ~ & 0 &~ \\
$-478.485$ &$+$ & $10.3323\,i$~ & $-343.902$ &$+$ & $11.1701\,i$~ & \multicolumn{3}{c}{~$\,-67.5556$}~ & \multicolumn{3}{c}{~$\,-14.2222$}~ & ~ & 0 &~ \\
$-2553.47$ &$+$ & $459.887\,i$~ & $412.809$ &$+$ & $417.017\,i$~ & $-776.143$ &$+$ & $44.6804\,i$~ & \multicolumn{3}{c}{~$\,33.1852$}~ & \multicolumn{3}{c}{~$\,-28.4444$}\\
$527.368$ &$-$ & $890.889\,i$~ & $-2505.67$ &$+$ & $1869.13\,i$~ & $1362.96$ &$+$ & $1519.13\,i$~ & \multicolumn{3}{c}{~$\,208.593$}~ & ~ & 0 &~ \\
$1675.61$ &$+$ & $810.709\,i$~ & $881.117$ &$-$ & $3916.98\,i$~ & \multicolumn{3}{c}{~$\,-1987.56$}~ & ~ & 0 &~ ~ & ~ & 0 &~ \\
\end{tabular}\right)
\end{scriptsize}
$
\\\\\\
$\kappa^{(9)}_{2,31lm} = 
\begin{scriptsize}
\left(\begin{tabular}{rcl rcl rcl rcl rcl}
~ & 0 &~ ~ & ~ & 0 &~ ~ & ~ & 0 &~ ~ & ~ & 0 &~ ~ & ~ & 0 &~ \\
~ & 0 &~ ~ & ~ & 0 &~ ~ & ~ & 0 &~ ~ & ~ & 0 &~ ~ & ~ & 0 &~ \\
~ & 0 &~ ~ & ~ & 0 &~ ~ & ~ & 0 &~ ~ & ~ & 0 &~ ~ & ~ & 0 &~ \\
\multicolumn{3}{c}{~$\,-0.131687$}~ & ~ & 0 &~ ~ & ~ & 0 &~ ~ & ~ & 0 &~ ~ & ~ & 0 &~ \\
$49.7778$ &$+$ & $33.5103\,i$~ & \multicolumn{3}{c}{~$\,21.3333$}~ & ~ & 0 &~ ~ & ~ & 0 &~ ~ & ~ & 0 &~ \\
$-421.619$ &$+$ & $297.87\,i$~ & $189.63$ &$+$ & $714.887\,i$~ & \multicolumn{3}{c}{~$\,227.556$}~ & ~ & 0 &~ ~ & ~ & 0 &~ \\
$1391.36$ &$-$ & $111.701\,i$~ & $-71.1111$ &$-$ & $1489.35\,i$~ & \multicolumn{3}{c}{~$\,-474.074$}~ & ~ & 0 &~ ~ & ~ & 0 &~ \\
\end{tabular}\right)
\end{scriptsize}
$
\\\\\\

\hrule
\noindent
Coefficients $\kappa^{(7)}_{2,ijlm}$ and $\rho^{(7)}_{2,ij}$ for the
decomposition of $f_2^{(7)}$
\begin{align}\rho^{(7)}_{2,00} = & -11.6973\,\hat{m}_c^{3}
& \rho^{(7)}_{2,10} = & -70.1839\,\hat{m}_c\nonumber\\\rho^{(7)}_{2,20} = & -421.103\,\hat{m}_c
& \rho^{(7)}_{2,30} = & 23.3946\,\hat{m}_c^{-1}-959.179\,\hat{m}_c\nonumber\end{align}
$\kappa^{(7)}_{2,00lm} =
\begin{scriptsize}
\left(\begin{tabular}{rcl rcl rcl rcl rcl}
~ & 0 &~ ~ & ~ & 0 &~ ~ & ~ & 0 &~ ~ & ~ & 0 &~ ~ & ~ & 0 &~ \\
~ & 0 &~ ~ & ~ & 0 &~ ~ & ~ & 0 &~ ~ & ~ & 0 &~ ~ & ~ & 0 &~ \\
~ & 0 &~ ~ & ~ & 0 &~ ~ & ~ & 0 &~ ~ & ~ & 0 &~ ~ & ~ & 0 &~ \\
$6.85597$ &$+$ & $3.10281\,i$~ & ~ & 0 &~ ~ & ~ & 0 &~ ~ & ~ & 0 &~ ~ & ~ & 0 &~ \\
$13.2214$ &$-$ & $9.55118\,i$~ & $31.3046$ &$-$ & $11.1701\,i$~ & $-3.55556$ &$-$ & $22.3402\,i$~ & \multicolumn{3}{c}{~$\,-2.37037$}~ & ~ & 0 &~ \\
$-11.182$ &$+$ & $18.3741\,i$~ & \multicolumn{3}{c}{~$\,27.9808$}~ & \multicolumn{3}{c}{~$\,-22.3402\,i$}~ & \multicolumn{3}{c}{~$\,-2.37037$}~ & ~ & 0 &~ \\
$7.26787$ &$-$ & $17.3757\,i$~ & $-17.9753$ &$+$ & $14.8935\,i$~ & \multicolumn{3}{c}{~$\,24.8889$}~ & ~ & 0 &~ ~ & ~ & 0 &~ \\
\end{tabular}\right)
\end{scriptsize}
$
\\\\\\
$\kappa^{(7)}_{2,01lm} = 0$
\\\\\\
$\kappa^{(7)}_{2,10lm} =
\begin{scriptsize}
\left(\begin{tabular}{rcl rcl rcl rcl rcl}
~ & 0 &~ ~ & ~ & 0 &~ ~ & ~ & 0 &~ ~ & ~ & 0 &~ ~ & ~ & 0 &~ \\
~ & 0 &~ ~ & ~ & 0 &~ ~ & ~ & 0 &~ ~ & ~ & 0 &~ ~ & ~ & 0 &~ \\
~ & 0 &~ ~ & ~ & 0 &~ ~ & ~ & 0 &~ ~ & ~ & 0 &~ ~ & ~ & 0 &~ \\
$12.4502$ &$-$ & $8.37758\,i$~ & $2.66667$ &$-$ & $5.58505\,i$~ & ~ & 0 &~ ~ & ~ & 0 &~ ~ & ~ & 0 &~ \\
$155.555$ &$-$ & $34.6839\,i$~ & $20.4061$ &$-$ & $78.1908\,i$~ & $26.9502$ &$-$ & $22.3402\,i$~ & \multicolumn{3}{c}{~$\,-2.37037$}~ & \multicolumn{3}{c}{~$\,2.37037$}\\
$-68.5374$ &$+$ & $91.4251\,i$~ & $204.484$ &$-$ & $67.0206\,i$~ & $-62.2222$ &$-$ & $111.701\,i$~ & \multicolumn{3}{c}{~$\,-14.2222$}~ & ~ & 0 &~ \\
$-70.5057$ &$-$ & $94.1903\,i$~ & $-113.738$ &$+$ & $148.935\,i$~ & \multicolumn{3}{c}{~$\,87.7037$}~ & ~ & 0 &~ ~ & ~ & 0 &~ \\
\end{tabular}\right)
\end{scriptsize}
$
\\\\\\
$\kappa^{(7)}_{2,11lm} =
\begin{scriptsize}
\left(\begin{tabular}{rcl rcl rcl rcl rcl}
~ & 0 &~ ~ & ~ & 0 &~ ~ & ~ & 0 &~ ~ & ~ & 0 &~ ~ & ~ & 0 &~ \\
~ & 0 &~ ~ & ~ & 0 &~ ~ & ~ & 0 &~ ~ & ~ & 0 &~ ~ & ~ & 0 &~ \\
~ & 0 &~ ~ & ~ & 0 &~ ~ & ~ & 0 &~ ~ & ~ & 0 &~ ~ & ~ & 0 &~ \\
\multicolumn{3}{c}{~$\,0.0987654$}~ & ~ & 0 &~ ~ & ~ & 0 &~ ~ & ~ & 0 &~ ~ & ~ & 0 &~ \\
$-6.22222$ &$-$ & $5.58505\,i$~ & \multicolumn{3}{c}{~$\,-3.55556$}~ & ~ & 0 &~ ~ & ~ & 0 &~ ~ & ~ & 0 &~ \\
\multicolumn{3}{c}{~$\,27.9808$}~ & \multicolumn{3}{c}{~$\,-44.6804\,i$}~ & \multicolumn{3}{c}{~$\,-14.2222$}~ & ~ & 0 &~ ~ & ~ & 0 &~ \\
$-40.4253$ &$-$ & $11.1701\,i$~ & $-7.11111$ &$+$ & $44.6804\,i$~ & \multicolumn{3}{c}{~$\,14.2222$}~ & ~ & 0 &~ ~ & ~ & 0 &~ \\
\end{tabular}\right)
\end{scriptsize}
$
\\\\\\
$\kappa^{(7)}_{2,20lm} =
\begin{scriptsize}
\left(\begin{tabular}{rcl rcl rcl rcl rcl}
~ & 0 &~ ~ & ~ & 0 &~ ~ & ~ & 0 &~ ~ & ~ & 0 &~ ~ & ~ & 0 &~ \\
~ & 0 &~ ~ & ~ & 0 &~ ~ & ~ & 0 &~ ~ & ~ & 0 &~ ~ & ~ & 0 &~ \\
\multicolumn{3}{c}{~$\,-0.0333333$}~ & ~ & 0 &~ ~ & ~ & 0 &~ ~ & ~ & 0 &~ ~ & ~ & 0 &~ \\
$116.815$ &$-$ & $9.54113\,i$~ & $70.0677$ &$-$ & $5.58505\,i$~ & \multicolumn{3}{c}{~$\,17.7778$}~ & \multicolumn{3}{c}{~$\,2.37037$}~ & ~ & 0 &~ \\
$542.972$ &$-$ & $88.6728\,i$~ & $-89.5971$ &$-$ & $134.041\,i$~ & $146.628$ &$-$ & $22.3402\,i$~ & \multicolumn{3}{c}{~$\,-7.11111$}~ & \multicolumn{3}{c}{~$\,7.11111$}\\
$-143.29$ &$+$ & $196.813\,i$~ & $496.749$ &$-$ & $234.572\,i$~ & $-193.778$ &$-$ & $268.083\,i$~ & \multicolumn{3}{c}{~$\,-35.5556$}~ & ~ & 0 &~ \\
$-228.849$ &$-$ & $209.21\,i$~ & $-231.862$ &$+$ & $484.038\,i$~ & \multicolumn{3}{c}{~$\,249.481$}~ & ~ & 0 &~ ~ & ~ & 0 &~ \\
\end{tabular}\right)
\end{scriptsize}
$
\\\\\\
$\kappa^{(7)}_{2,21lm} =
\begin{scriptsize}
\left(\begin{tabular}{rcl rcl rcl rcl rcl}
~ & 0 &~ ~ & ~ & 0 &~ ~ & ~ & 0 &~ ~ & ~ & 0 &~ ~ & ~ & 0 &~ \\
~ & 0 &~ ~ & ~ & 0 &~ ~ & ~ & 0 &~ ~ & ~ & 0 &~ ~ & ~ & 0 &~ \\
~ & 0 &~ ~ & ~ & 0 &~ ~ & ~ & 0 &~ ~ & ~ & 0 &~ ~ & ~ & 0 &~ \\
\multicolumn{3}{c}{~$\,0.0987654$}~ & ~ & 0 &~ ~ & ~ & 0 &~ ~ & ~ & 0 &~ ~ & ~ & 0 &~ \\
$-14.2222$ &$-$ & $11.1701\,i$~ & \multicolumn{3}{c}{~$\,-7.11111$}~ & ~ & 0 &~ ~ & ~ & 0 &~ ~ & ~ & 0 &~ \\
$83.9424$ &$-$ & $22.3402\,i$~ & $-14.2222$ &$-$ & $134.041\,i$~ & \multicolumn{3}{c}{~$\,-42.6667$}~ & ~ & 0 &~ ~ & ~ & 0 &~ \\
$-165.257$ &$-$ & $22.3402\,i$~ & $-14.2222$ &$+$ & $178.722\,i$~ & \multicolumn{3}{c}{~$\,56.8889$}~ & ~ & 0 &~ ~ & ~ & 0 &~ \\
\end{tabular}\right)
\end{scriptsize}
$
\\\\\\
$\kappa^{(7)}_{2,30lm} =
\begin{scriptsize}
\left(\begin{tabular}{rcl rcl rcl rcl rcl}
~ & 0 &~ ~ & ~ & 0 &~ ~ & ~ & 0 &~ ~ & ~ & 0 &~ ~ & ~ & 0 &~ \\
$0.000646678$ &$-$ & $0.015514\,i$~ & ~ & 0 &~ ~ & ~ & 0 &~ ~ & ~ & 0 &~ ~ & ~ & 0 &~ \\
$-5.68087$ &$+$ & $0.15514\,i$~ & \multicolumn{3}{c}{~$\,-2.93333$}~ & \multicolumn{3}{c}{~$\,-0.592593$}~ & ~ & 0 &~ ~ & ~ & 0 &~ \\
$251.971$ &$-$ & $9.82039\,i$~ & $181.255$ &$-$ & $5.58505\,i$~ & \multicolumn{3}{c}{~$\,37.3333$}~ & \multicolumn{3}{c}{~$\,7.11111$}~ & ~ & 0 &~ \\
$1136.13$ &$-$ & $154.918\,i$~ & $-255.94$ &$-$ & $186.168\,i$~ & $346.59$ &$-$ & $22.3402\,i$~ & \multicolumn{3}{c}{~$\,-16.5926$}~ & \multicolumn{3}{c}{~$\,14.2222$}\\
$-271.07$ &$+$ & $314.524\,i$~ & $871.089$ &$-$ & $532.442\,i$~ & $-425.481$ &$-$ & $491.485\,i$~ & \multicolumn{3}{c}{~$\,-66.3704$}~ & ~ & 0 &~ \\
$-464.161$ &$-$ & $325.499\,i$~ & $-350.695$ &$+$ & $1109.56\,i$~ & \multicolumn{3}{c}{~$\,576.593$}~ & ~ & 0 &~ ~ & ~ & 0 &~ \\
\end{tabular}\right)
\end{scriptsize}
$
\\\\\\
$\kappa^{(7)}_{2,31lm} =
\begin{scriptsize}
\left(\begin{tabular}{rcl rcl rcl rcl rcl}
~ & 0 &~ ~ & ~ & 0 &~ ~ & ~ & 0 &~ ~ & ~ & 0 &~ ~ & ~ & 0 &~ \\
~ & 0 &~ ~ & ~ & 0 &~ ~ & ~ & 0 &~ ~ & ~ & 0 &~ ~ & ~ & 0 &~ \\
~ & 0 &~ ~ & ~ & 0 &~ ~ & ~ & 0 &~ ~ & ~ & 0 &~ ~ & ~ & 0 &~ \\
\multicolumn{3}{c}{~$\,0.0987654$}~ & ~ & 0 &~ ~ & ~ & 0 &~ ~ & ~ & 0 &~ ~ & ~ & 0 &~ \\
$-23.1111$ &$-$ & $16.7552\,i$~ & \multicolumn{3}{c}{~$\,-10.6667$}~ & ~ & 0 &~ ~ & ~ & 0 &~ ~ & ~ & 0 &~ \\
$164.329$ &$-$ & $78.1908\,i$~ & $-49.7778$ &$-$ & $268.083\,i$~ & \multicolumn{3}{c}{~$\,-85.3333$}~ & ~ & 0 &~ ~ & ~ & 0 &~ \\
$-416.697$ &$-$ & $11.1701\,i$~ & $-7.11111$ &$+$ & $446.804\,i$~ & \multicolumn{3}{c}{~$\,142.222$}~ & ~ & 0 &~ ~ & ~ & 0 &~ \\
\end{tabular}\right)
\end{scriptsize}
$
\\\\\\

\end{appendix}

\end{document}